\documentclass[twoside,onecolumn]{article}

\usepackage[backend=bibtex,style=numeric,sorting=none]{biblatex}
\usepackage[ruled]{algorithm}
\usepackage[english]{babel}
\usepackage{algpseudocode}
\usepackage{subcaption}
\usepackage{blindtext}
\usepackage{amsfonts}
\usepackage{csquotes}
\usepackage{graphicx}
\usepackage{hyperref}
\usepackage{amsmath}
\usepackage{titling}
\usepackage{float}
\usepackage{bm}

\usepackage[hmarginratio=1:1,left=20mm,right=20mm,top=20mm,bottom=20mm,columnsep=20pt]{geometry}

\usepackage{lettrine}
\usepackage{enumitem}
\setlist[itemize]{noitemsep}

\usepackage{siunitx}
\usepackage{abstract}

\usepackage{titlesec}
\renewcommand\thesection{\arabic{section}}
\renewcommand\thesubsection{\arabic{subsection}}
\titleformat{\section}[block]{\bfseries}{\thesection.}{1em}{}
\titleformat{\subsection}[block]{}{\thesection.\thesubsection.}{1em}{}

\newcommand{\comment}[1]{}

\pretitle{\begin{center}\LARGE\bfseries}
\posttitle{\end{center}}
\title{Application of the Newton Time-Extracting Wavelet Transform as a chirp filter}
\author{\large Alejandro Silva$^\text{1,2}$\comment{\thanks{A thank you or further information}} \\[1ex]
\normalsize $^\text{1}$Department of Applied Mathematics \\
\normalsize Escuela Técnica Superior de Ingenieros Industriales, Universidad Politécnica de Madrid \\
\normalsize Calle de José Gutiérrez Abascal, 2, 28006, Madrid, Spain \\
\normalsize $^\text{2}$Department of Mechanical Engineering, Politecnico di Milano \\
\normalsize Via Privata Giuseppe La Masa, 1, 20156, Milano MI, Italy \\
\normalsize Email: \href{Mailto:alejandro.silva@upm.es}{alejandro.silva@upm.es}
}
\date{}
\renewcommand{\maketitlehookd}{
\begin{abstract}
\noindent The problem of detecting chirps is present in many applications of Signal Processing. Proper denoising, which involves filtering the signals after their acquisition, improves the efficacy of their detection. This manuscript describes how a recently-published method of Time-Frequency Analysis (TFA) with reassignment, namely the Newton Time-Extracting Wavelet Transform (NTEWT), can be employed as a highly-performing chirp filter. The proposed methodology has the advantage of denoising chirps without distorting their instantaneous phases, as linear convolutional filters do. Numerical experiments have proven the efficacy of the proposed filter. After NTEWT-based filtering, the resolution of chirp detection with matched filtering is notably improved, even when the signals contain white noise. The computation times of the proposed numerical implementation of the NTEWT are lower than those reported in its seminar paper.
\end{abstract}
\textbf{Keywords: }Time-frequency analysis, Time reassignment operator, Newton time-extracting wavelet transform, Chirps, Filtering, Denoising
}

\begin{document}

\maketitle

\section{Introduction}

Linear and nonlinear Frequency Modulation Chirps (FMCs) are commonly defined as a composition of a smooth and positive instantaneous amplitude and an instantaneous phase, provided that the instantaneous amplitude evolves \textit{slowly} with respect to the oscillations of the instantaneous phase \cite{flandrin2002chirps}. They are found in a myriad of real systems and technical applications such as radar \cite{klauder1960theory}, sonar \cite{schock1990some}, laser amplification \cite{maine1988generation,delfyett2012chirped}, electroencephalograms \cite{feltane2013analyzing} and gravitational waves \cite{chassande1999gravitational}. The problem of chirp detection is present in all of these applications. As chirps appear as curves in the Time-Frequency (TF) domain, a TF domain-based methodology turns out to be the ideal approach for chirp detection in signals \cite{gonccalves1997time,chassande1999time,flandrin2001time,morvidone2003time}. There are many formulations available for TFA of non-stationary signals. The most popular linear TFA methods are the Short-Time Fourier Transform, the Wavelet Transform (WA) and the Chirplet Transform \cite{mallat1999wavelet,mann1995chirplet}. Nonlinear TFA methods include the Local Polynomial Transform \cite{li2011local} and the Matched Demodulation Transform \cite{wang2014matching}. These methods are relatively simple to implement and provide results that are easy to interpret, however their resolution is bounded by the Heisenberg uncertainty principle. This limitation can be bypassed by applying bilinear (i.e. quadratic) TFA methods \cite{hlawatsch1992linear}, of which the Wigner-Ville Distribution (WVD) is the most fundamental. In practice, the WVD offers excellent resolution for mono-component signals at the cost of a very blurred representation due to interference terms, especially in multi-component signals. More generally, the problem of choosing a TFA method boils down to finding a trade-off between the resolution of the Time-Frequency Representations (TFRs) and the attenuation of the interferences between signal components at their crossing points.

The Reassignment Method (RM) \cite{kodera1976new,kodera1978analysis,auger1995improving} has been developed to improve the resolution of the individual signal components in the TFRs without compromising their resolution at the crossing points. This algorithm redistributes the energy of the TFR around its ridges in both time and frequency directions. The locus of ridge points calculated in the frequency direction give estimation of the Instantaneous Frequencies (IFs) of the signal components, while the ridge points obtained in the time direction approximate the Group Delays (GDs) of its components. As a result the reassignment method leads to enhanced TFR resolution. Unfortunately, the RM is not invertible, hence the signal cannot be reconstructed from the reassigned TFR. A solution to this problem is to perform reassignment in either the time or the frequency direction (but not in both). For time-varying signals (signals whose components are horizontal or almost horizontal on their TFR, due to low-varying IFs), reassignment in the frequency direction, which is based on estimation of the IFs of the signal components, is the best choice. Methods that implement frequency-direction reassignment are the Synchrosqueezing Wavelet Transform \cite{daubechies2011synchrosqueezed}, the Synchroextracting Transform \cite{yu2017synchroextracting}, the Matched Synchrosqueezing Wavelet Transform \cite{wang2018matching}, and the Self-Matched Extracting Wavelet Transform \cite{li2022self}. Instead, reassignment in the time direction, based on GD estimation, works better for frequency-varying signals (those whose components are vertical or almost vertical on the TFR due to fast-varying IFs). Examples of time-direction reassignment algorithms are the Time-Reassigned Synchrosqueezing Transform \cite{he2019time}, the Transient-Extracting Transform \cite{yu2019concentrated}, the Time-Reassigned Synchrosqueezing Wavelet Transform \cite{li2022theoretical} and the Time-Extracting Wavelet Transform \cite{li2023time}. It is known that under certain conditions the GD and the IF trajectories of a signal component coincide up to a small tolerance factor. If the instantaneous amplitude varies slowly with respect to the oscillations of the instantaneous phase the stationary phase principle guarantees that the signal's IFs are a good approximation of the first derivative of the instantaneous phases of their components. If this condition holds (as in FMCs) and the bandwidth-time product is large enough then its GD trajectory is a good estimate of its instantaneous phase and vice versa \cite{boashash1992estimatingI}.

A paper recently published by Li et al \cite{li2023newton} has proposed a novel time-reassignment method named Newton Time-Extracting Wavelet Transform (NTEWT), which combines the WT with a reassignment step in the time direction. It was formulated to improve the TFR resolution of signals with weakly-separated \textit{frequency-varying linear chirps} (FVLCs): components that behave locally in the frequency domain as linear chirps. Impulsive signals in the time domain, such as chirp bursts, are examples of such signals. Estimation of the signal components' ridge points is based on the calculation of the fixed points of a time-reassignment operator. Once the ridge points are estimated, the TFR energy is reassigned in the time direction around those points, improving the resolution of the TFR and allowing for a more accurate GD estimation of FVLCs. Moreover, it is invertible for such signals: given the reassigned TFA they can be reconstructed as a sum of FVLCs. The fact that the NTEWT fixed point operator reminds the formula for the Newton's iteration method inspires the algorithm's name.

Since the fixed point reassignment operator provides GD trajectory estimation only for signals holding the theoretical assumptions that characterize LVFCs, it was hypothesized that this discriminating property of the NTEWT reassignment operator could enable its application as a signal filter for highly-impulsive signals in the time domain. Given the aforementioned importance of FMCs in so many Signal Processing fields, particularly short-length chirp bursts, we studied the application of the NTEWT algorithm as a filter for such signals. The results are promising, and the proposed filtering methodology performs well even if the input signals are contaminated with low or moderate white noise. It is shown that, unlike linear convolutional filters, NTEWT-based filtering does not distort the instantaneous phases of the signal components, nor causes any time delay between the input and the output (save for the filter's execution time). As a result, the efficacy of FMC detection with a matched filter is greatly improved when the signals are pre-processed with the proposed NTEWT-based method. Efficient algorithms for the numerical calculation of NTEWT scalograms and for NTEWT-based filtering are presented, and the real-time applicability of these algorithms is studied. The calculation times reported in this manuscript are much lower that those reported by Li et al \cite{li2023newton}, and further algorithmic improvements would pave the way to the filter's application on high-performing radar and sonar systems, where ultra-high sampling frequencies and chirp rates are required for a good range and resolution.

The manuscript is structured as follows: Section~\ref{Matandmeth} presents the continuous formulation of the NTEWT as originally presented by Li et al \cite{li2023newton} and provides a discrete formulation of the NTEWT and efficient algorithms for its numerical implementation. Section~\ref{numexpers} shows the validation of NTEWT-based chirp filtering on a number of synthetic signals and reports filtering times and bounding sampling frequencies for real-time processing as a function of signal length. The manuscript ends up with the conclusions (Section~\ref{concls}).

\vspace{5pt}

\section{Material and methods}\label{Matandmeth}

\subsection{Continuous formulation of the NTEWT}

According to the notation and conventions of \cite{li2023newton}, the following definition of the WT of a function $x(t)\in L_2(\mathbb{R})$ is used:

\begin{equation}
	\label{eq:wt}
	W_x^\psi(b,a)=a^{-1}\int_{-\infty}^\infty x(t)\psi^*\bigg(\frac{t-b}{a}\bigg)~dt,
\end{equation}

\vspace{10pt}

where $\psi(t)$ is an admissible wavelet in $L_2(\mathbb{R})$ satisfying the admissibility condition \cite{mallat1999wavelet}, and $a\in \mathbb{R}^+$ and $b\in \mathbb{R}$ are scale and time parameters, respectively. $*$ represents complex conjugation.

\clearpage

The WT outputs a TFR of the analyzed signal, $W_x^\psi$: a two-dimensional function that assigns a complex value to each real tuple $(b,a)$. Application of the Plancherel's theorem allows to re-write Eq.~\ref{eq:wt} as an integral in the frequency domain:

\vspace{-15pt}

\begin{equation}
	\label{eq:plancherelwt}
	\begin{split}
		W_x^\psi(b,a)=(2\pi)^{-1}\int_{-\infty}^\infty\hat{x}(\omega)\hat{\psi}^*(a\omega)e^{i\omega b}~d\omega=\\
		=(2\pi)^{-1}\mathcal{F}^{-1}(\hat{x}(\omega)\hat{\psi}^*(a\omega))(b),
	\end{split}
\end{equation}

where $\hat{x}$ and $\hat{\psi}$ are the Fourier Transforms (FTs) of $x$ and $\psi$, and $\mathcal{F}^{-1}$ denotes the inverse FT.

Given the WT of a signal $x$, its NTEWT is defined as \cite{li2022theoretical}:

\vspace{-10pt}

\begin{equation}
	\label{eq:4}
	NTe(b,a)=W_x^\psi(b,a)\cdot\delta(b-\bar{t}_x(b,a)),
\end{equation}

where $\delta$ denotes the Kronecker delta function. Therefore, the NTEWT-reassigned TFR has non-vanishing terms only at points holding the condition $b=\bar{t}_x(b,a)$: the so-called \textit{fixed points} of the TFR. \cite{li2022theoretical} shows that the locus of those points form the GD trajectories of FVLC components. The Newton GD (NGD) estimator $\bar{t}_x(b,a)$ is calculated as

\vspace{-10pt}

\begin{equation}
	\label{eq:ngd}
	\bar{t}_x(b,a)=b-\frac{b-\tilde{t}_x(b,a)}{1-\partial_b\tilde{t}_x(b,a)},
\end{equation}

where $\partial_b$ denotes partial derivation with respect to $b$ and $\tilde{t}_x(b,a)$ is a complex time reassignment operator given by the formula

\vspace{-10pt}

\begin{equation}
	\label{eq:ctro}
	\tilde{t}_x(b,a)=b+a\frac{W_x^{t\psi}(b,a)}{W_x^\psi(b,a)}.
\end{equation}

By the properties of the Fourier Transform, the FT of $t\psi$ ($t\in\mathbb{R}$) is $iD_\omega\hat{\psi}$, with $D_\omega$ denoting derivation with respect to the frequency $\omega$. Again, applying the Plancherel's theorem we have

\vspace{-5pt}

\begin{equation}
	\label{eq:plancherelwtt}
	W_x^{t\psi}(b,a)=(2\pi)^{-1}\mathcal{F}^{-1}(i\hat{x}(\omega)D_\omega\hat{\psi}^*(a\omega))(b),
\end{equation}

and differentiating Eq.~\ref{eq:ctro} with respect to $b$ we get

\vspace{-8pt}

\begin{equation}
	%\label{eq:4}
	\partial_b\tilde{t}_x(b,a)=1+a\frac{(\partial_bW_x^{t\psi})W_x^\psi-W_x^{t\psi}(\partial_bW_x^\psi)}{(W_x^\psi)^2}.
\end{equation}

Partial derivatives of the TFRs $W_x^\psi$ and $W_x^{t\psi}$ with respect to $b$ follow from the properties of the FT:

\vspace{-10pt}

\begin{align}
	%\label{eq:4}
	&\partial_bW_x^\psi(b,a)=(2\pi)^{-1}\frac{\partial}{\partial b}\int_{-\infty}^\infty\hat{x}(\omega)\hat{\psi}^*(a\omega)e^{i\omega b}~d\omega=\nonumber\\
	&=(2\pi)^{-1}\mathcal{F}^{-1}(i\omega\hat{x}(\omega)\hat{\psi}^*(a\omega))(b)~\text{and}\\
	&\partial_bW_x^{t\psi}(b,a)=(2\pi)^{-1}\mathcal{F}^{-1}(-\omega\hat{x}(\omega)D_\omega\hat{\psi}^*(a\omega))(b).
\end{align}

\vspace{5pt}

Li et al \cite{li2023newton} proves theoretically the equivalence between TFR fixed points and GD trajectories of FVLCs. One such function is the Morlet wavelet, defined in the frequency domain as

\vspace{-3pt}

\begin{equation}
	\label{eq:mwavspect}
	\hat{\psi}(a\omega)=(4\pi\sigma^2)^{\frac{1}{4}}e^{-\sigma^2(a\omega-\omega_\psi)^2/2},
\end{equation}

\vspace{3pt}

with $a$ being the scale parameter. Its derivative with respect to $\omega$ is given by

\vspace{-10pt}

\begin{equation}
	%\label{eq:4}
	\begin{split}
		D_{\omega}\hat{\psi}(a\omega)=-a\sigma^5(a\omega-\omega_\psi)\hat{\psi}(a\omega).
	\end{split}
\end{equation}

\subsection{Discrete formulation and numerical implementation of the NTEWT}

The numerical calculation of a signal's NTEWT requires its discretization. Let $x[j]$ be a discrete signal with $n$ samples and $j=0,1,\ldots,n-1$. The discrete FT and inverse discrete FT are defined as

\vspace{-5pt}

\begin{align}
	%\label{eq:4}
	&\hat{x}[k]=\mathcal{F}(x[j])=\sum_{j=0}^{n-1}x[j]W_n^{jk}~\text{and}\\ &x[j]=\mathcal{F}^{-1}(\hat{x}[k])=\frac{1}{n}\sum_{k=0}^{n-1}\hat{x}[k]W_n^{-kj},\label{eq:idft}
\end{align}

\vspace{5pt}

where $W_n=e^{-2\pi i/n}$ and $j$, $k=0,\ldots,n-1$.

A discrete version of the NTEWT must be defined to deal with discrete signals. Let $W_x^\psi[j,k]$ be a two-dimensional array containing the discrete wavelet TFR of $x[j]$, with $j=0,\ldots,n-1$ and $k=0,\ldots,n/2-1$. Scaling and time parameters $a[k]=1/(k+1)$ and $b[j]=j/n$ are associated to each array element $[j,k]$ and the vector of frequencies $\omega[l]=2\pi l$, $l=0\ldots,n-1$ is defined. Then, the Plancherel's theorem can be applied to calculate $W_x^\psi$ row-wise as in Eq.~\ref{eq:plancherelwt}. For each value of $k$, the vector $\hat{W}_x^\psi[\cdot,k]$ is evaluated in the frequency domain and subsequently transformed to the time domain:

\vspace{-5pt}

\begin{align}
	&\hat{W}_x^\psi[\cdot,k]=(\hat{x}[l]\hat{\psi}^*(a[k]\omega[l]))_{l=0}^{n-1}~\text{and}\\
	&W_x^\psi[\cdot,k]=\mathcal{F}^{-1}(\hat{W}_x^\psi[\cdot,k]),%\label{eq:3}
\end{align}

\vspace{5pt}

where $(\hat{\psi}^*(a[k]\omega[l]))_{l=0}^{n-1}$ is a vector of length $n$ obtained by sampling $\hat{\psi}(a\omega)$ (Eq.~\ref{eq:mwavspect}) at $a\omega=a[k]\omega[l]$, $l=0,\ldots,n-1$. These two steps are repeated iteratively from $k=0$ to $n/2-1$ until the whole TFR of $x[j]$ is calculated.

Given a discrete wavelet TFR, the signal $x[j]$ can be reconstructed coefficient-wise (up to a scaling constant) using the following expression:

\vspace{-5pt}

\begin{equation}
	\label{eq:wtrec}
	(x_{\text{rec}})_{l=0}^{n-1}=2\text{Re}\bigg(\sum_{j=0}^{n-1}\sum_{k=1}^{n/2}a[k]W_x^\psi[j,k]\psi[j,k]\bigg),
\end{equation}

\vspace{5pt}

where $(x_{\text{rec}})_{l=0}^{n-1}$ is the reconstructed discrete signal with length $n$ and $l=0,\ldots,n-1$, $W_x^\psi[j,k]$ denotes each wavelet coefficient, and $\psi[j,k]$ is the sampled time-domain Morlet wavelet, dilated and translated according to the parameters $a[k]$ and $b[j]$. It is obtained by taking the inverse discrete FT (Eq.~\ref{eq:idft}) of the vector $\hat{\psi}(0,a[k])=\hat{\psi}(a[k]\omega[l]),~l=0,\ldots,n-1$ (Eq.~\ref{eq:mwavspect}) and then shifting the result leftwards by $j$ positions. The $\text{Re}(\cdot)$ operator takes the real part of each vector element.

Discrete versions of the NGD estimator (Eq.~\ref{eq:ngd}) and the complex time reassignment operator (Eq.~\ref{eq:ctro}) are

\vspace{-5pt}

\begin{align}
	%\label{eq:3}
	&\bar{t}_x[j,k]=b[k]-\frac{b[k]-\tilde{t}_x[j,k]}{1-\partial_b\tilde{t}_x[j,k]}~\text{and}\\
	&\tilde{t}_x[j,k]=b[k]+a[j]\frac{W_x^{t\psi}[j,k]}{W_x^\psi[j,k]},%\label{eq:3}
\end{align}

\vspace{5pt}

where

\vspace{-10pt}

\begin{align}
	%\label{eq:4}
	&W_x^\psi[j,k]=W_x^\psi(b[j],a[k])~\text{and}\\ &W_x^{t\psi}[j,k]=W_x^{t\psi}(b[j],a[k]).
\end{align}

\vspace{10pt}

With all the aforementioned definitions, the NTEWT of a discrete signal $x[j]$ can be defined as

\vspace{5pt}

\begin{equation}
	%\label{eq:4}
	NTe[j,k]=W_x^\psi[j,k]\cdot\delta(|b[j]-\bar{t}_x[j,k]|),
\end{equation}

where the Kronecker delta function $\delta$ is substituted with a function that takes the value one when the module of $t$ is smaller than a predefined tolerance $\epsilon$:

\vspace{-5pt}

\begin{align}
	%\label{eq:4}
	&\delta(t)=\begin{cases}
			1 & \text{if $|t|<\epsilon$}, \\
			0 & \text{otherwise}.
		\end{cases}
\end{align}

The formula for signal reconstruction given its NTEWT is analogous to the WT reconstruction formula (Eq.~\ref{eq:wtrec}) and is given by

\vspace{-10pt}

\begin{equation}
	\label{eq:xfil}
	(x_{\text{fil}})_{l=0}^{n-1}=2\text{Re}\bigg(\sum_{j=0}^{n-1}\sum_{k=1}^{n/2}a[k]NTe[j,k]\psi[j,k]\bigg).
\end{equation}

Theoretically, $x_{\text{fil}}(t)=x_{\text{rec}}(t)$ (up to a scaling constant) if the signal $x(t)$ is expressible as a sum of weakly-separated FVLCs \cite{li2023newton}. Otherwise, as it will be seen in the following section, $x_{\text{fil}}(t)$ contains the components of $x(t)$ behaving as FVLCs.

A proposed algorithm for the calculation of the NTEWT of a discrete signal (and as a byproduct, its CWT) is shown in Algorithm~\ref{alg:one}. The input parameters are the Morlet wavelet's width parameter $\sigma$ and central frequency $\omega_\psi$ and the Kronecker delta's tolerance $\epsilon$. Algorithm~\ref{alg:two} shows the algorithm for NTEWT-based filtering, which adds to the row-wise calculation of the NTEWT a nested loop for the calculation of the reconstructed (filtered) signal $x_{\text{fil}}[j]$ (Eq.~\ref{eq:xfil}).

\begin{algorithm}[ht]
	\caption{Calculation of CWT's and NTEWT's time-frequency representations}\label{alg:one}
	\begin{algorithmic}[1]
		\State \textbf{Input:}
		\Statex $\quad x[j]\in\mathbb{R}^n$, $j=0,\ldots,n-1$,
		\Statex $\quad\sigma>0$, $\omega_\psi>0$, $\epsilon>0$
		\State \textbf{Output:}
		\Statex $\quad NTe[j,k]$, $j=0,\ldots,n-1$, $k=0,\ldots,n/2-1$
		\State Set $a[k] \gets 1/(k+1),~k=0\ldots,n/2-1$
		\State Set $b[j] \gets j/n,~j=0\ldots,n-1$
		\State Set $\omega[l] \gets 2\pi l,~l=0\ldots,n-1$
		\State Initialize $NTe[j,k]$. $j=0,\ldots,n-1$,
		\Statex $\quad k=0,\ldots,n/2-1$
		\State Calculate DFT/FFT of $x[j]$: $\hat{x}[j]$
		\For{$k=0$ \text{\normalfont\textbf{to}} $n/2-1$}
			\State Sample $\hat{\psi}(a[k]\omega[l])$ and $D_\omega\hat{\psi}(a[k]\omega[l])$,
			\Statex $\quad\quad~~~l=0,\ldots,n-1$;
			\State Calculate $W_x^\psi[\cdot,k]$ and $\bar{t}_x[\cdot,k]$
			\For{$j=0$ \text{\normalfont\textbf{to}} $n-1$}
				\If{$|\bar{t}_x[j,k]-b[j]|<\epsilon$}
					\State $NTe[j,k] \gets W_x^\psi[j,k]$
				\EndIf
			\EndFor
			\State $NTe[\cdot,k] \gets NTe[\cdot,k]\cdot\frac{||W_x^\psi[\cdot,k]||_2}{||NTe[\cdot,k]||_2}$
		\EndFor
	\end{algorithmic}
\end{algorithm}

\begin{algorithm}[ht]
	\caption{An NTEWT-based filter for $x[j]$}\label{alg:two}
	\begin{algorithmic}[1]
		\State \textbf{Input:}
		\Statex $\quad x[j]\in\mathbb{R}^n$, $j=0,\ldots,n-1$,
		\Statex $\quad\sigma>0$, $\omega_\psi>0$, $\epsilon>0$
		\State \textbf{Output:}
		\Statex $\quad x_{\text{fil}}[j]$, $j=0,\ldots,n-1$
		\State Set $a[k] \gets 1/(k+1),~k=0\ldots,n/2-1$
		\State Set $b[j] \gets j/n,~j=0\ldots,n-1$
		\State Set $\omega[l] \gets 2\pi l,~l=0\ldots,n-1$
		\State Initialize $NTe[j,k]$, $x_{\text{fil}}[j]$, $j=0,\ldots,n-1$,
		\Statex $\quad j=0,\ldots,n/2-1$
		\State Calculate DFT/FFT of $x[j]$: $\hat{x}[j]$
		\For{$k=0$ \text{\normalfont\textbf{to}} $n/2-1$}
			\State Sample $\hat{\psi}(a[k]\omega[l])$ and $D_\omega\hat{\psi}(a[k]\omega[l])$,
			\Statex $\quad\quad~~~l=0,\ldots,n-1$
			\State Calculate IDFT/IFFT of $\hat{\psi}$, $\psi[0,k]$
			\State Calculate $W_x^\psi[\cdot,k]$ and $\bar{t}_x[\cdot,k]$
			\For{$j=0$ \text{\normalfont\textbf{to}} $n-1$}
				\If{$|\bar{t}_x[j,k]-b[j]|<\epsilon$}
					\State $NTe[j,k] \gets W_x^\psi[j,k]$
				\EndIf
			\EndFor
			\State $NTe[\cdot,k] \gets NTe[\cdot,k]\cdot\frac{||W_x^\psi[\cdot,k]||_2}{||NTe[\cdot,k]||_2}$
			\For{$j=0$ \text{\normalfont\textbf{to}} $n-1$}
				\If{$NTe[j,k]\neq 0$}
					\State Circular shift of $\psi$ by $j$ positions:
					\Statex $\quad\quad\quad\quad\quad~~\psi[0,k]\Rightarrow\psi[j,k]$
					\State $x_{\text{fil}}[j] \gets x_{\text{fil}}[j]+a[k]W_x^\psi[j,k]\psi[j,k]$
				\EndIf
			\EndFor
		\EndFor
		\State $x_{\text{fil}}[j] \gets 2\text{Re}(x_{\text{fil}}[j])$, $j=0,\ldots,n-1$
	\end{algorithmic}
\end{algorithm}

\section{Numerical experiments}\label{numexpers}

All the numerical tests described in this section have been carried out on Matlab R2023a in an ASUS ZenBook UX325UA G3 with a 64-bit AMD Ryzen 7 5700U processor (1.8 GHz) and 16 GB of RAM. Algorithms~\ref{alg:one} and \ref{alg:two} were implemented as Matlab functions in the homonimous programming language. The Morlet wavelet's central frequency was set to $\omega_\psi=6$~rad/s in all of the experiments. A sampling frequency parameter $f_\text{s}$ was set to 180~kHz. Discrete signals were generated sample by sample according to the following formula:

\begin{equation}
	x[j]=x_{\text{chirp}}[j]+\sum\cos(2\pi f_i\cdot j/f_\text{s})+\mathcal{N}(s),
\end{equation}

where $f_i$ are frequencies of stationary harmonics added as deterministic noise, and $\mathcal{N}(s)$ represents random Gaussian noise with standard deviation $s$. A linear chirp was added to the discrete signal by sampling from the following expression:

\vspace{-5pt}

\begin{equation}
	x_{\text{chirp}}(t)=A(t)\sin(\phi(t)),
\end{equation}

where $t\in[0,T]$, being $T$ the chirp length, $A(t)$ is an instantaneous amplitude function and $\phi(t)$ is the instantaneous phase. In linear chirps, the instantaneous phase is quadratic and reads

\vspace{-5pt}

\begin{equation}
	\phi(t)=2\pi tf_{c,1}+\pi t^2(f_{c,2}-f_{c,1})/T,
\end{equation}

where $f_{c,1}$ and $f_{c,2}$ are the bounding instantaneous chirp frequencies.

The \textit{slow evolution} property that defines chirp signals ensures that the first time derivative of the instantaneous phase, $\phi^\prime(t)$, approximates well the signal's IFs \cite{boashash1992estimatingI}. This property holds easily by simply setting a constant amplitude $A(t)=1$. Additionally, large bandwidth-time products are required for the analyzed linear chirps so that the GD trajectories provided by the NTEWT's fixed point operator accurately estimates the chirps' IFs, and therefore, their instantaneous phases in the time domain. Since in many practical applications such as radar and sonar short signal periods are a key for performance, a large bandwidth is necessary to maximize the efficacy of the proposed NTEWT-based filter.

\subsection{Test 1: long signal containing a long-duration, low-rate chirp pulse mixed with stationary harmonics}

\begin{figure}[ht]
	\centering
	\includegraphics[width=0.45\textwidth]{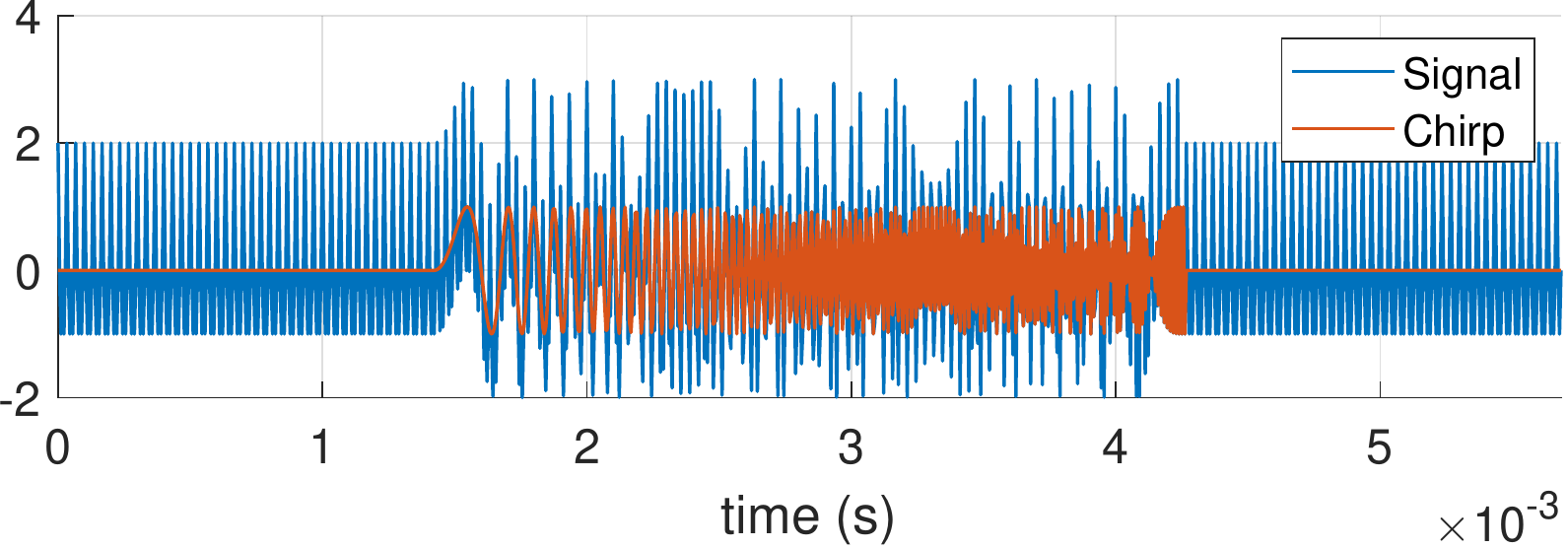}
	\caption{(Test 1) The chirp pulse (in red) plus the stationary harmonics (in blue).}
	\label{Test1Signal}
\end{figure}

A discrete signal was generated with length $n=1024$ samples ($\approx$ 0.00569 s), two stationary harmonics with frequencies $30$~kHz and $60$~kHz and no random noise ($s=0$). The parameters of the chirp signal $x_{chirp}[j]$ are shown in the table below. Both $x[j]$ and $x_{chirp}[j]$ are depicted in Fig.~\ref{Test1Signal}.

\begin{table}[H]
	\centering
	\begin{tabular}{ll}
		\hline
		Frequency lower bound (kHz) & 0                                    \\
		Frequency upper bound (kHz) & 90                                   \\
		Pulse length (samples)      & 512 ($\approx$ 0.02844 s) \\ \hline
	\end{tabular}
\end{table}

Fig.~\ref{Test1NTEWT} shows CWT and NTEWT \textit{scalograms} of the signal $x[j]$ (i.e. the modules of their TFRs) for three different values of the Morlet wavelet's width $\sigma$. The NTEWT reassignment tolerance parameter $\epsilon$ was set to an optimal value of \num{1e-3} during these computations. As expected, lower values of $\sigma$ produce scalograms with better time resolution and worse frequency resolution, especially at the highest frequencies, while higher values of $\sigma$ have the opposite effect: scalograms with better frequency resolution and worse time resolution, mostly at lower frequencies. NTEWT reassignment consists in zeroing out all CWT coefficients except those that are \textit{approximately} fixed points of the TFR: points holding the bound $|b-\bar{t}_x(b,a)|<\epsilon$. Then, the TFR is re-scaled frequency-wise so that the two-norm of the NTEWT scalogram's $k$-th row coincides with the two-norm of the CWT scalogram's $k$-th row (see Algorithms~\ref{alg:one} and \ref{alg:two}).

\begin{figure}[H]
	\centering
	\includegraphics[width=0.45\textwidth]{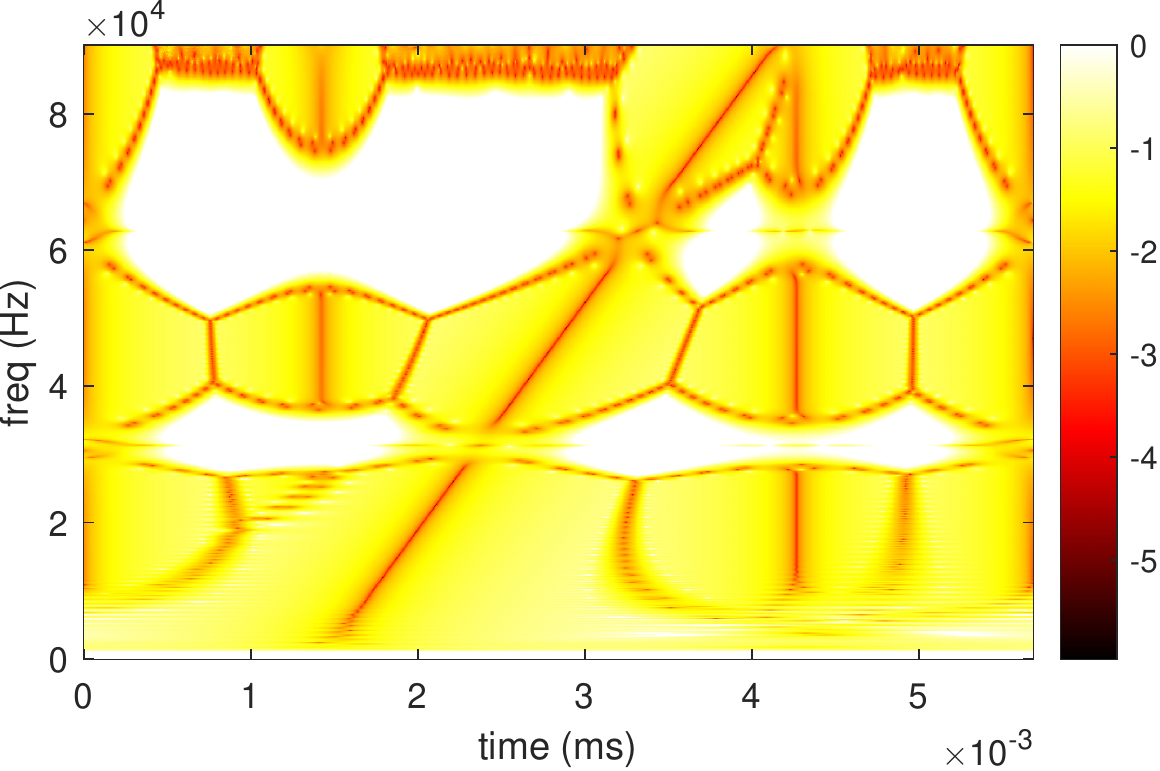}
	\caption{(Test 1) The log of the fixed point metric of the CWT of $x[j]$, $|\bar{t}_x[j,k]-b[j]|$, with $\sigma=5$.}
	\label{Test1fixed}
\end{figure}

It was observed that both the NTEWT scalogram's resolution and the efficacy of the NTEWT reassignment improve under high values of the wavelet width $\sigma$: Fig.~\ref{Test1NTEWTf} ($\sigma=5$) shows less crossing-terms interference and slightly higher resolution than Fig.~\ref{Test1NTEWTd} ($\sigma=3$). Fig.~\ref{Test1fixed} depicts the values of the fixed point metric $|\bar{t}_x[j,k]-b[j]|$ across the NTEWT scalogram with $\sigma=5$. The discriminating features of this metric are evident. Even if it cannot be exactly zero at discrete points of the TFR, there exist some $\epsilon$ such that the TFR points satisfying $|\bar{t}_x[j,k]-b[j]|$ belong or are close to the chirp's GD trajectory. The value of this tolerance parameter should be properly fine-tuned in Algorithms~\ref{alg:one} and \ref{alg:two} so that most of the TFR energy not associated to the chirp (stationary harmonics, components with slow-varying frequencies in the time domain and eventually the signal noise) are filtered out during the NTEWT reassignment step.

\vspace{40pt}

\begin{figure}[ht]
	\centering
	\begin{subfigure}{0.45\textwidth}
		\includegraphics[width=\textwidth]{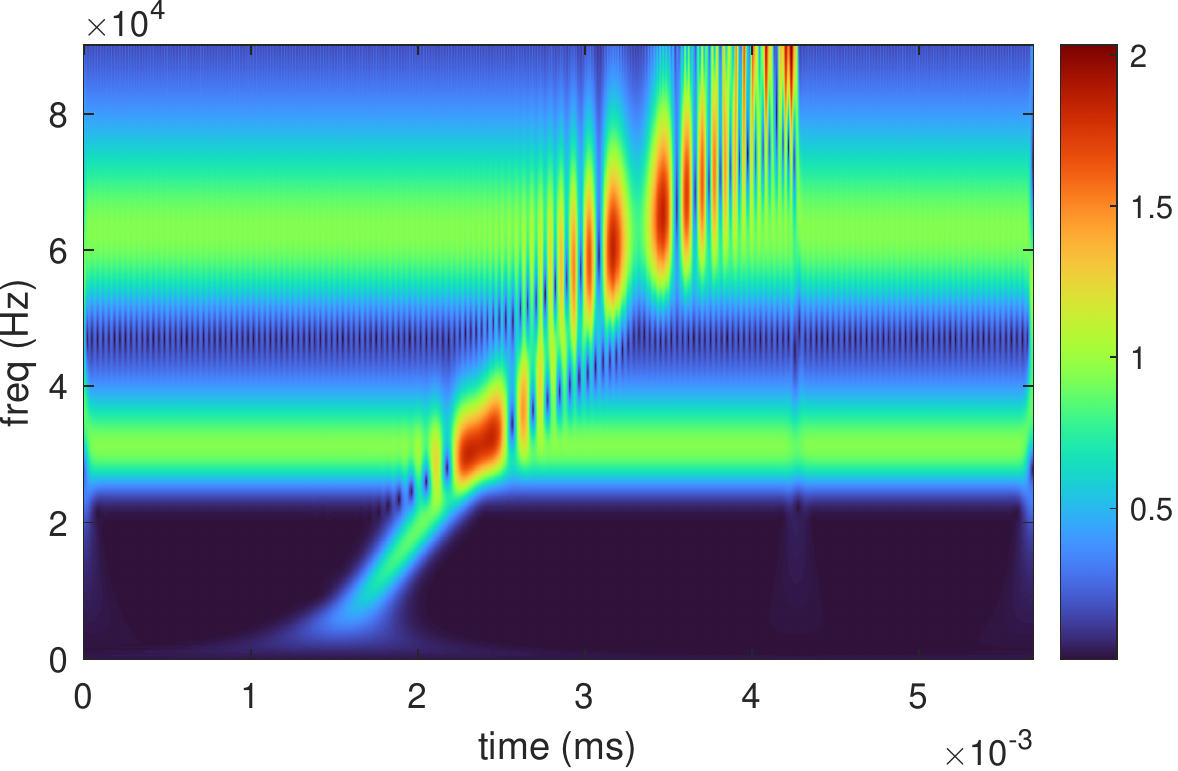}
		\caption{CWT scalogram with $\sigma=1$.}
	\end{subfigure}
	\hfill
	\begin{subfigure}{0.45\textwidth}
		\includegraphics[width=\textwidth]{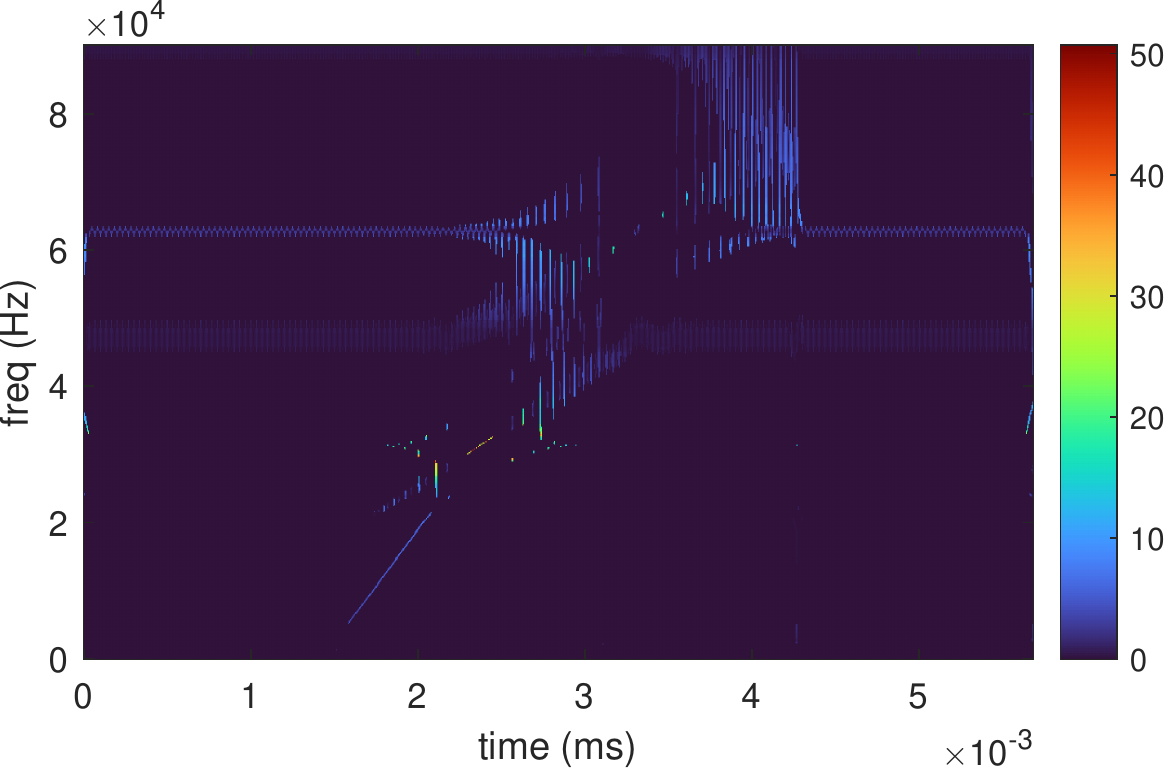}
		\caption{NTEWT scalogram with $\sigma=1$.}
	\end{subfigure}
	\begin{subfigure}{0.45\textwidth}
		\includegraphics[width=\textwidth]{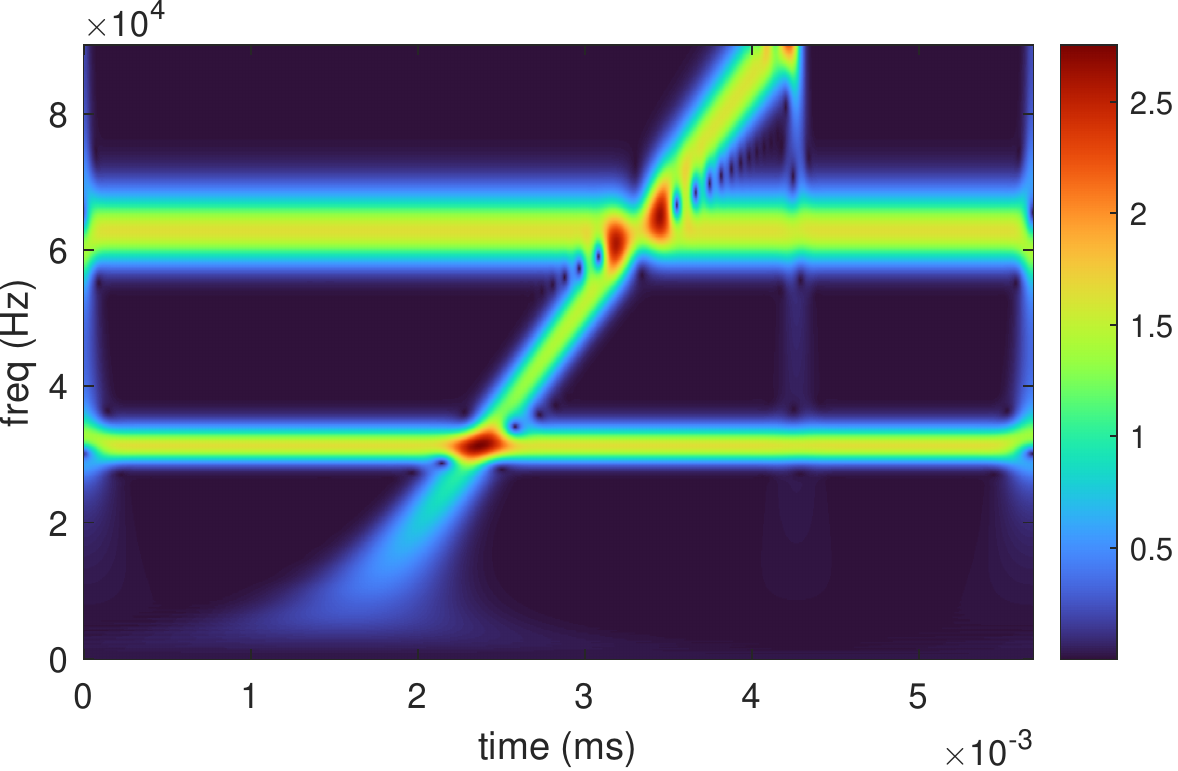}
		\caption{CWT scalogram with $\sigma=3$.}
	\end{subfigure}
	\hfill
	\begin{subfigure}{0.45\textwidth}
		\includegraphics[width=\textwidth]{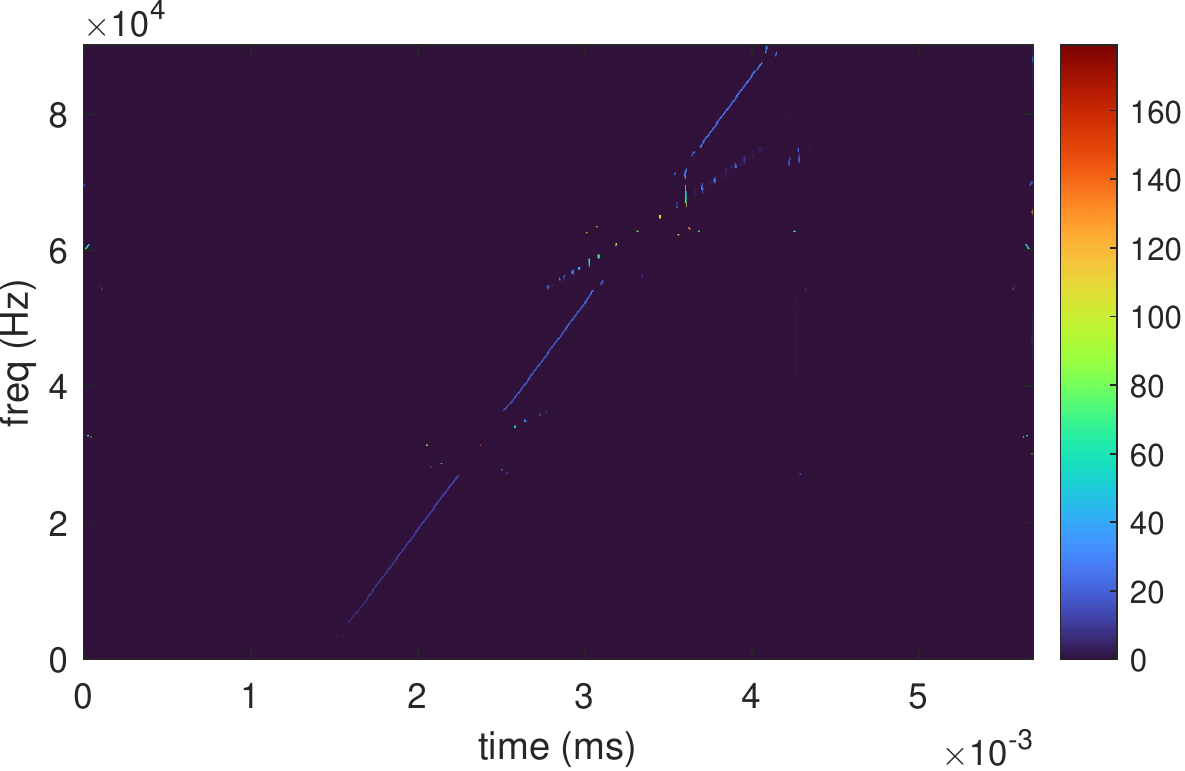}
		\caption{NTEWT scalogram with $\sigma=3$.}
		\label{Test1NTEWTd}
	\end{subfigure}
\end{figure}
\clearpage
\begin{figure}[H]\ContinuedFloat
	\begin{subfigure}{0.45\textwidth}
		\includegraphics[width=\textwidth]{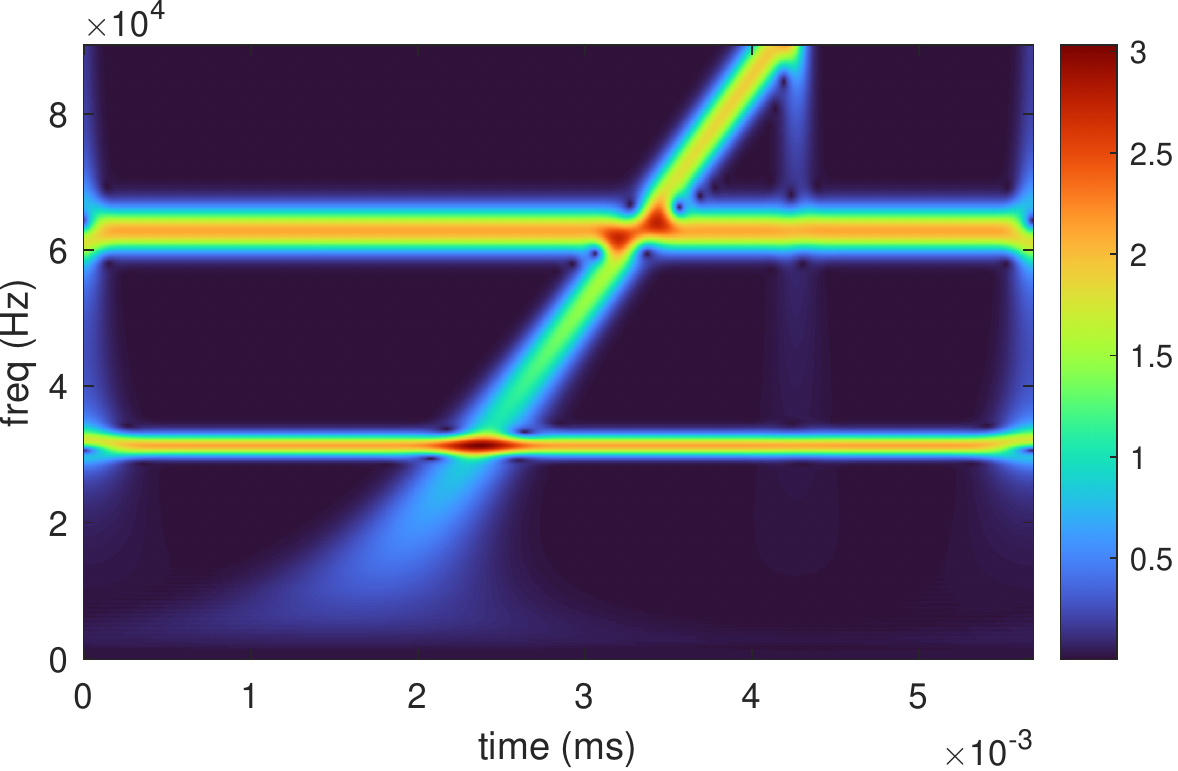}
		\caption{CWT scalogram with $\sigma=5$.}
	\end{subfigure}
	\hfill
	\begin{subfigure}{0.45\textwidth}
		\includegraphics[width=\textwidth]{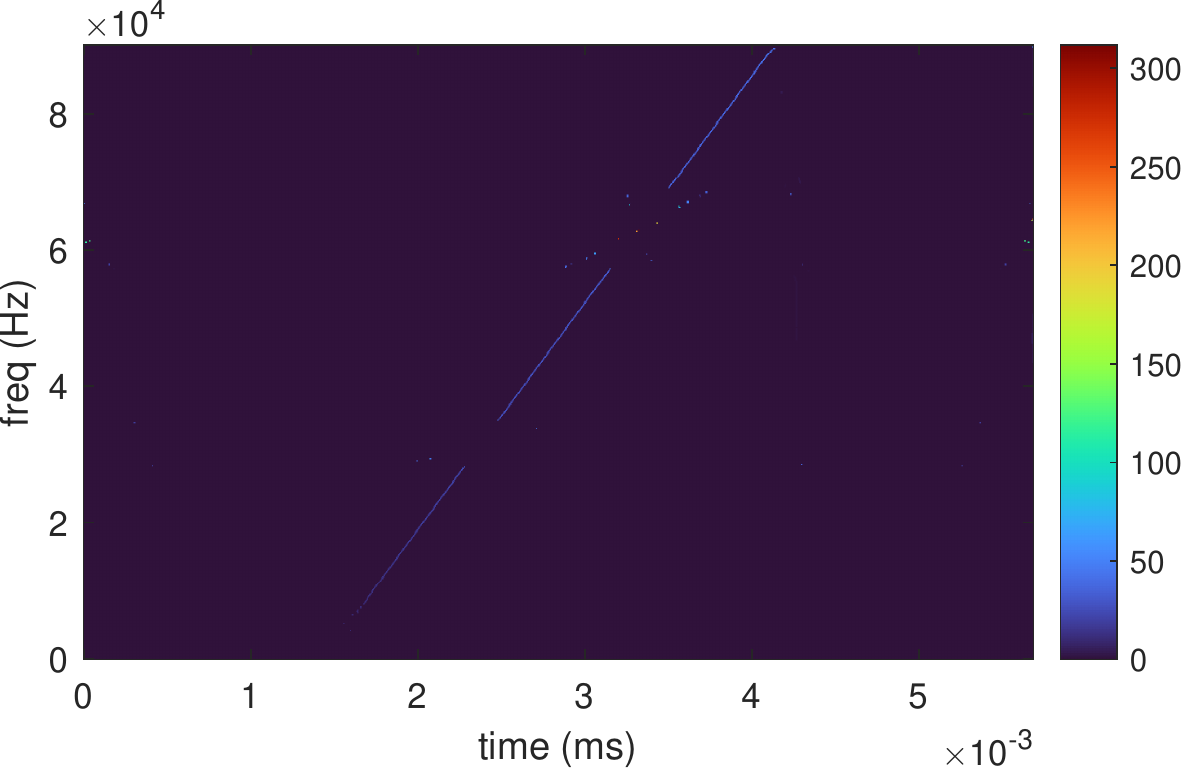}
		\caption{NTEWT scalogram with $\sigma=5$.}
		\label{Test1NTEWTf}
	\end{subfigure}
	\caption{(Test 1) CWT and NTEWT scalograms of the signal in Fig~\ref{Test1Signal} with different values of $\sigma$.}
	\label{Test1NTEWT}
\end{figure}

Fig.~\ref{Test1fil} compares the outputs of the NTEWT-based filters with the input, noiseless chirps. The most distinctive feature of the NTEWT-based filtering is the lack of phase distortion in the output with respect to the input signals. As suggested by Fig.~\ref{Test1NTEWT}, the best filtering performance is achieved when $\sigma$ is set to five, even though the border effects are greater. Figs.~\ref{Test1filb}, \ref{Test1fild} and \ref{Test1filf} compared the output of a matched filter applied on the original noisy signal of Fig.~\ref{Test1Signal} (blue) with the outputs of the same matched filter as applied on the reconstructed (NTEWT-filtered) chirp signals (red). Matched filtering consists in convoluting the original signal and NTEWT-filter outputs with a template of the chirp signal to detect (Blue, Figs.~\ref{Test1fila}, \ref{Test1filc} and \ref{Test1file}). Ultimately, the performance of the signal filtering should be evaluated as the degree of improvement on matched filter output's resolution achieved. The resolution gain of the chirp detection is maximized with high wavelet widths ($\sigma=5$), despite border effects and lower time resolution.

\vspace{10pt}

\subsection{Test 2: long signal containing a short-duration, high-rate chirp pulse mixed with stationary harmonics and Gaussian noise}

In this experiment, another signal with length 1024 samples was generated to test the performance of the NTEWT-based filter for shorter, higher-rate chirps and its robustness to the presence of white noise. The parameters of the new chirp signal are shown here:

\vspace{5pt}

\begin{table}[H]
	\centering
	\begin{tabular}{ll}
		\hline
		Frequency lower bound (kHz) & 0                                    \\
		Frequency upper bound (kHz) & 90                                   \\
		Pulse length (samples)      & 32 ($\approx$ 0.000178 s) \\ \hline
	\end{tabular}
\end{table}

\begin{figure}[H]
	\centering
	\begin{subfigure}{0.45\textwidth}
		\includegraphics[width=\textwidth]{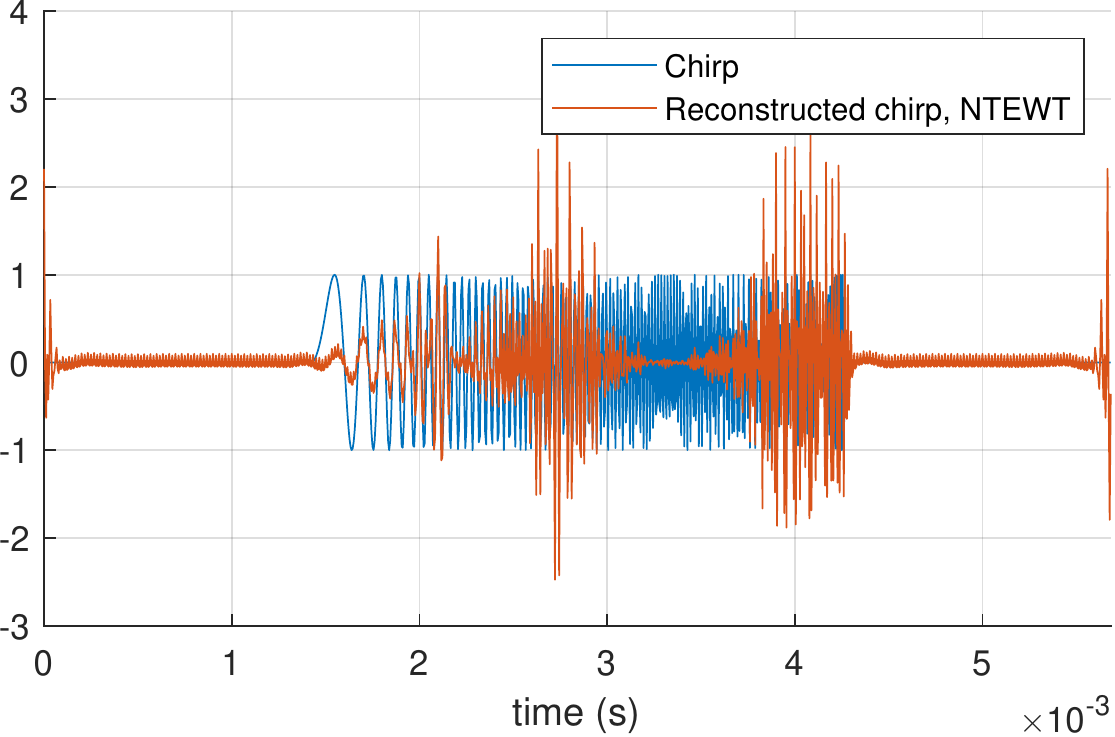}
		\caption{$x_{\text{fil}}[j]$ with $\sigma=1$.}
		\label{Test1fila}
	\end{subfigure}
	\hfill
	\begin{subfigure}{0.45\textwidth}
		\includegraphics[width=\textwidth]{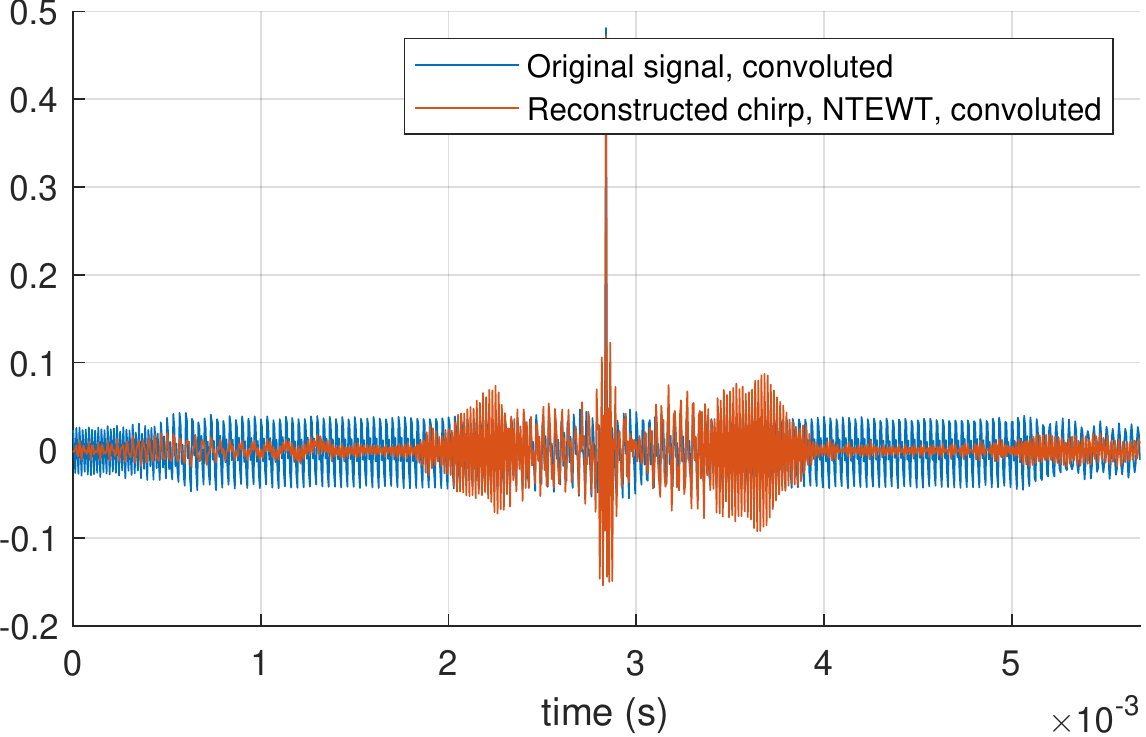}
		\caption{Matched filter output with $\sigma=1$.}
		\label{Test1filb}
	\end{subfigure}
	\begin{subfigure}{0.45\textwidth}
		\includegraphics[width=\textwidth]{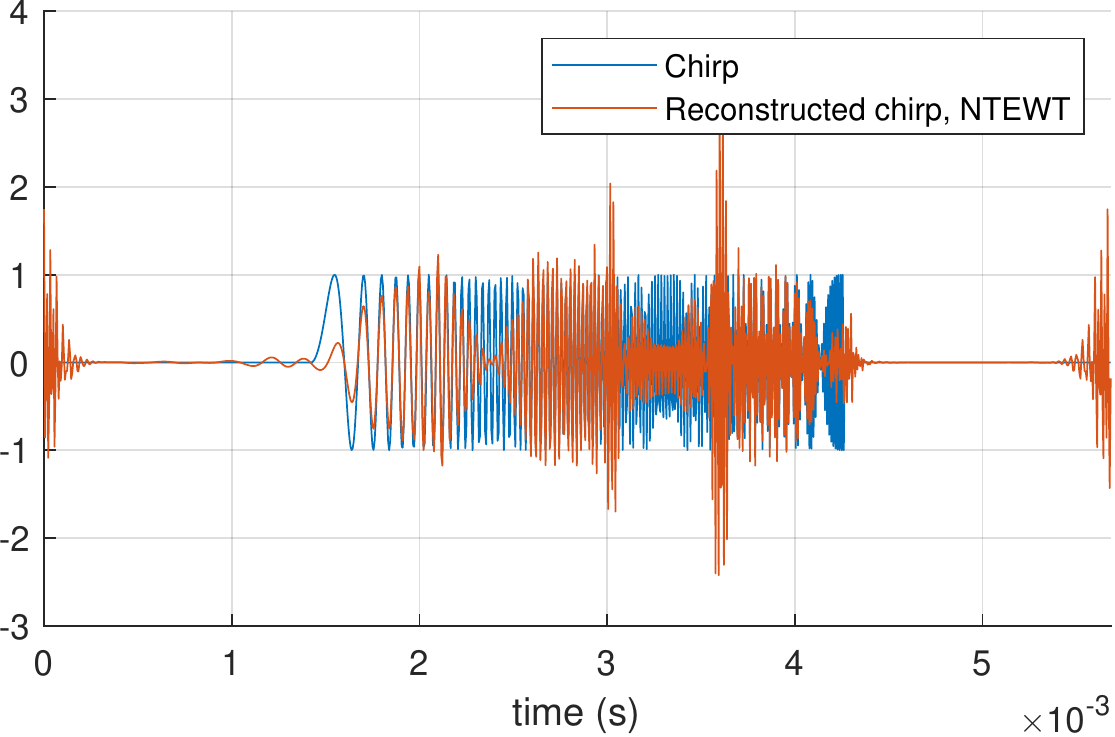}
		\caption{$x_{\text{fil}}[j]$ with $\sigma=3$.}
		\label{Test1filc}
	\end{subfigure}
	\hfill
	\begin{subfigure}{0.45\textwidth}
		\includegraphics[width=\textwidth]{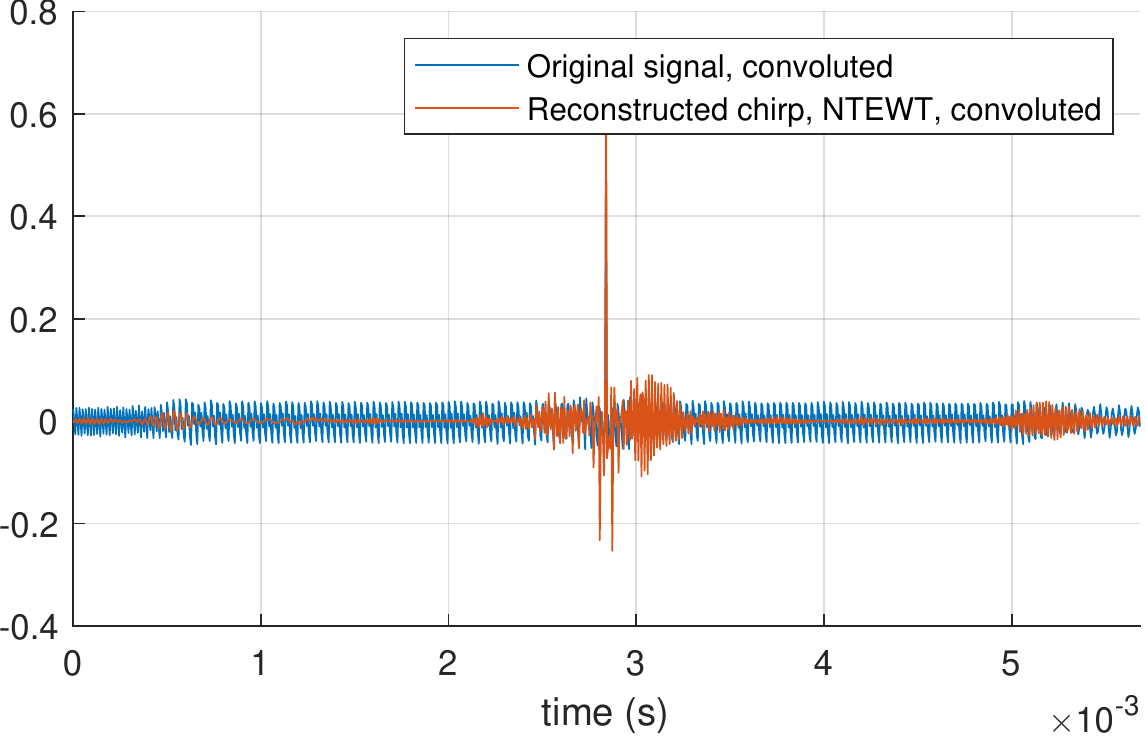}
		\caption{Matched filter output with $\sigma=3$.}
		\label{Test1fild}
	\end{subfigure}
	\begin{subfigure}{0.45\textwidth}
		\includegraphics[width=\textwidth]{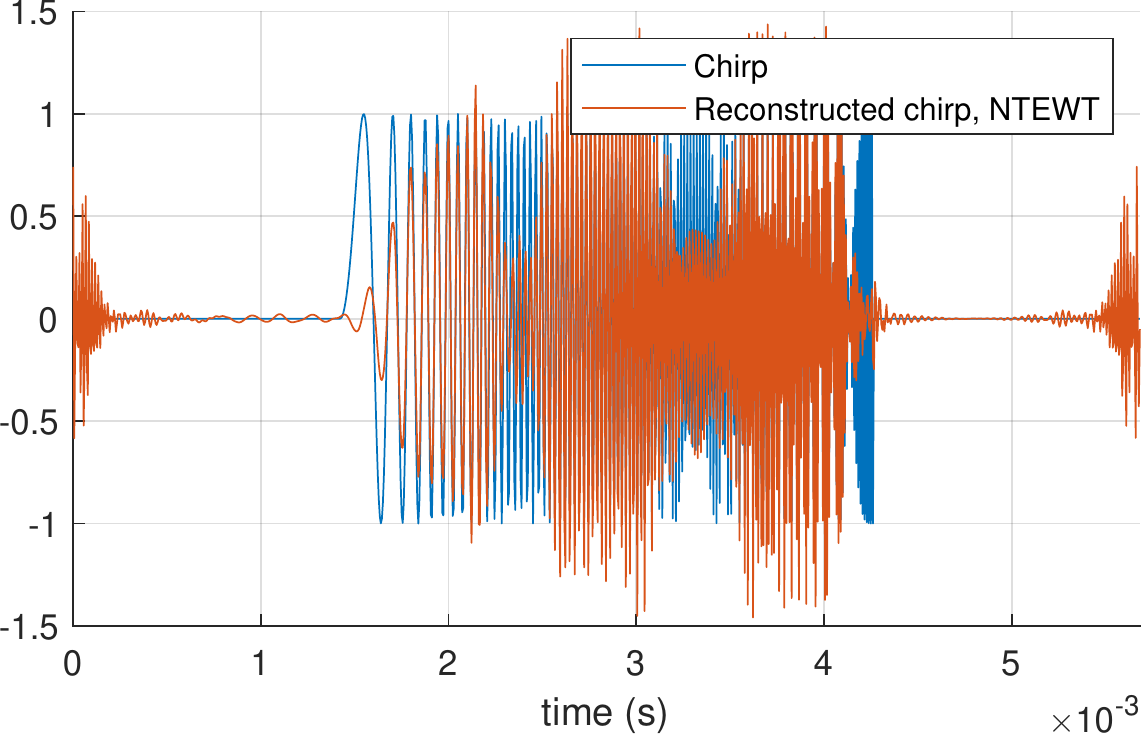}
		\caption{$x_{\text{fil}}[j]$ with $\sigma=5$.}
		\label{Test1file}
	\end{subfigure}
	\hfill
	\begin{subfigure}{0.45\textwidth}
		\includegraphics[width=\textwidth]{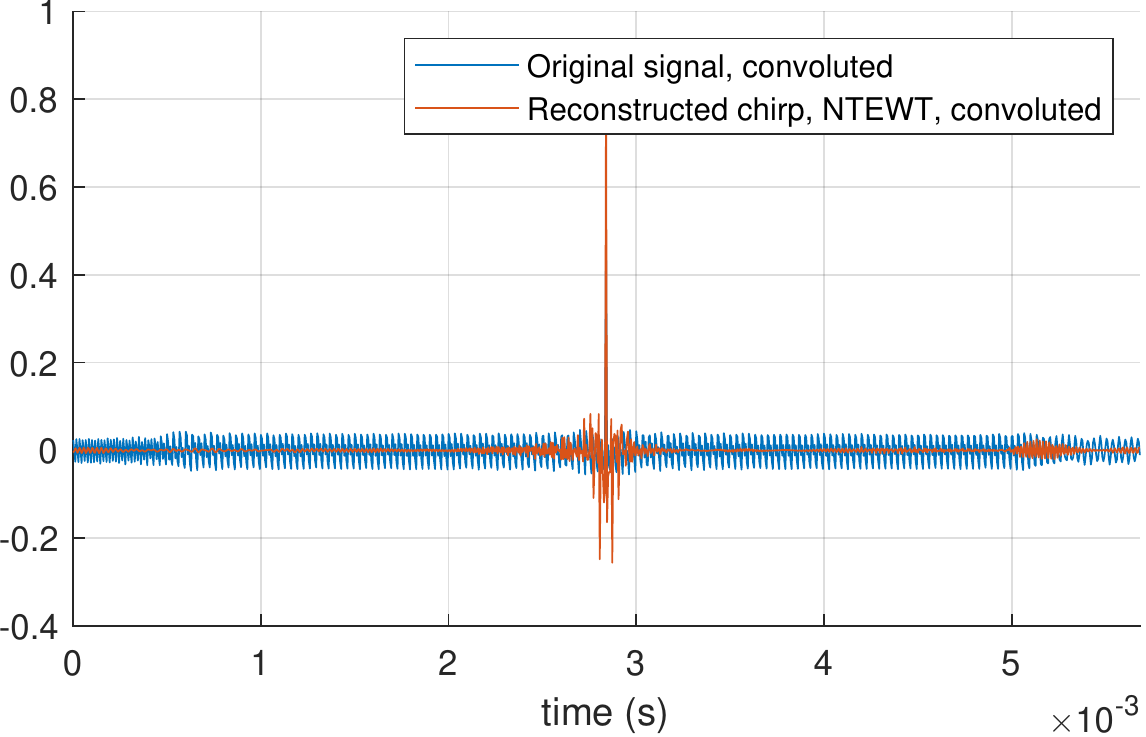}
		\caption{Matched filter output with $\sigma=5$.}
		\label{Test1filf}
	\end{subfigure}
	\caption{(Test 1) (a), (c) and (d) The outputs $x_{\text{fil}}[j]$ of the NTEWT-based filter (reconstructed chirp) compared with the noiseless chirp pulse. (b), (d) and (f) Outputs of a matched chirp filter applied on the signal in Fig.~\ref{Test1Signal} (in blue) and on the NTEWT-filter output (in red).}
	\label{Test1fil}
\end{figure}

\begin{figure}[H]
	\centering
	\begin{subfigure}{0.45\textwidth}
		\includegraphics[width=\textwidth]{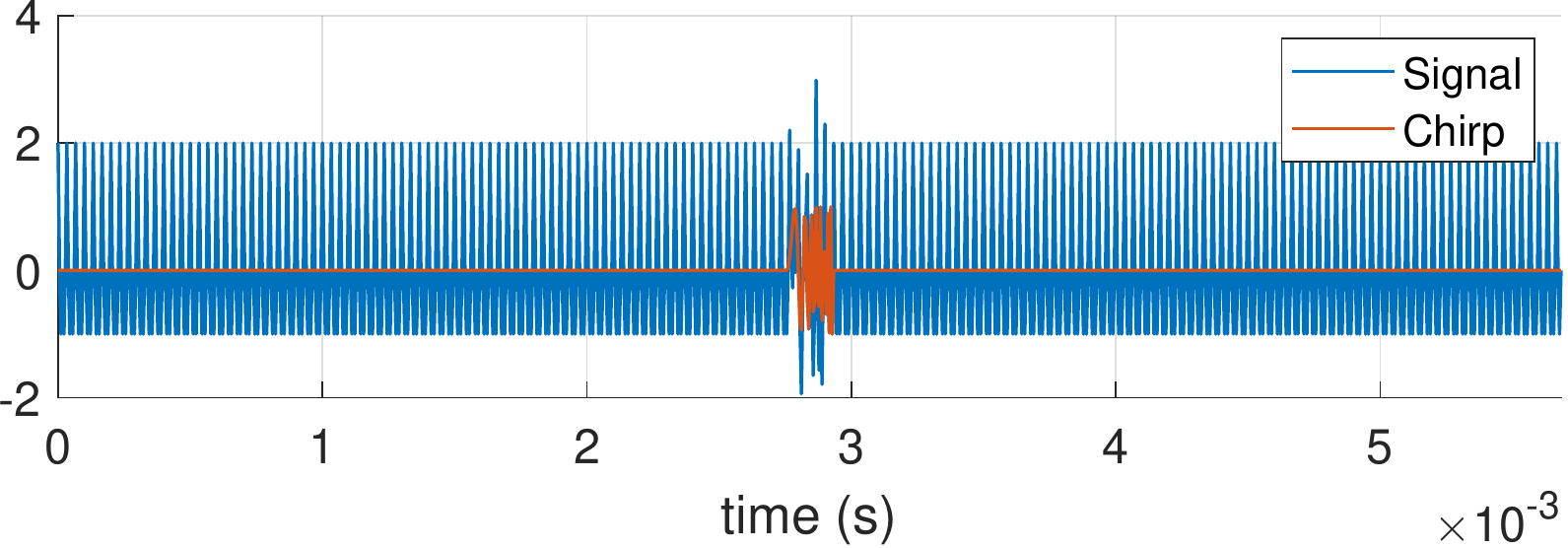}
		\caption{No Gaussian noise.}
	\end{subfigure}
	\hfill
	\begin{subfigure}{0.45\textwidth}
		\includegraphics[width=\textwidth]{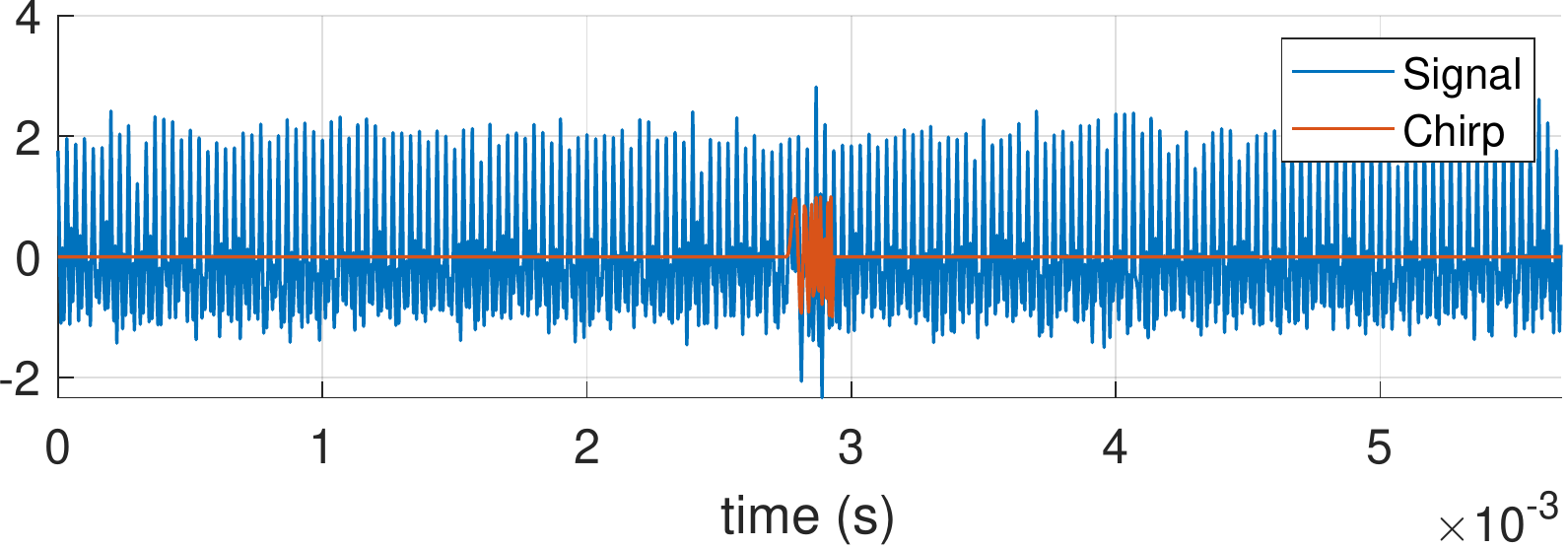}
		\caption{Gaussian noise with $s=0.2$.}
	\end{subfigure}
	\begin{subfigure}{0.45\textwidth}
		\includegraphics[width=\textwidth]{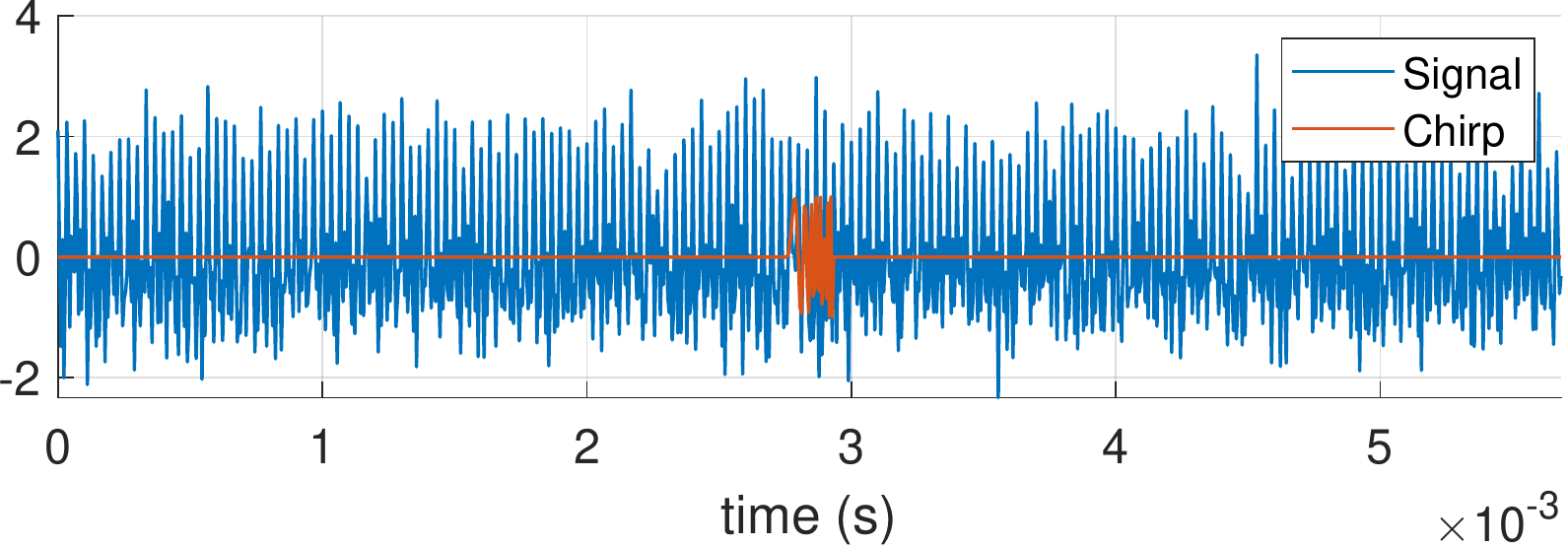}
		\caption{Gaussian noise with $s=0.4$.}
	\end{subfigure}
	\caption{(Test 2) The chirp pulse (in red) plus the stationary harmonics and the Gaussian noise (in blue).}
	\label{Test2Signal}
\end{figure}

Three synthetic signals shown in Fig.~\ref{Test2Signal} were generated with three different intensities of white Gaussian noise. Two stationary harmonics with frequencies 30 kHz and 60 kHz were added to the signals as in Test~1. The optimal values found for the reassignment tolerance $\epsilon$ and the wavelet width $\sigma$ for these new signals were respectively $\num{2e-3}$ and 5.

\begin{figure}[H]
	\centering
	\begin{subfigure}{0.45\textwidth}
		\includegraphics[width=\textwidth]{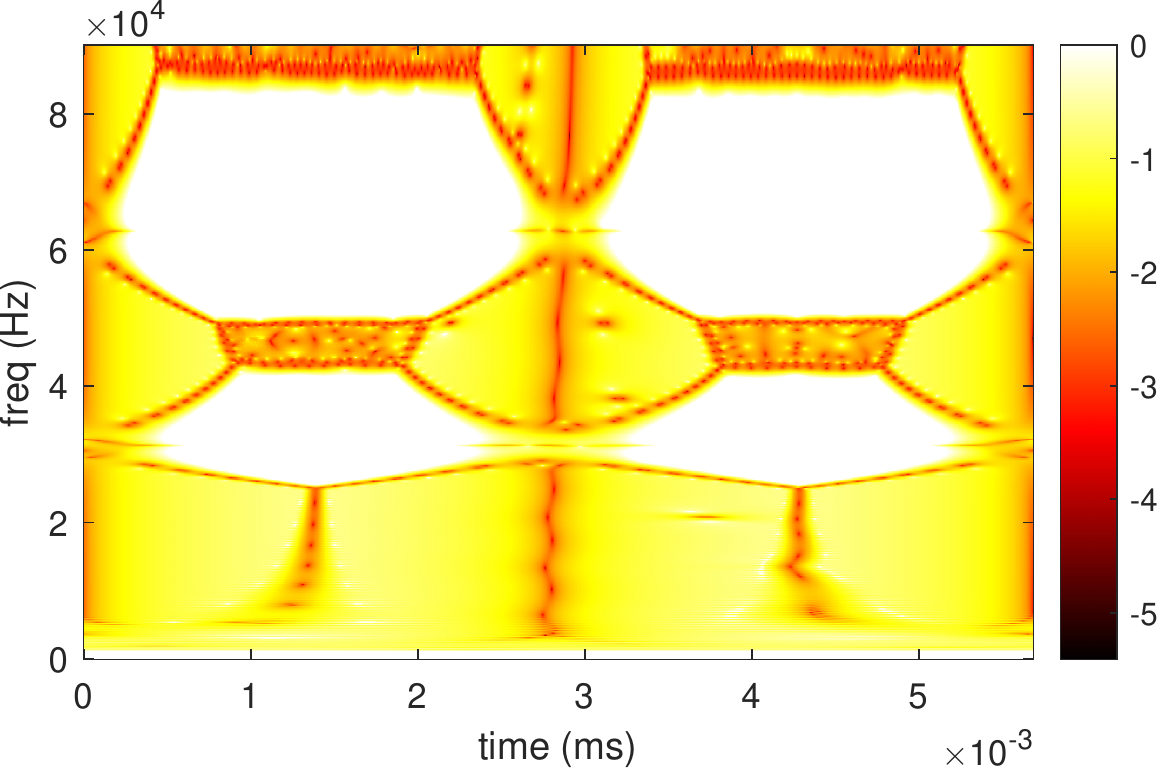}
		\caption{No Gaussian noise.}
		\label{Test2fixeda}
	\end{subfigure}
	\hfill
	\begin{subfigure}{0.45\textwidth}
		\includegraphics[width=\textwidth]{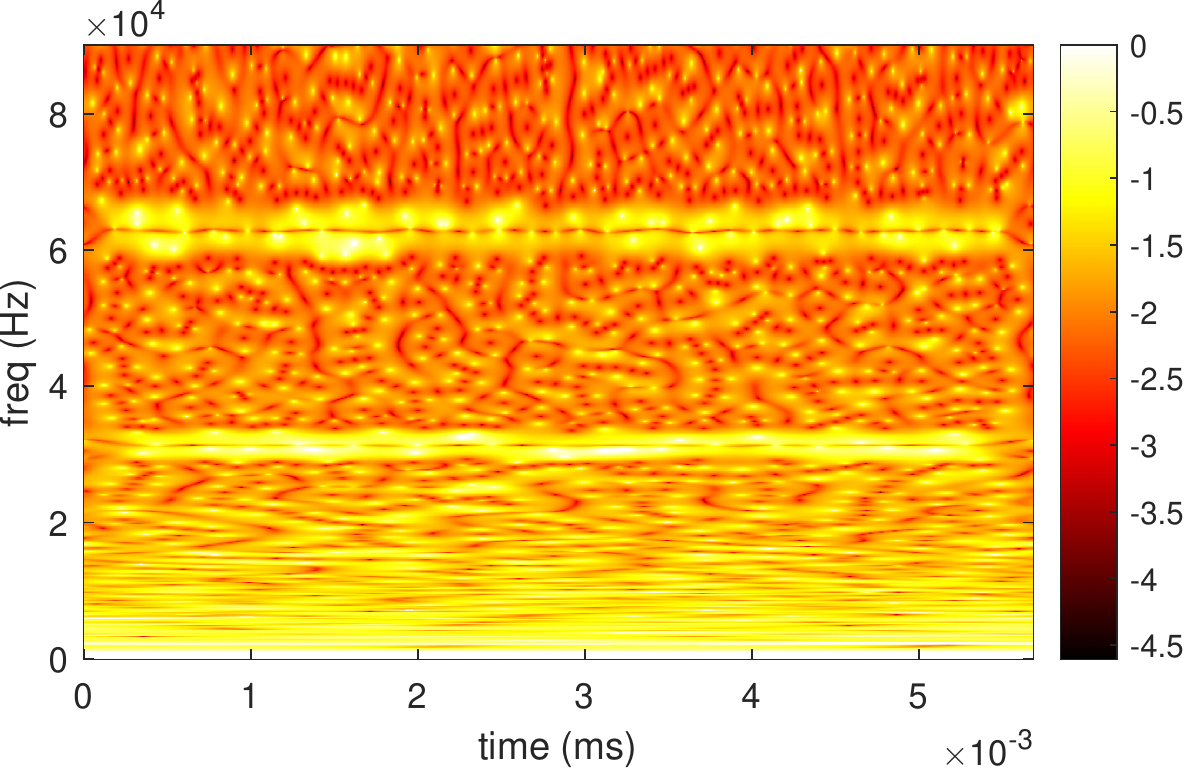}
		\caption{Gaussian noise with $s=0.4$.}
		\label{Test2fixedb}
	\end{subfigure}
	\caption{(Test 2) The log of the fixed point metric of the CWT of $x[j]$, $|\bar{t}_x[j,k]-b[j]|$, with $\sigma=3$.}
	\label{Test2fixed}
\end{figure}

CWT and NTEWT scalograms of the synthetic signals in Fig.~\ref{Test2Signal}, fixed point metrics for the clean signal and for $s=0.4$, NTEWT-filter outputs and matched filter outputs are shown in Figs.~\ref{Test2NTEWT}, \ref{Test2fixed} and \ref{Test2fil}. NTEWT reassignment without random noise (Figs.~\ref{Test2NTEWTa}, \ref{Test2NTEWTb}, \ref{Test2fixeda} and \ref{Test2fixedb}) performs as well as in the long-duration chirp (Figs.~\ref{Test1NTEWTf}, \ref{Test1fixed}, \ref{Test1file} and \ref{Test1filf}). Obviously, the performance of the reassignment degrades as the noise level increases, but the chirp's GD trajectory is still traceable at the high end of the frequency axis. NTEWT-based pre-filtering improves chirp detection performance also under low and moderate white noise (Figs.~\ref{Test2fild} and \ref{Test2filf}).

\vspace{50pt}

\begin{figure}[H]
	\centering
	\begin{subfigure}{0.45\textwidth}
		\includegraphics[width=\textwidth]{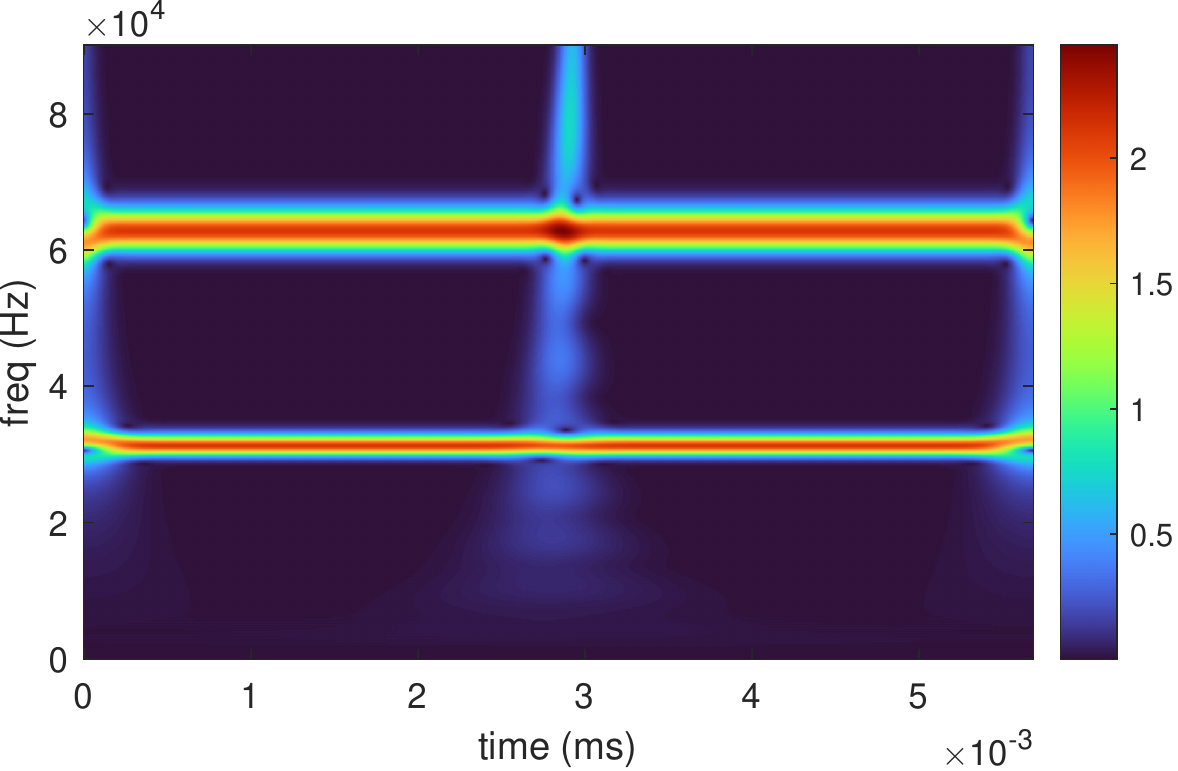}
		\caption{CWT scalogram. No Gaussian noise.}
		\label{Test2NTEWTa}
	\end{subfigure}
	\hfill
	\begin{subfigure}{0.45\textwidth}
		\includegraphics[width=\textwidth]{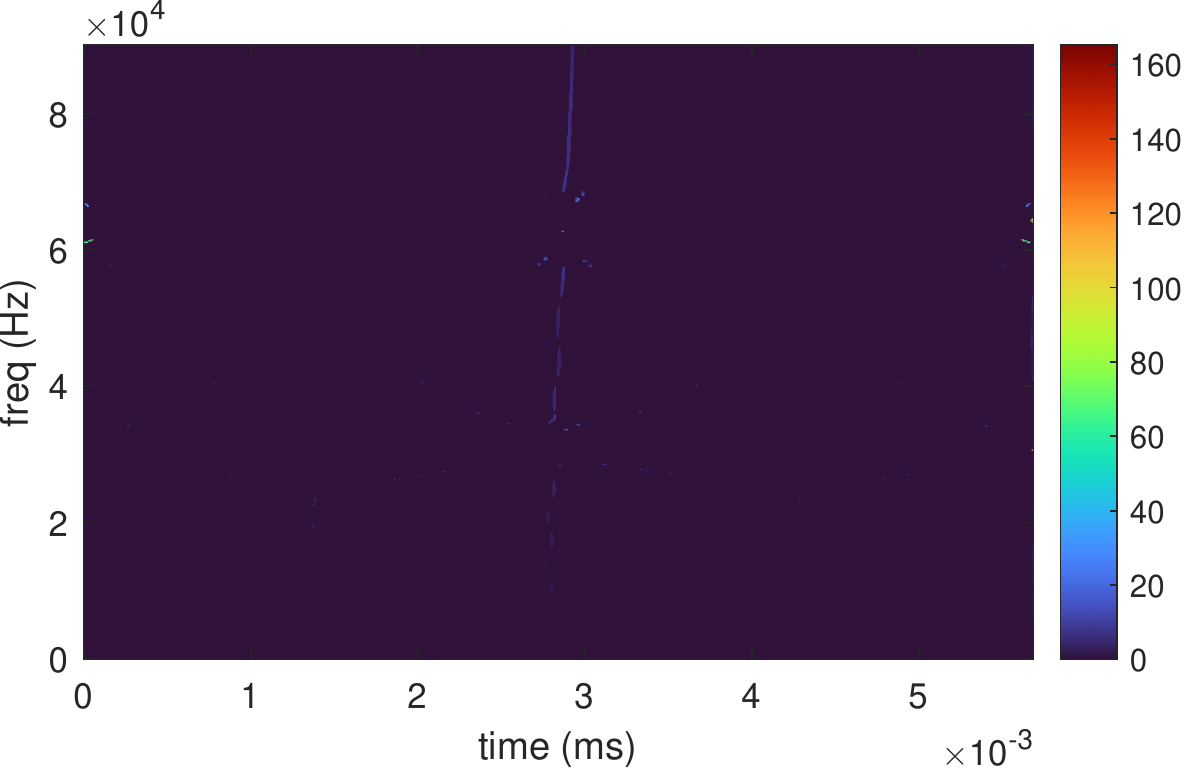}
		\caption{NTEWT scalogram. No Gaussian noise.}
		\label{Test2NTEWTb}
	\end{subfigure}
	\begin{subfigure}{0.45\textwidth}
		\includegraphics[width=\textwidth]{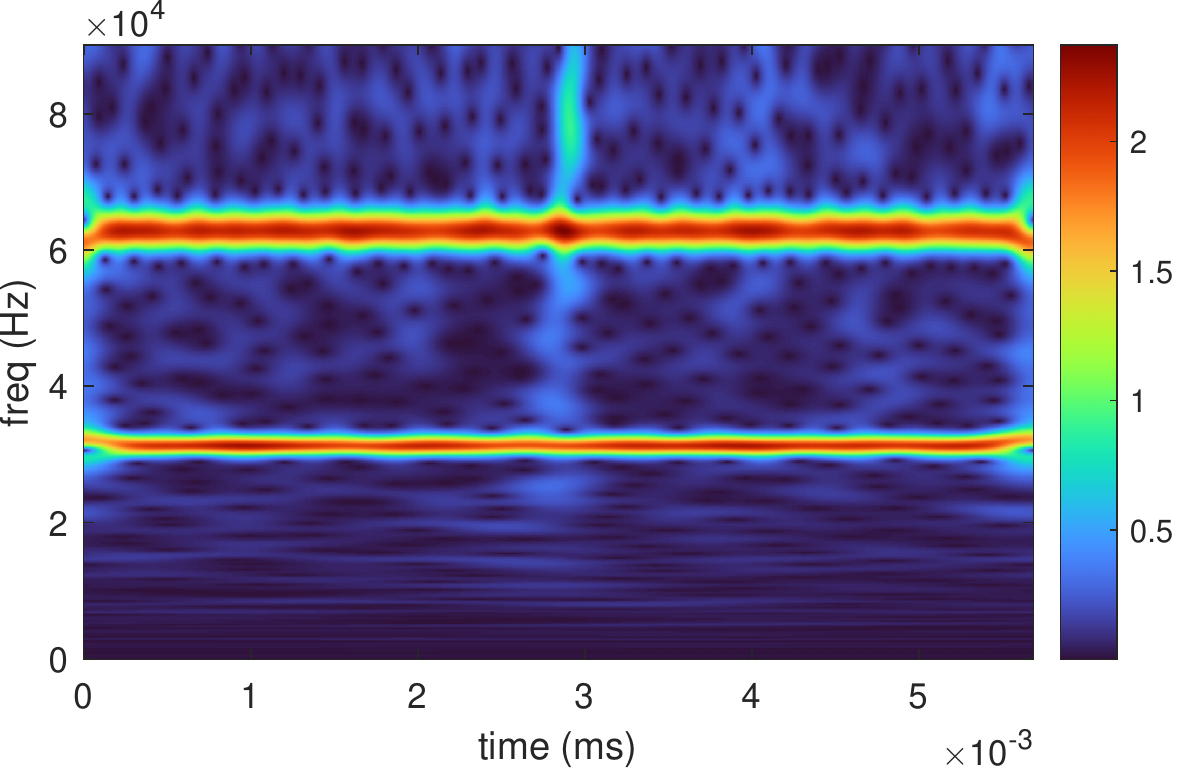}
		\caption{CWT scalogram. Gaussian noise with $s=0.2$.}
		\label{Test2NTEWTc}
	\end{subfigure}
	\hfill
	\begin{subfigure}{0.45\textwidth}
		\includegraphics[width=\textwidth]{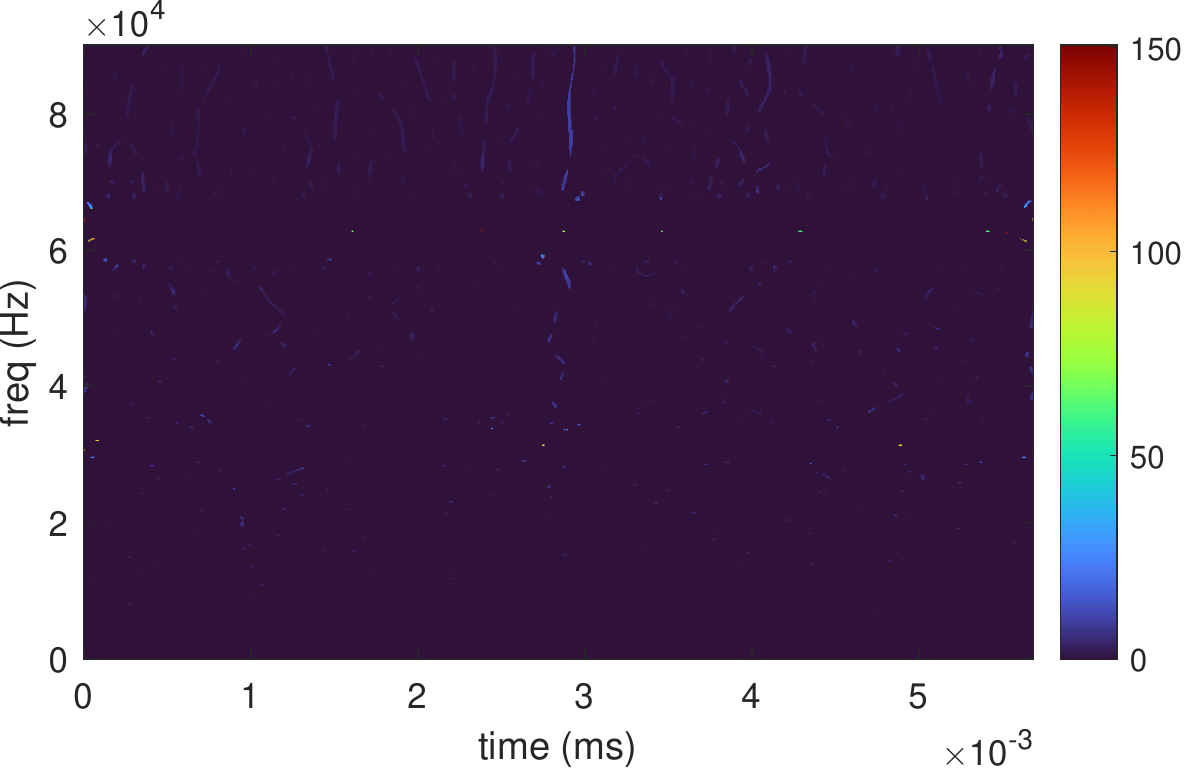}
		\caption{NTEWT scalogram. Gaussian noise with $s=0.2$.}
		\label{Test2NTEWTd}
	\end{subfigure}
 \end{figure}
 \clearpage
\begin{figure}[H]\ContinuedFloat
	\begin{subfigure}{0.45\textwidth}
		\includegraphics[width=\textwidth]{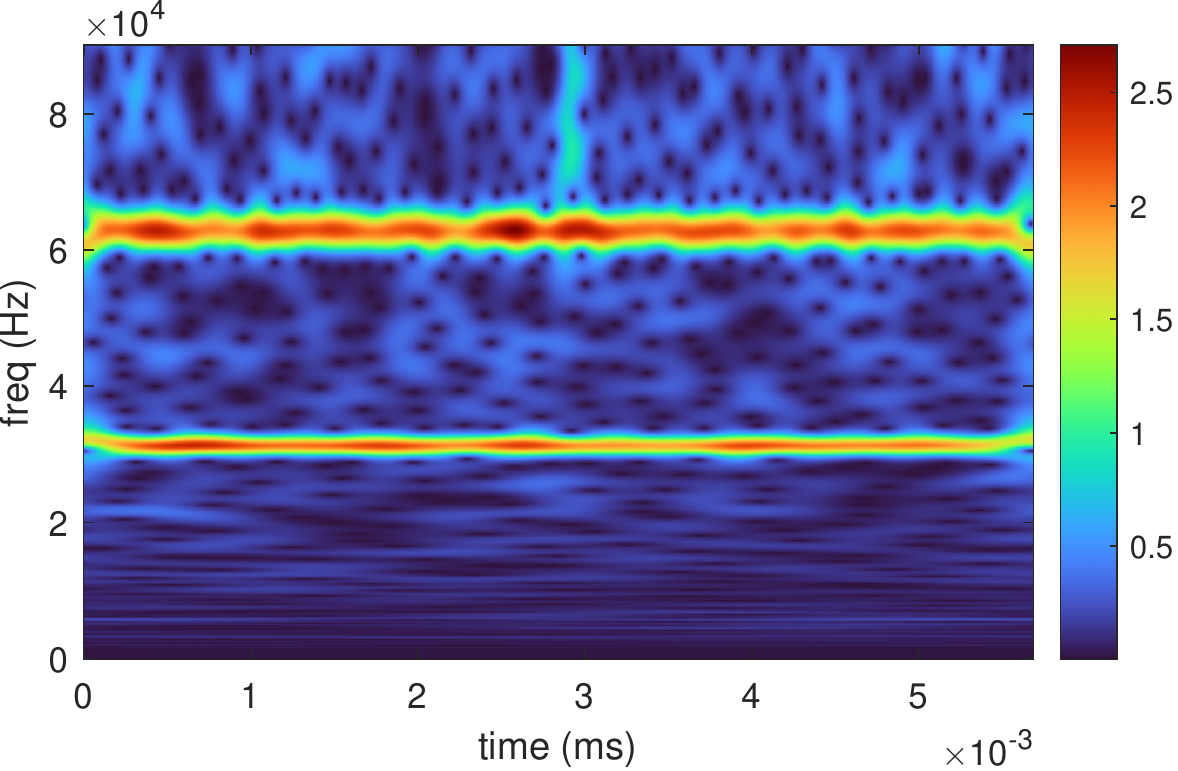}
		\caption{CWT scalogram. Gaussian noise with $s=0.4$.}
		\label{Test2NTEWTe}
	\end{subfigure}
	\hfill
	\begin{subfigure}{0.45\textwidth}
		\includegraphics[width=\textwidth]{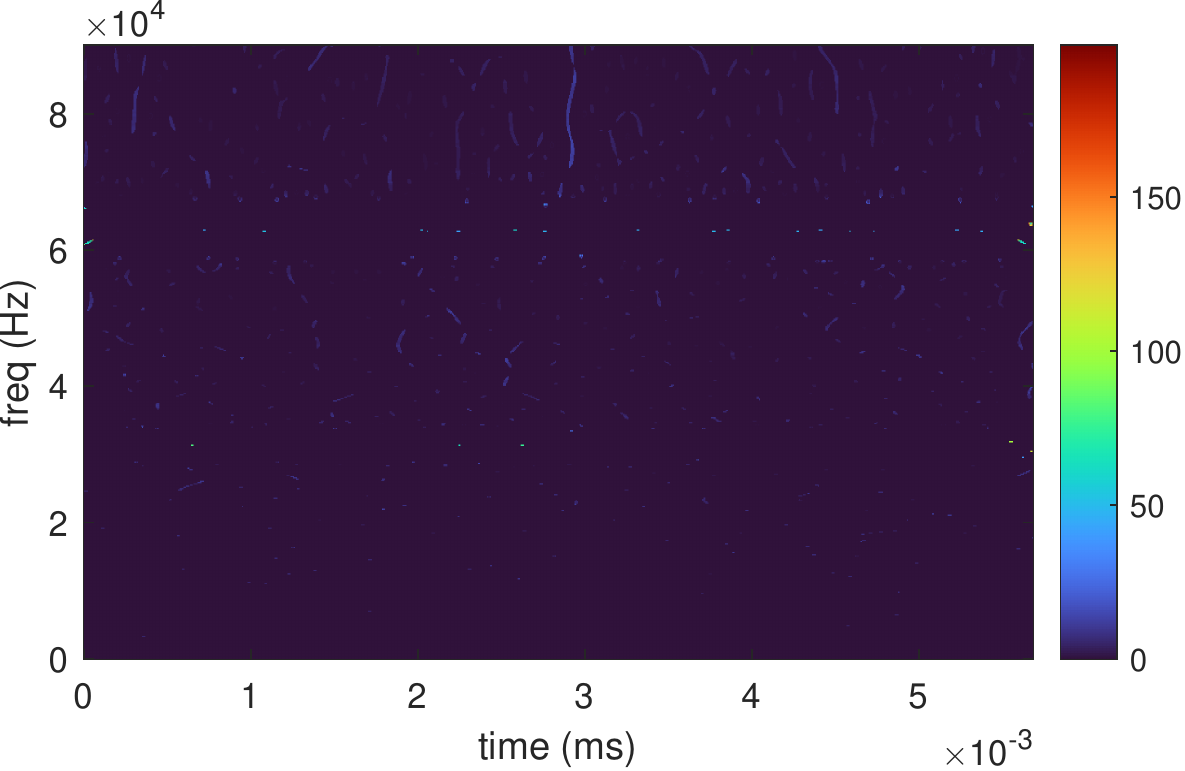}
		\caption{NTEWT scalogram. Gaussian noise with $s=0.4$.}
		\label{Test2NTEWTf}
	\end{subfigure}
	\caption{(Test 2) CWT and NTEWT scalograms of the signals in Fig~\ref{Test2Signal} with $\sigma=5$.}
	\label{Test2NTEWT}
\end{figure}

\vspace{40pt}

\vspace{40pt}

\begin{figure}[H]
	\centering
	\begin{subfigure}{0.45\textwidth}
		\includegraphics[width=\textwidth]{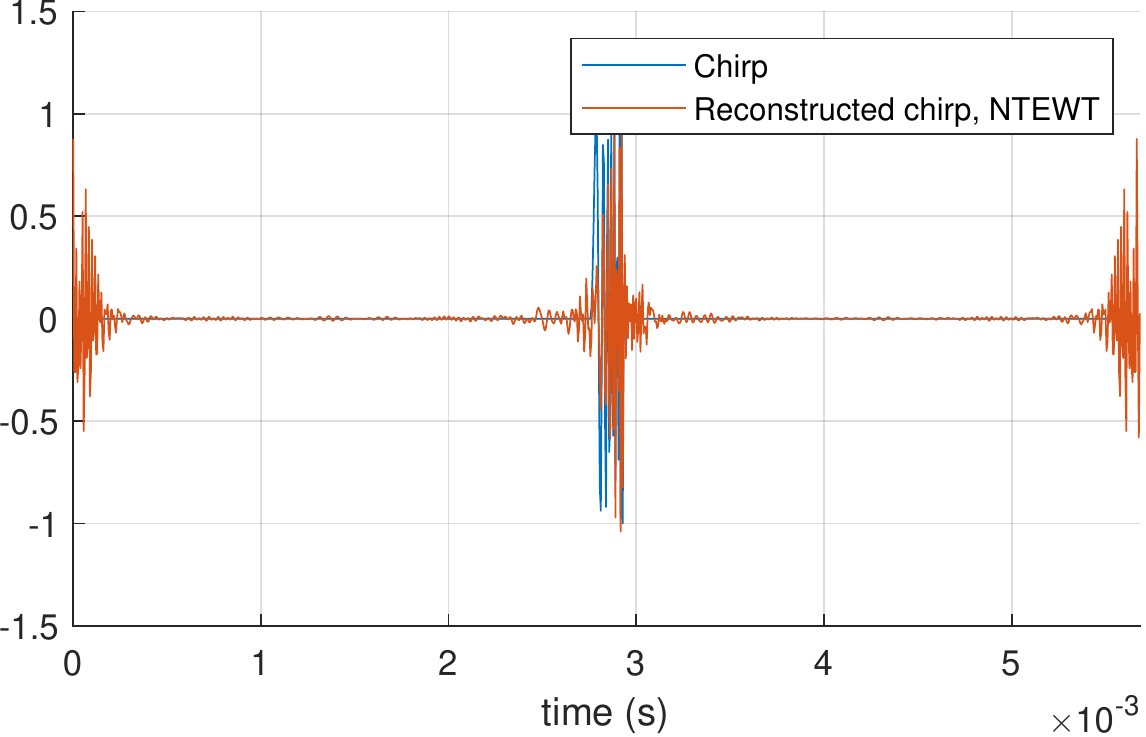}
		\caption{$x_{\text{fil}}[j]$. No Gaussian noise.}
		\label{Test2fila}
	\end{subfigure}
	\hfill
	\begin{subfigure}{0.45\textwidth}
		\includegraphics[width=\textwidth]{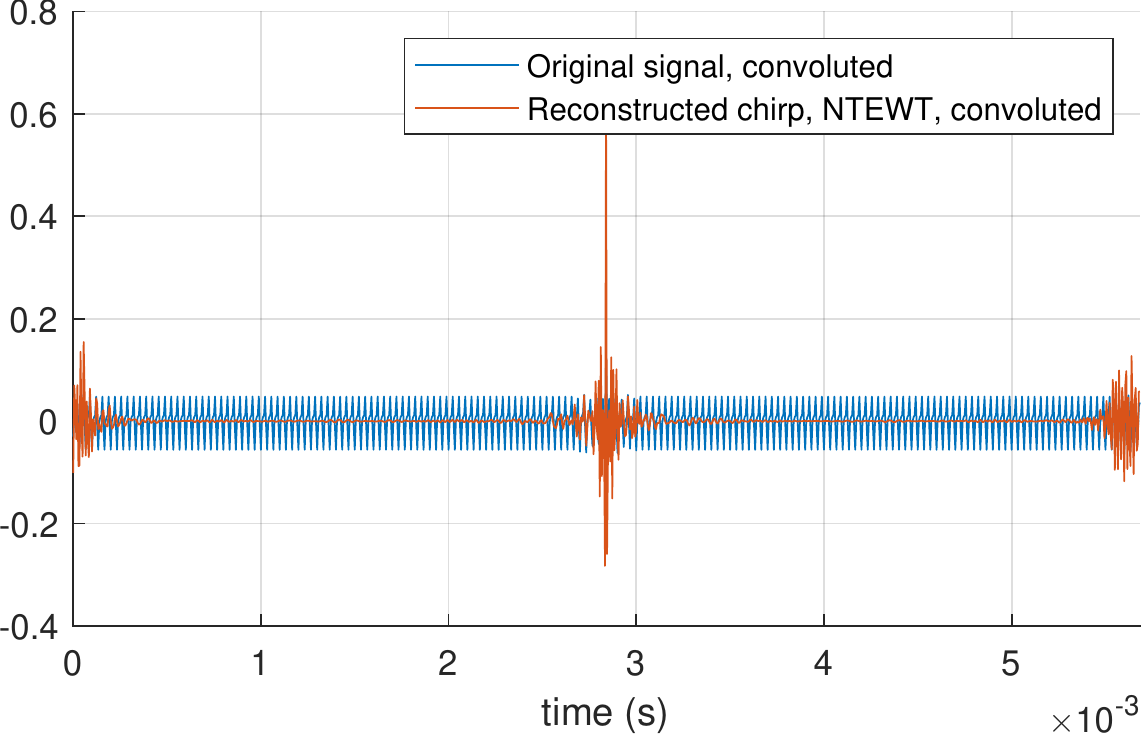}
		\caption{Matched filter output. No Gaussian noise.}
		\label{Test2filb}
	\end{subfigure}
 \end{figure}
 \clearpage
\begin{figure}[H]\ContinuedFloat
	\begin{subfigure}{0.45\textwidth}
		\includegraphics[width=\textwidth]{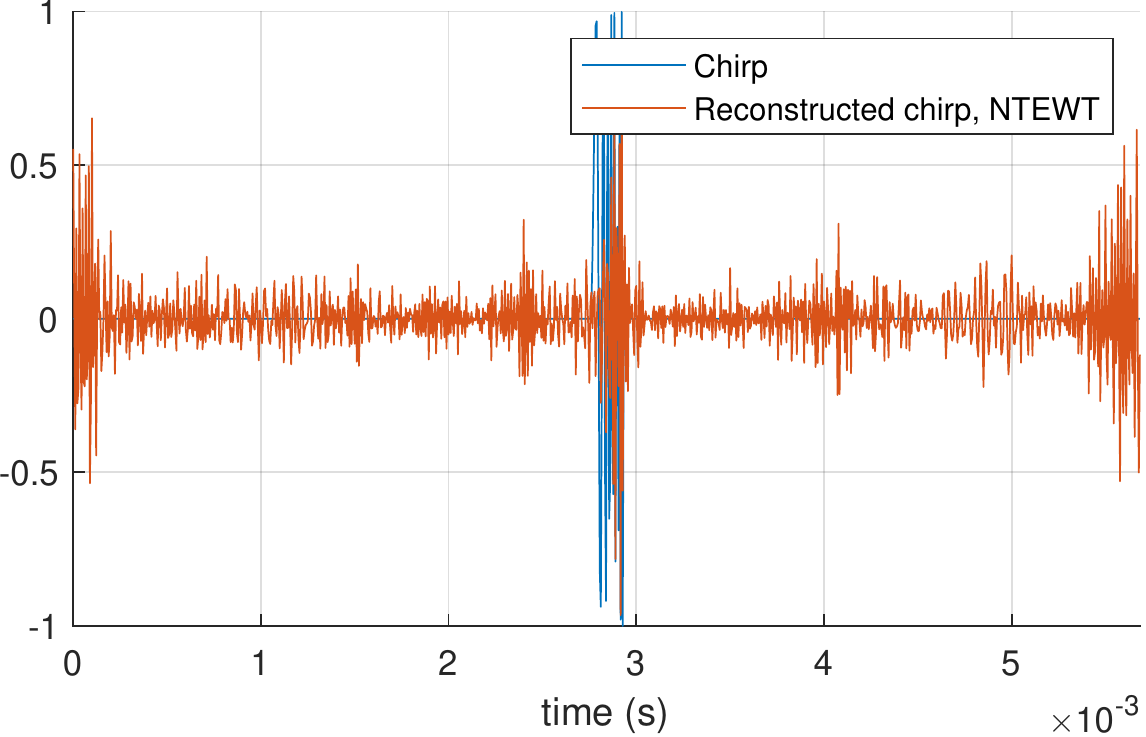}
		\caption{$x_{\text{fil}}[j]$. Gaussian noise with $s=0.2$.}
		\label{Test2filc}
	\end{subfigure}
	\hfill
	\begin{subfigure}{0.45\textwidth}
		\includegraphics[width=\textwidth]{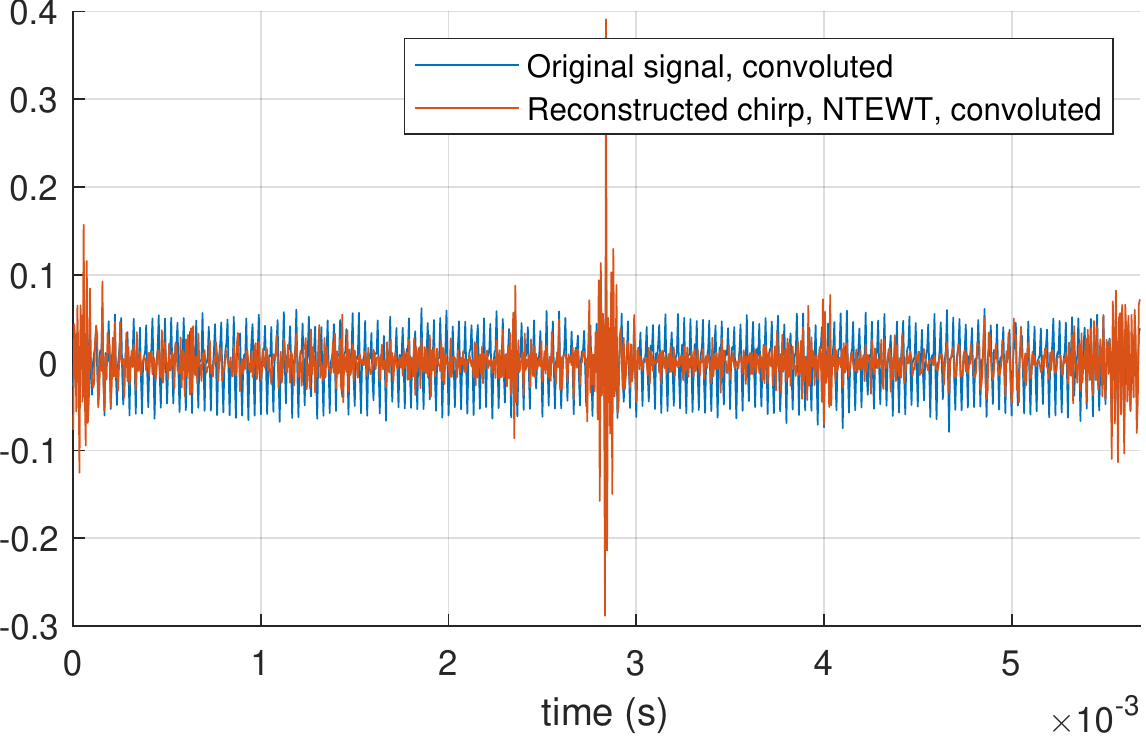}
		\caption{Matched filter output. Gaussian noise with $s=0.2$.}
		\label{Test2fild}
	\end{subfigure}
	\begin{subfigure}{0.45\textwidth}
		\includegraphics[width=\textwidth]{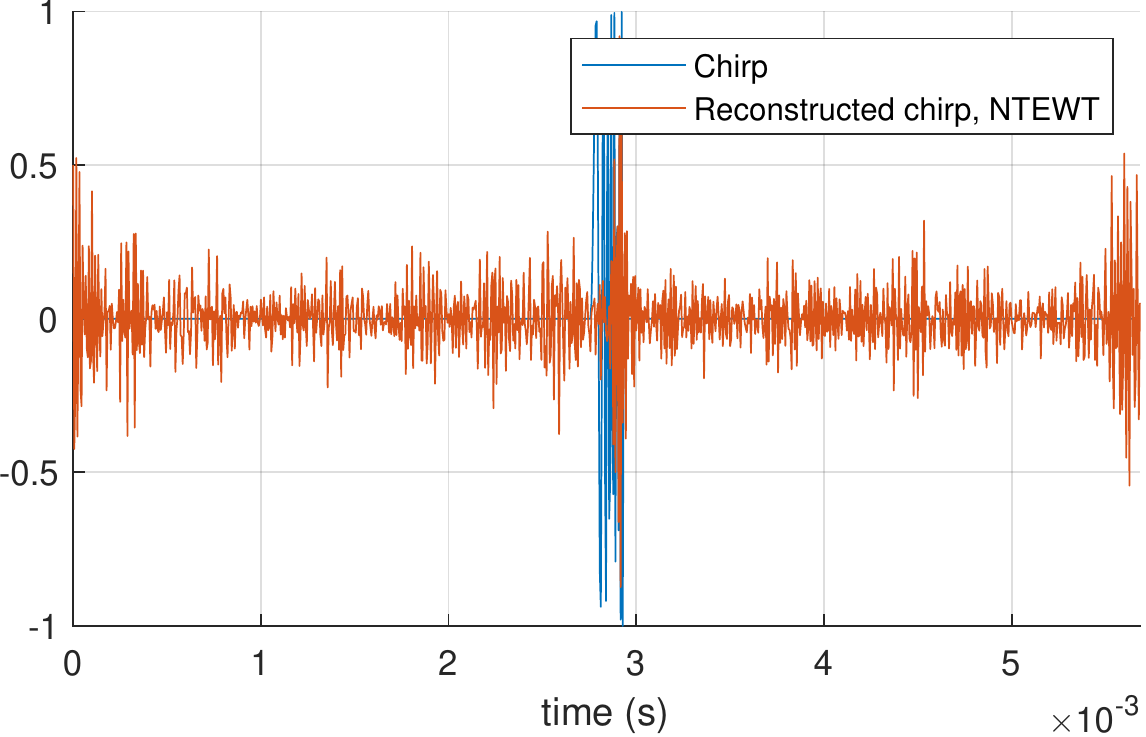}
		\caption{$x_{\text{fil}}[j]$. Gaussian noise with $s=0.4$.}
		\label{Test2file}
	\end{subfigure}
	\hfill
	\begin{subfigure}{0.45\textwidth}
		\includegraphics[width=\textwidth]{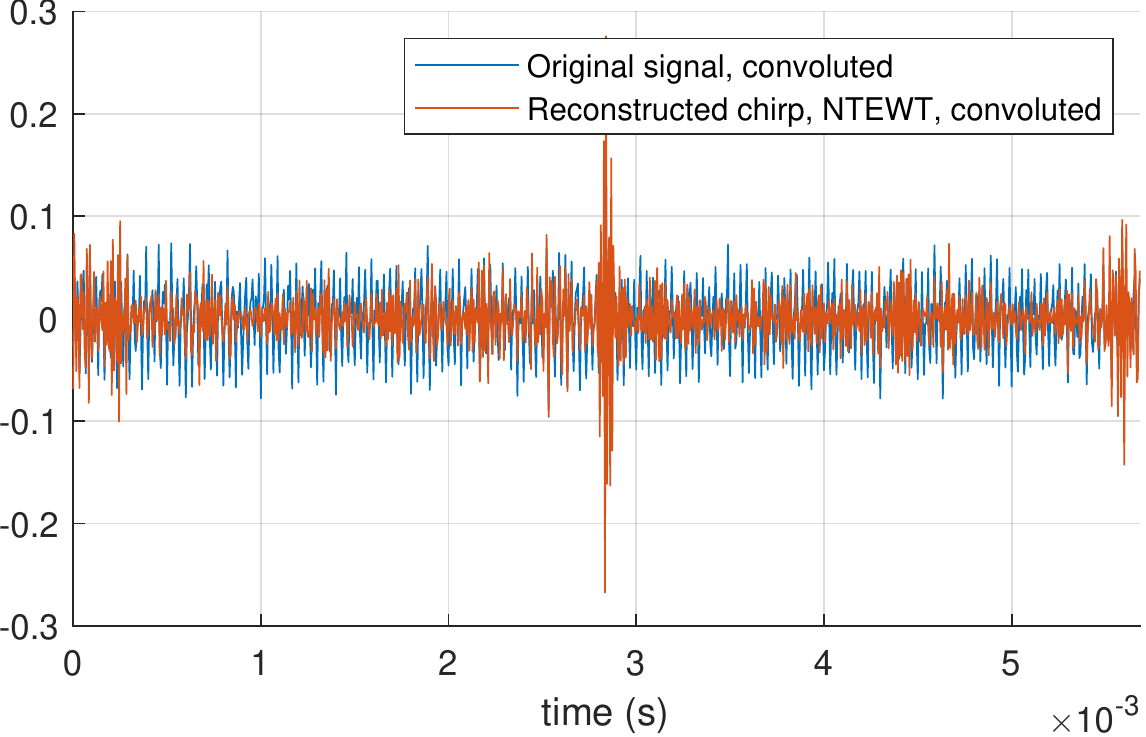}
		\caption{Matched filter output. Gaussian noise with $s=0.4$.}
		\label{Test2filf}
	\end{subfigure}
	\caption{(Test 2) (a), (c) and (d) The outputs $x_{\text{fil}}[j]$ of the NTEWT-based filter (reconstructed chirp) compared with the noiseless chirp pulse. (b), (d) and (f) Outputs of a matched chirp filter applied on the signals in Fig.~\ref{Test2Signal} (in blue) and on the NTEWT-filter outputs (in red). Analytic Morlet with $\sigma=5$.}
	\label{Test2fil}
\end{figure}

\vspace{-15pt}

\subsection{Test 3: short signal containing a short-duration, high-rate chirp pulse mixed with stationary harmonics and Gaussian noise}

\begin{figure}[H]
	\centering
	\begin{subfigure}{0.45\textwidth}
		\includegraphics[width=\textwidth]{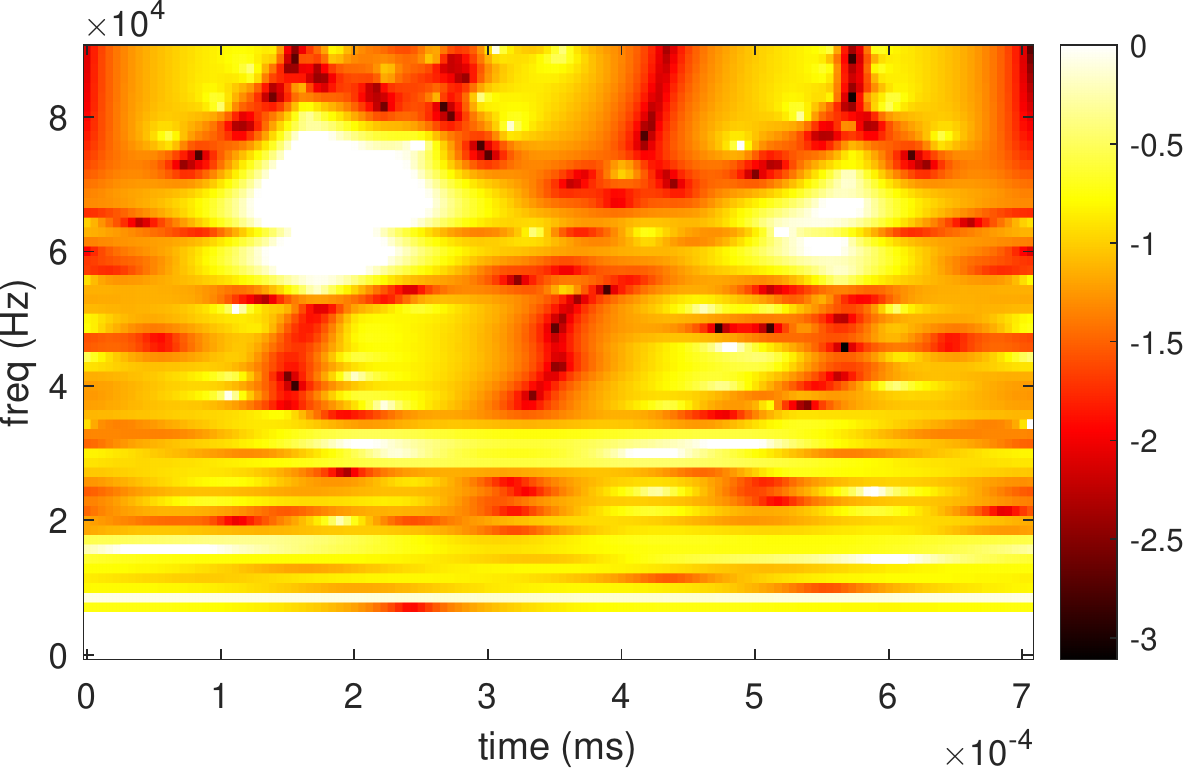}
		\caption{No Gaussian noise.}
	\end{subfigure}
	\hfill
	\begin{subfigure}{0.45\textwidth}
		\includegraphics[width=\textwidth]{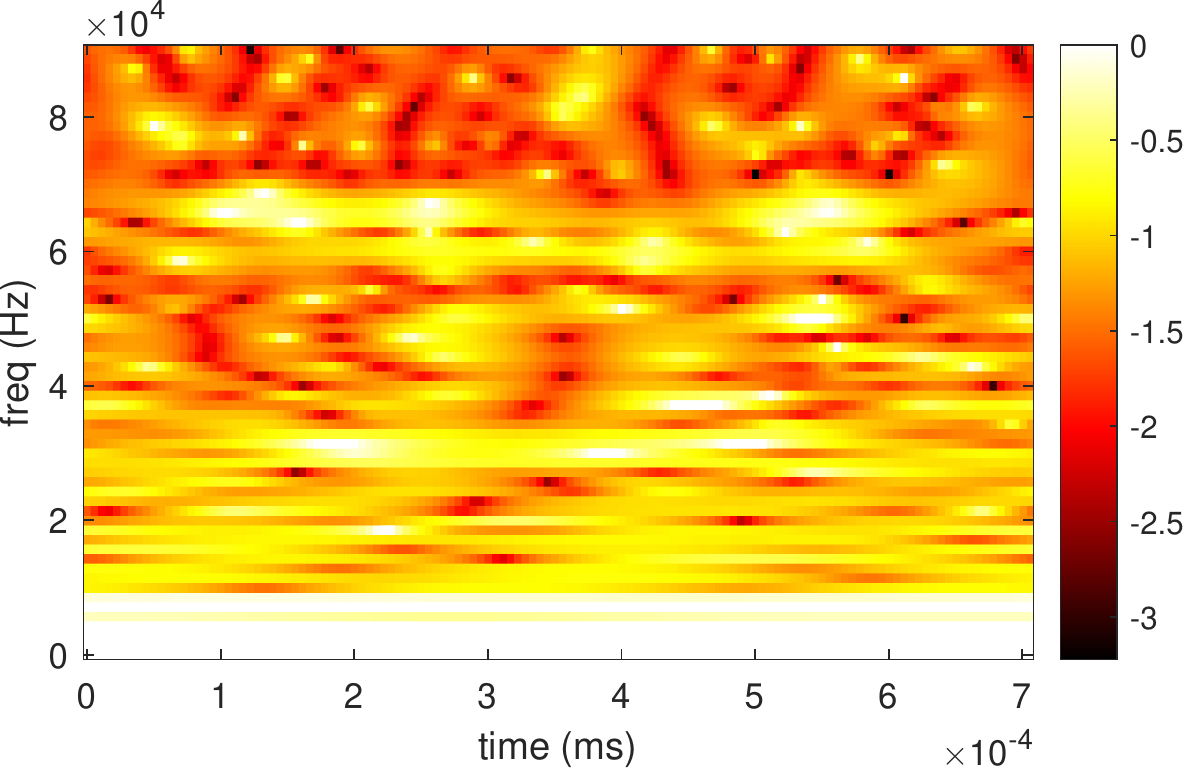}
		\caption{Gaussian noise with $s=0.4$.}
	\end{subfigure}
	\caption{(Test 3) The log of the fixed point metric of the CWT of $x[j]$, $|\bar{t}_x[j,k]-b[j]|$, with $\sigma=3$.}
	\label{Test3fixed}
\end{figure}

In this experiment, synthetic signals with only 128 samples ($\approx$ 0.000711 s) were generated to test the performance of the NTEWT filter when algorithm speed requirements constrain the length of the signals. A chirp with the same frequency bounds and duration as in Test~2 was added to the signal, as well as stationary harmonics with 30 and 60~kHz and white Gaussian noise. The three analyzed signals are shown in Fig.~\ref{Test3Signal}. The optimal values found for the reassignment tolerance $\epsilon$ and the wavelet width $\sigma$ for the new signals were respectively $\num{1e-2}$ and $\sigma=3$. As seen in Figs.~\ref{Test3NTEWT}, \ref{Test3fixed} and \ref{Test3fil}, shorter signal lengths lead to lower NTEWT-based filtering performances. This occurs because of the lower frequency resolution of the TFRs and because the border effect masks a larger proportion of them. Nevertheless, the results in Fig.~\ref{Test3fil} prove that with a proper selection of the NTEWT parameters chirp detection through matched filtering is more performing after NTEWT pre-filtering when the random noise levels are not too high.

\vspace{-5pt}

\subsection{Test 4: short signal containing a chirp pulse train mixed with Gaussian noise}

\begin{figure}[H]
	\centering
	\begin{subfigure}{0.45\textwidth}
		\includegraphics[width=\textwidth]{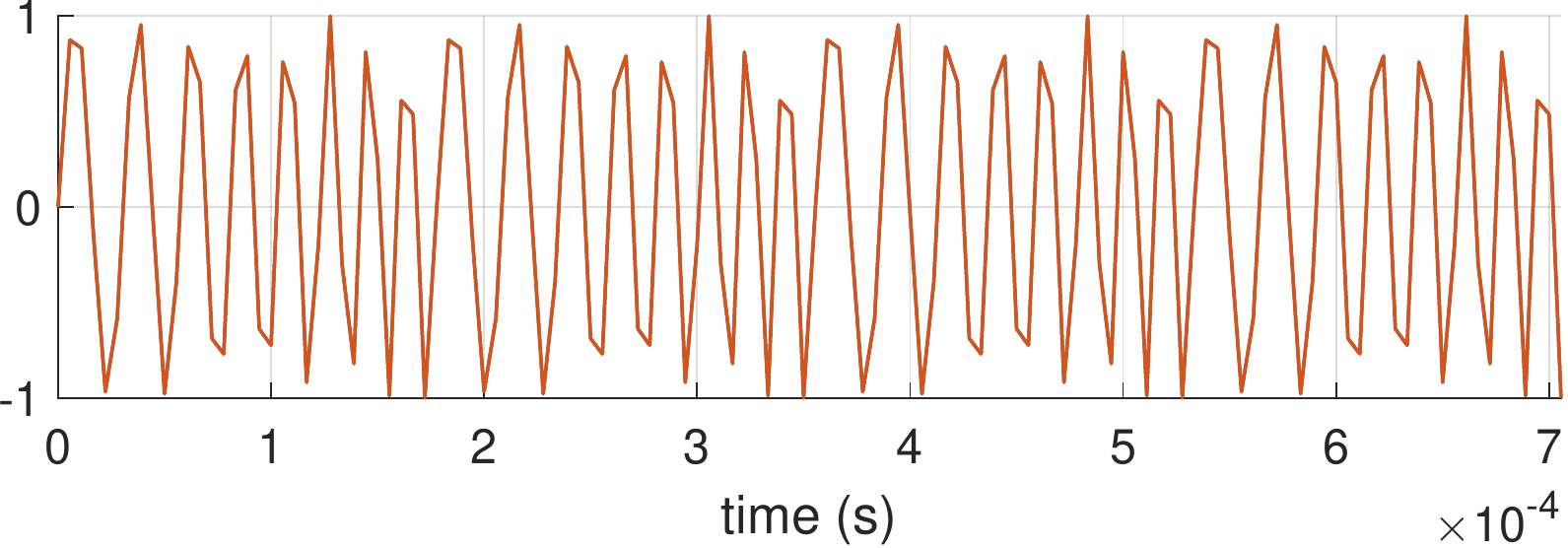}
		\caption{No Gaussian noise.}
		\label{Test4aSignal}
	\end{subfigure}
	\hfill
	\begin{subfigure}{0.45\textwidth}
		\includegraphics[width=\textwidth]{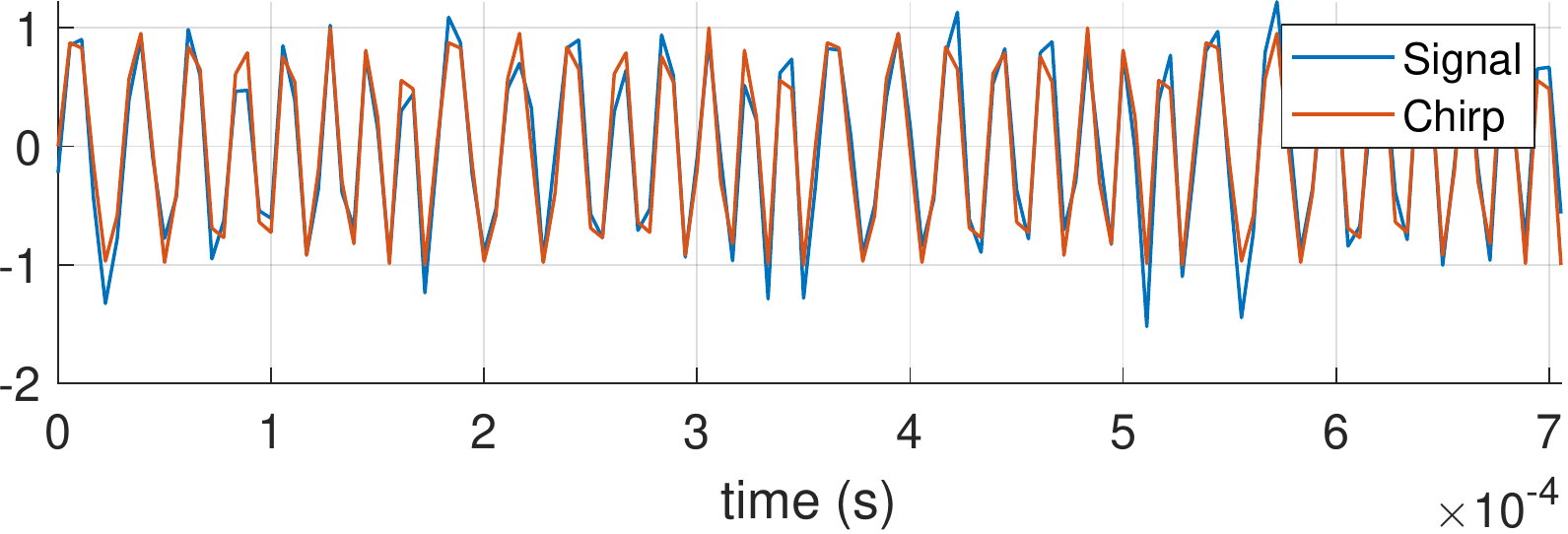}
		\caption{Gaussian noise with $s=0.2$.}
		\label{Test4bSignal}
	\end{subfigure}
	\begin{subfigure}{0.45\textwidth}
		\includegraphics[width=\textwidth]{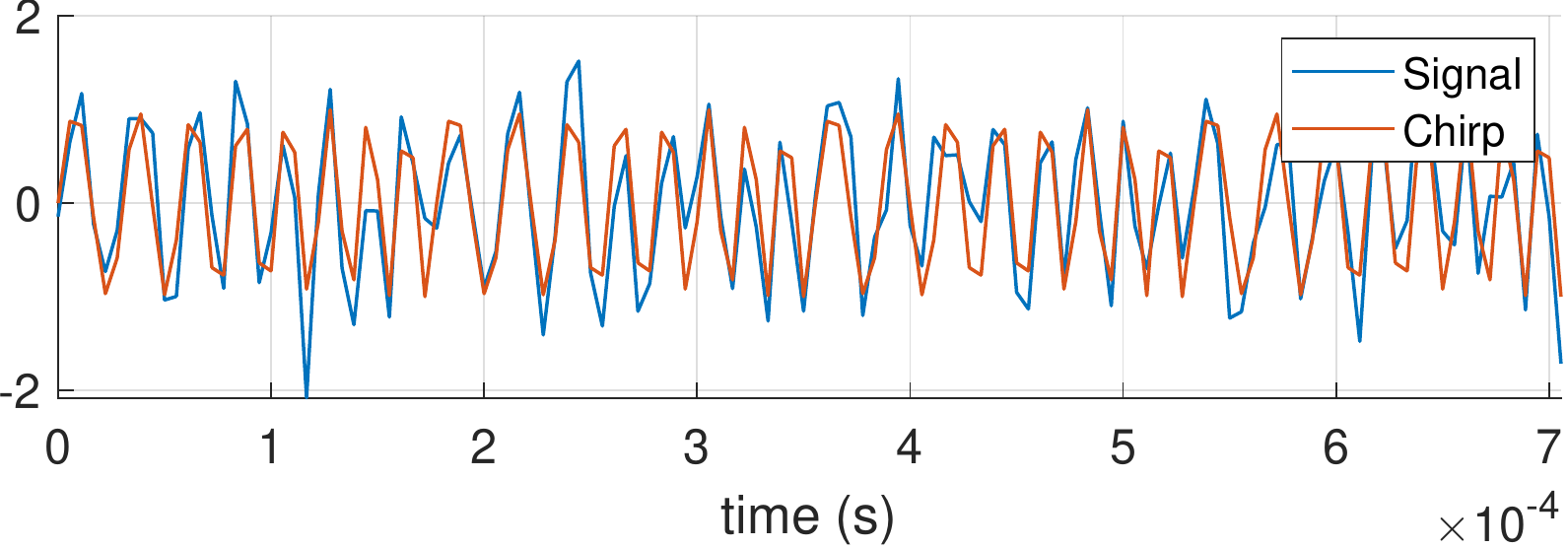}
		\caption{Gaussian noise with $s=0.4$.}
		\label{Test4cSignal}
	\end{subfigure}
	\caption{(Test 4) The chirp pulse train (in red) plus the Gaussian noise (in blue).}
	\label{Test4Signal}
\end{figure}

In practical radar and sonar applications, a train of chirp pulses with a predefined separation interval (i.e. a distance between consecutive pulses) is often emitted \cite{richards2012principles}. If some of those pulses encounter an obstacle, they bounce back to a receiver. The receiver's task is to detect those pulses and to calculate their time delay and frequency shift with respect to the emitted pulses in order to estimate target's distance and velocity. A low separation interval allows for a higher spatial resolution but may be challenging from the point of view of signal processing. Here, the worst case scenario will be considered for the assessment of the proposed filtering methodology: that a train of chirp pulses with a pulse repetition interval equal to the chirp pulses' lengths is received. Fig.~\ref{Test4Signal} shows three synthetic signals with a length of 128 samples ($\approx$ 0.000711 s) containing four concatenated chirp pulses. The resulting chirp train has been mixed with three different levels of Gaussian noise: $s=0$ (Fig.~\ref{Test4aSignal}), $s=0.2$ (Fig.~\ref{Test4bSignal}) and $s=0.4$ (Fig.~\ref{Test4cSignal}). The parameters of the individual pulses are these:

\begin{table}[H]
	\centering
	\begin{tabular}{ll}
		\hline
		Frequency lower bound (kHz) & 30                                   \\
		Frequency upper bound (kHz) & 60                                   \\
		Pulse length (samples)      & 32 ($\approx$ 0.000178 s) \\ \hline
	\end{tabular}
\end{table}

\vspace{-5pt}

\begin{figure}[h]
	\centering
	\begin{subfigure}{0.45\textwidth}
		\includegraphics[width=\textwidth]{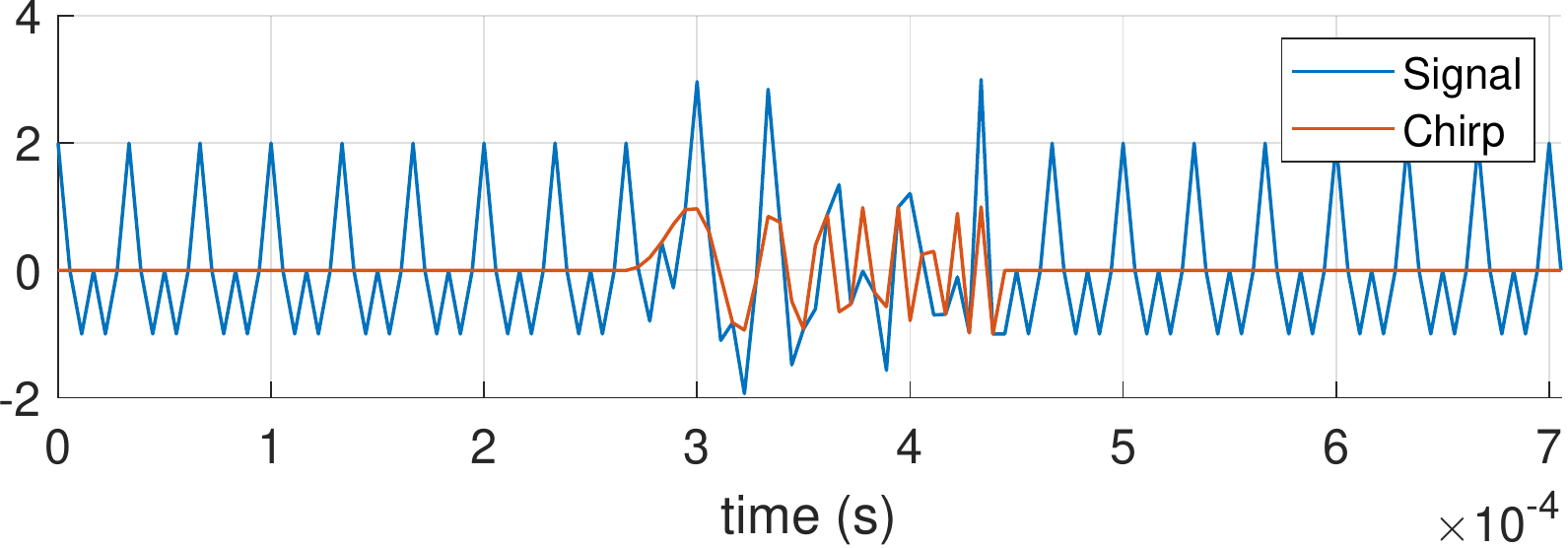}
		\caption{No Gaussian noise.}
		\label{Test3aSignal}
	\end{subfigure}
	\hfill
	\begin{subfigure}{0.45\textwidth}
		\includegraphics[width=\textwidth]{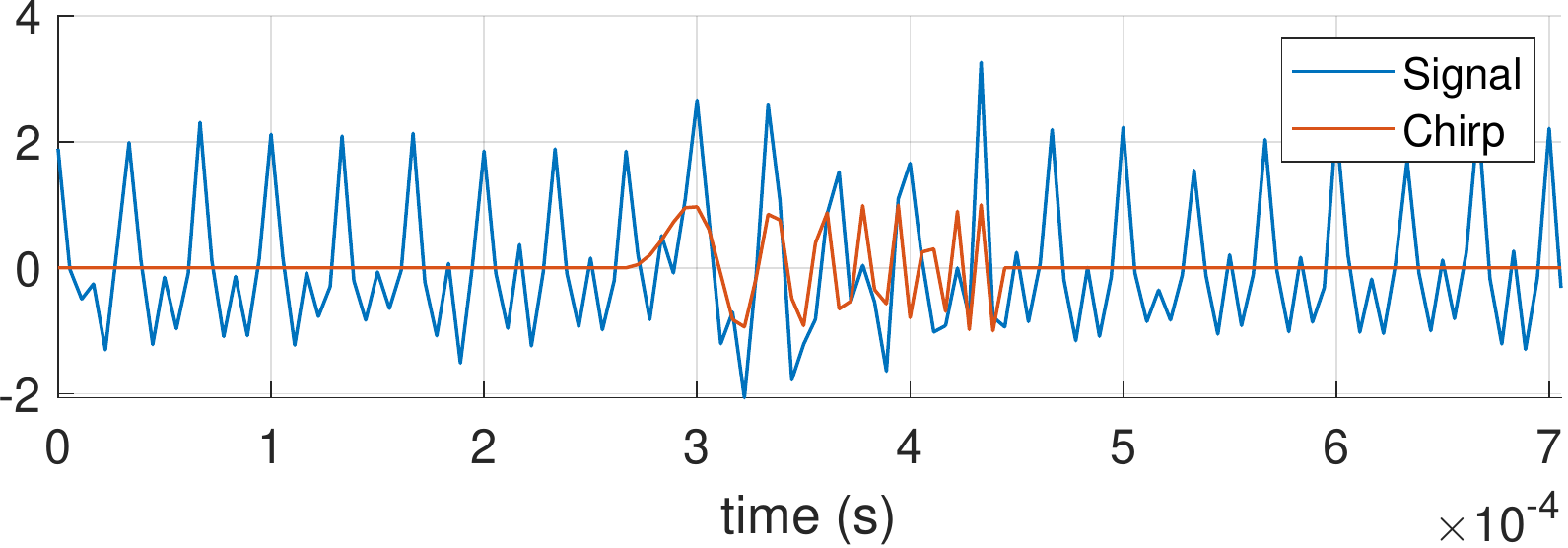}
		\caption{Gaussian noise with $s=0.2$.}
	\end{subfigure}
	\begin{subfigure}{0.45\textwidth}
		\includegraphics[width=\textwidth]{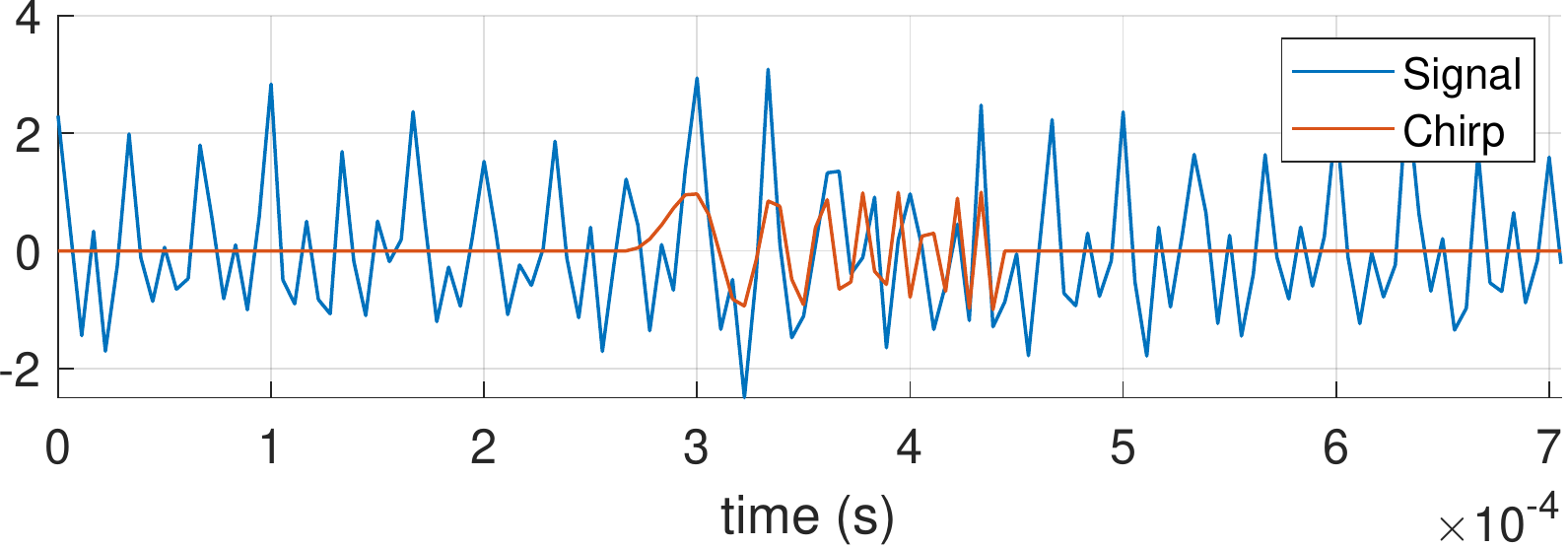}
		\caption{Gaussian noise with $s=0.4$.}
	\end{subfigure}
	\caption{(Test 3) The chirp pulse (in red) plus the stationary harmonics and the Gaussian noise (in blue).}
	\label{Test3Signal}
\end{figure}

\vspace{25pt}

\begin{figure}[H]
	\centering
	\begin{subfigure}{0.45\textwidth}
		\includegraphics[width=\textwidth]{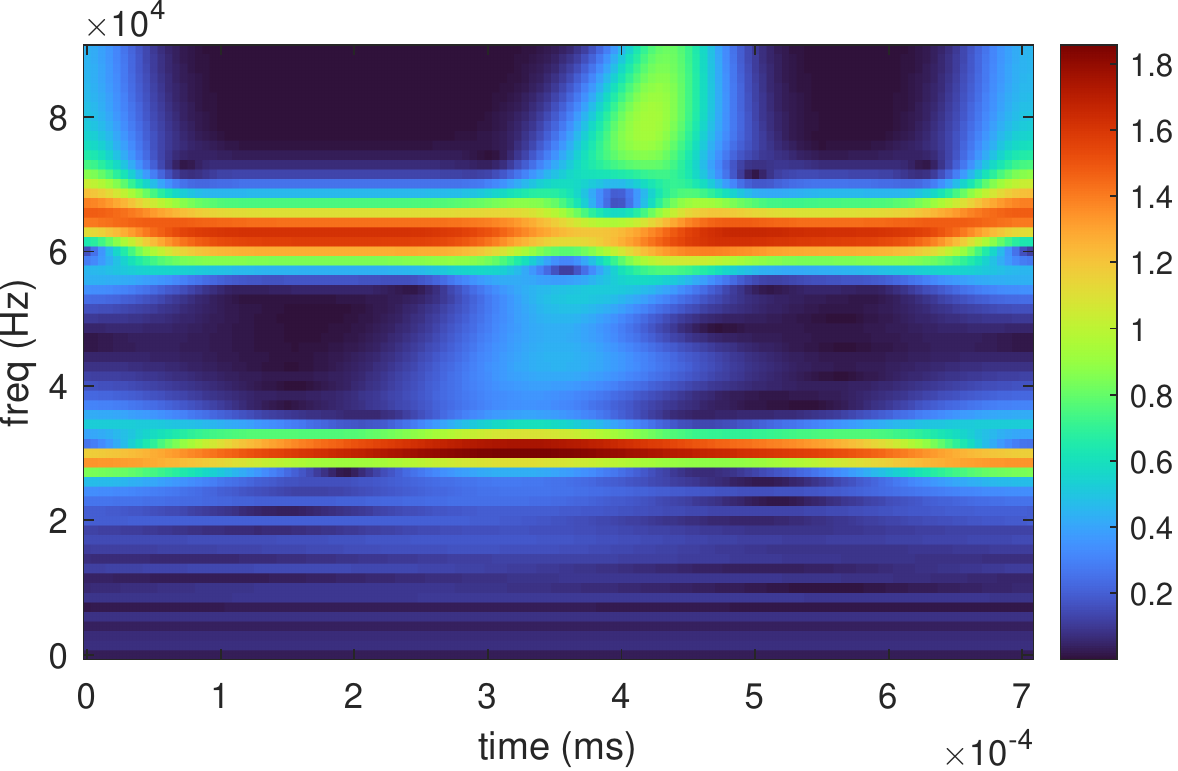}
		\caption{CWT scalogram. No Gaussian noise.}
	\end{subfigure}
	\hfill
	\begin{subfigure}{0.45\textwidth}
		\includegraphics[width=\textwidth]{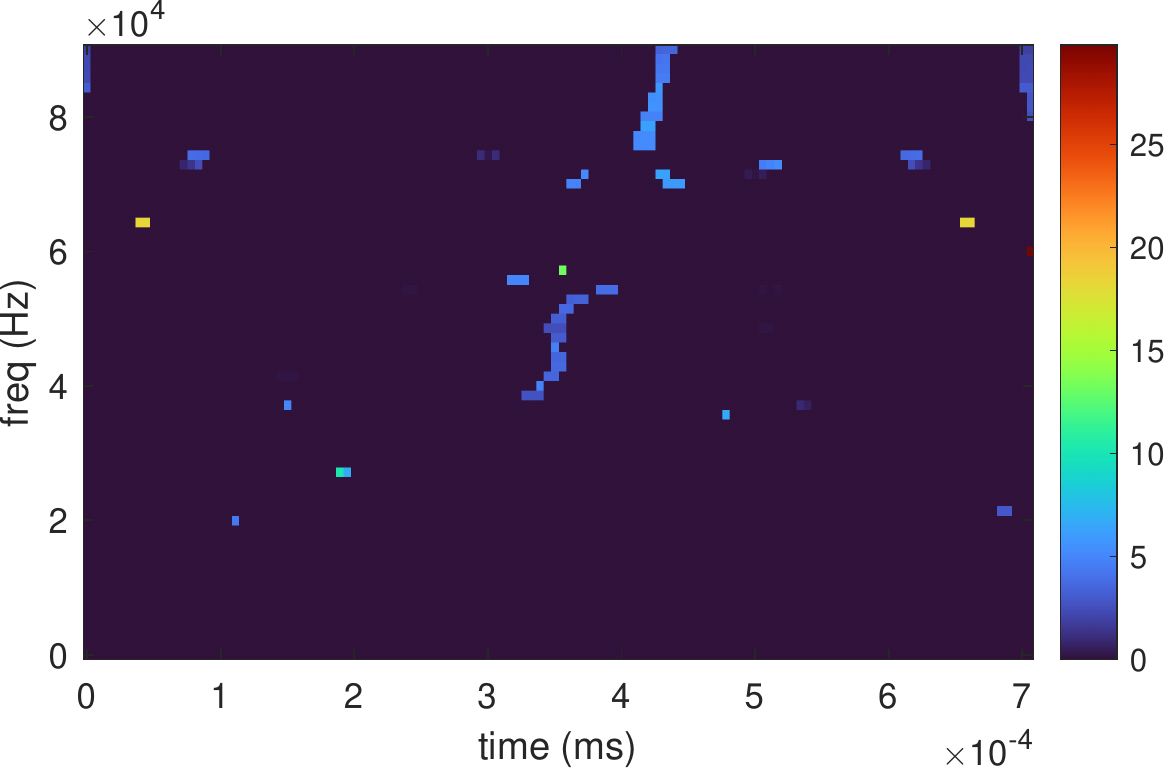}
		\caption{NTEWT scalogram. No Gaussian noise.}
	\end{subfigure}
	\begin{subfigure}{0.45\textwidth}
		\includegraphics[width=\textwidth]{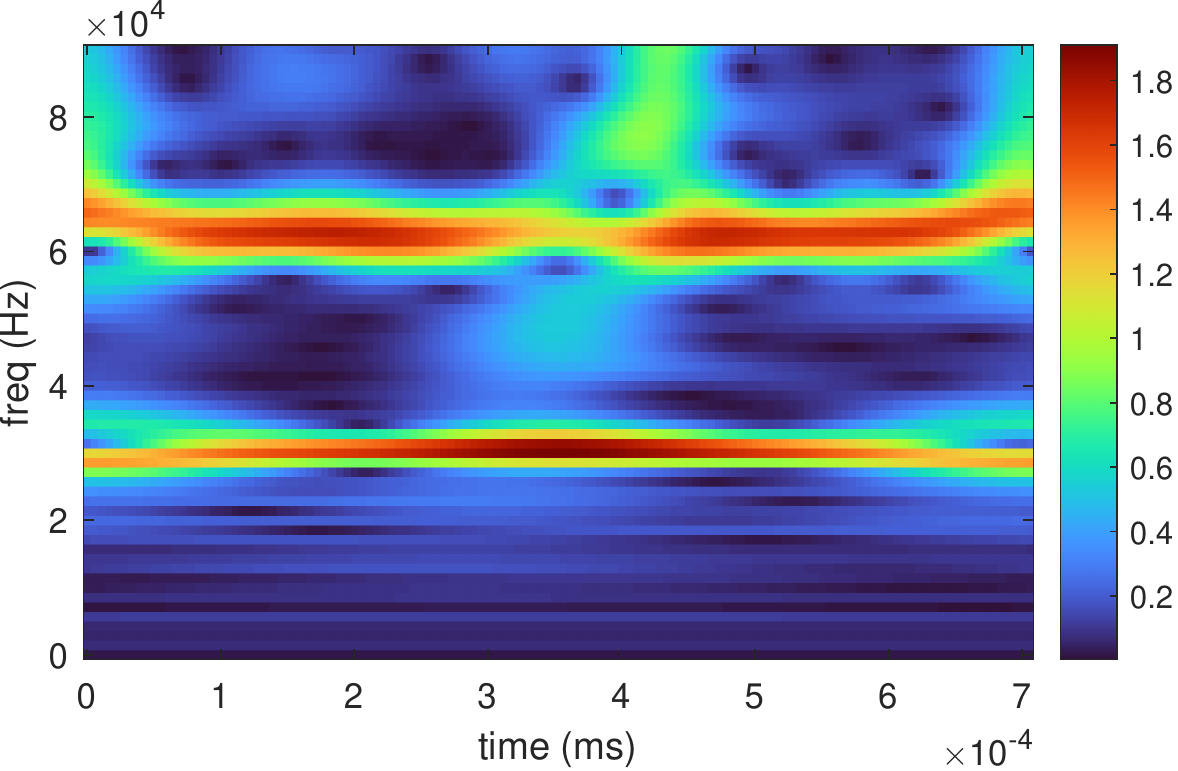}
		\caption{CWT scalogram. Gaussian noise with $s=0.2$.}
	\end{subfigure}
	\hfill
	\begin{subfigure}{0.45\textwidth}
		\includegraphics[width=\textwidth]{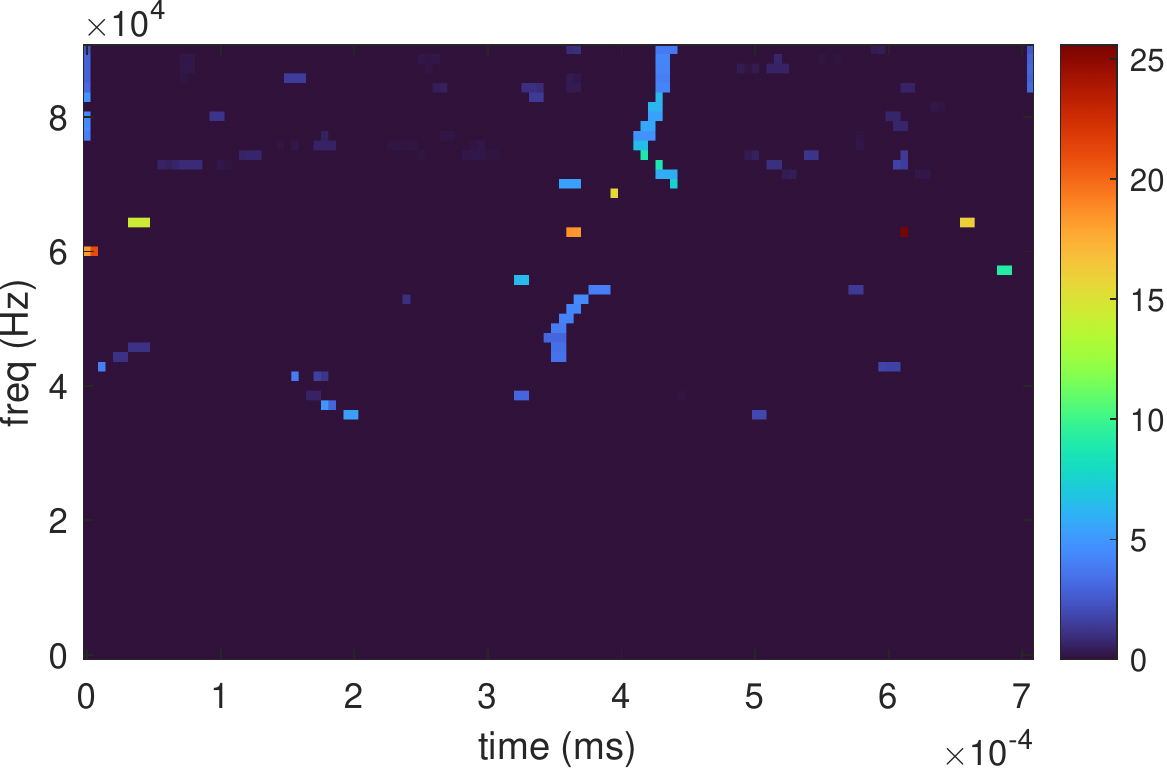}
		\caption{NTEWT scalogram. Gaussian noise with $s=0.2$.}
	\end{subfigure}
\end{figure}
\clearpage
\begin{figure}[H]\ContinuedFloat
	\begin{subfigure}{0.45\textwidth}
		\includegraphics[width=\textwidth]{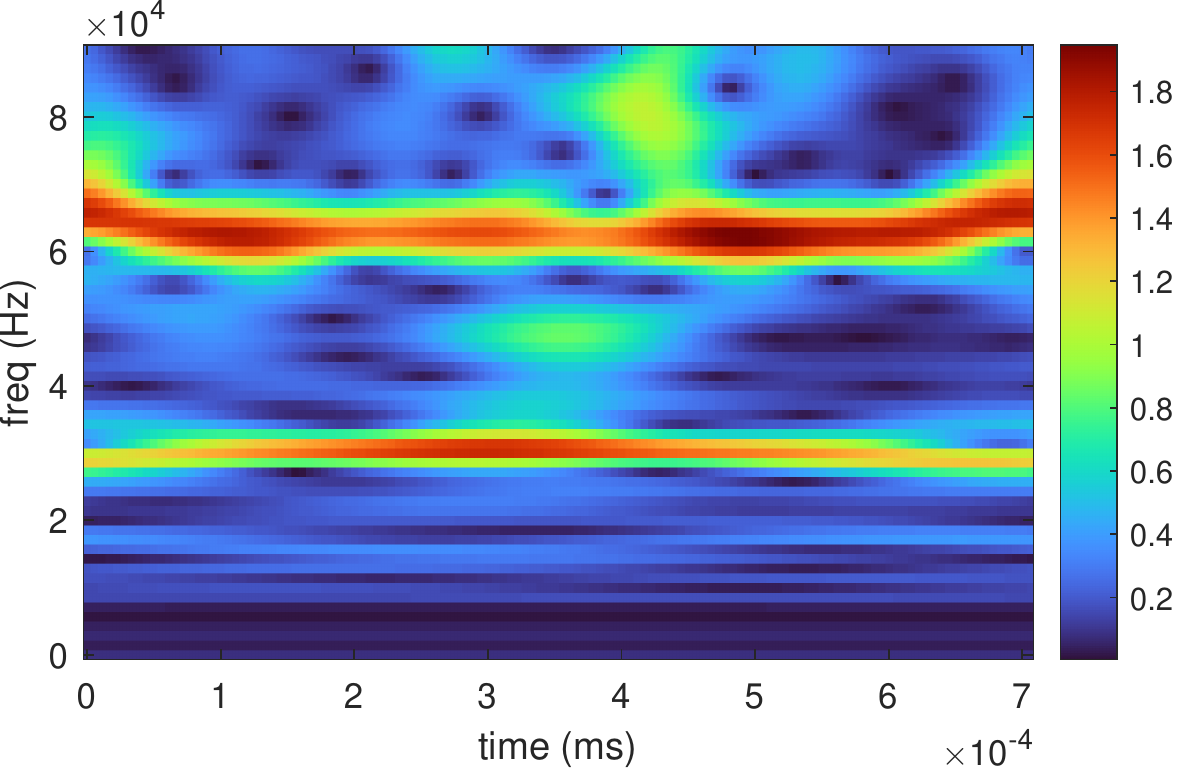}
		\caption{CWT scalogram. Gaussian noise with $s=0.4$.}
	\end{subfigure}
	\hfill
	\begin{subfigure}{0.45\textwidth}
		\includegraphics[width=\textwidth]{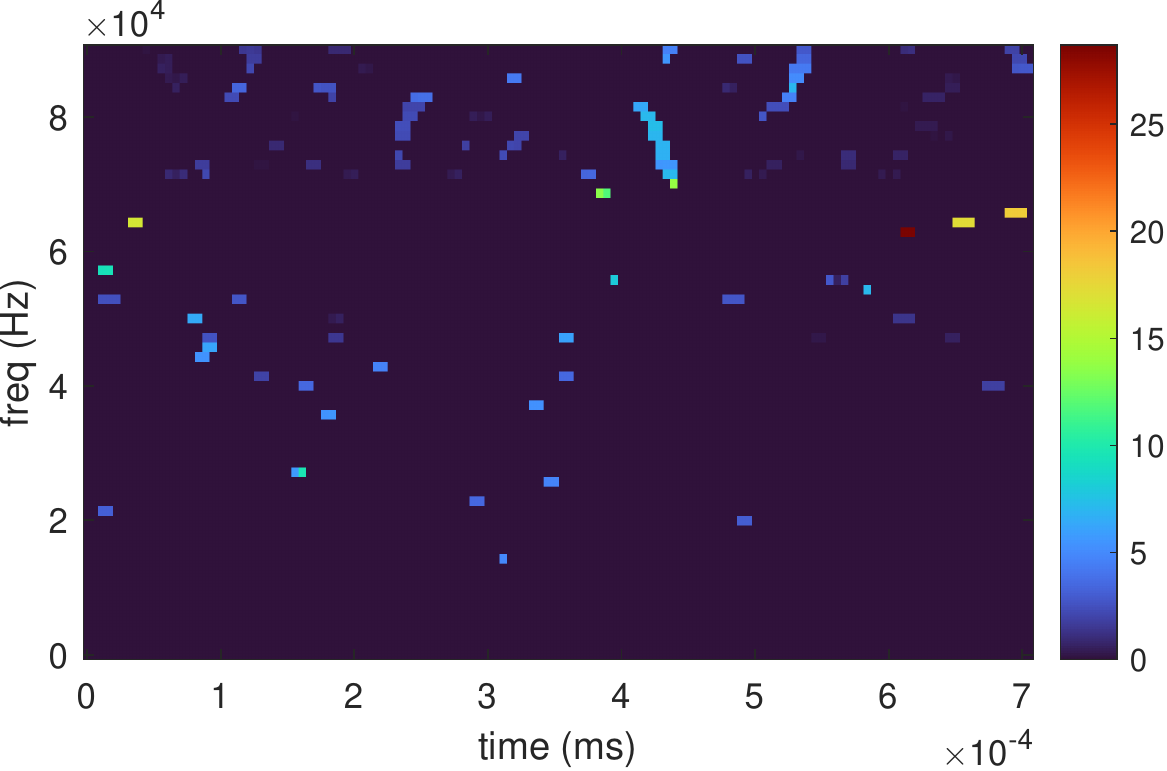}
		\caption{NTEWT scalogram. Gaussian noise with $s=0.4$.}
	\end{subfigure}
	\caption{(Test 3) CWT and NTEWT scalograms of the signals in Fig~\ref{Test3Signal} with $\sigma=3$.}
	\label{Test3NTEWT}
\end{figure}

\vspace{50pt}

\vspace{50pt}

\begin{figure}[H]
	\centering
	\begin{subfigure}{0.45\textwidth}
		\includegraphics[width=\textwidth]{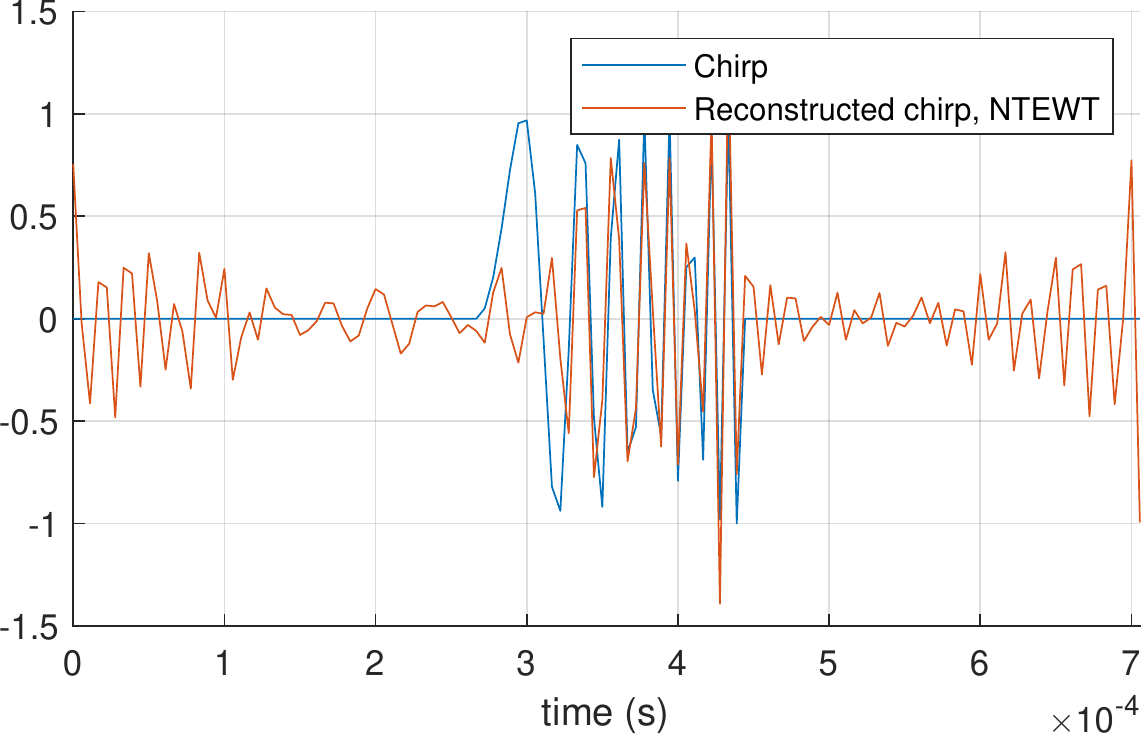}
		\caption{$x_{\text{fil}}[j]$. No Gaussian noise.}
	\end{subfigure}
	\hfill
	\begin{subfigure}{0.45\textwidth}
		\includegraphics[width=\textwidth]{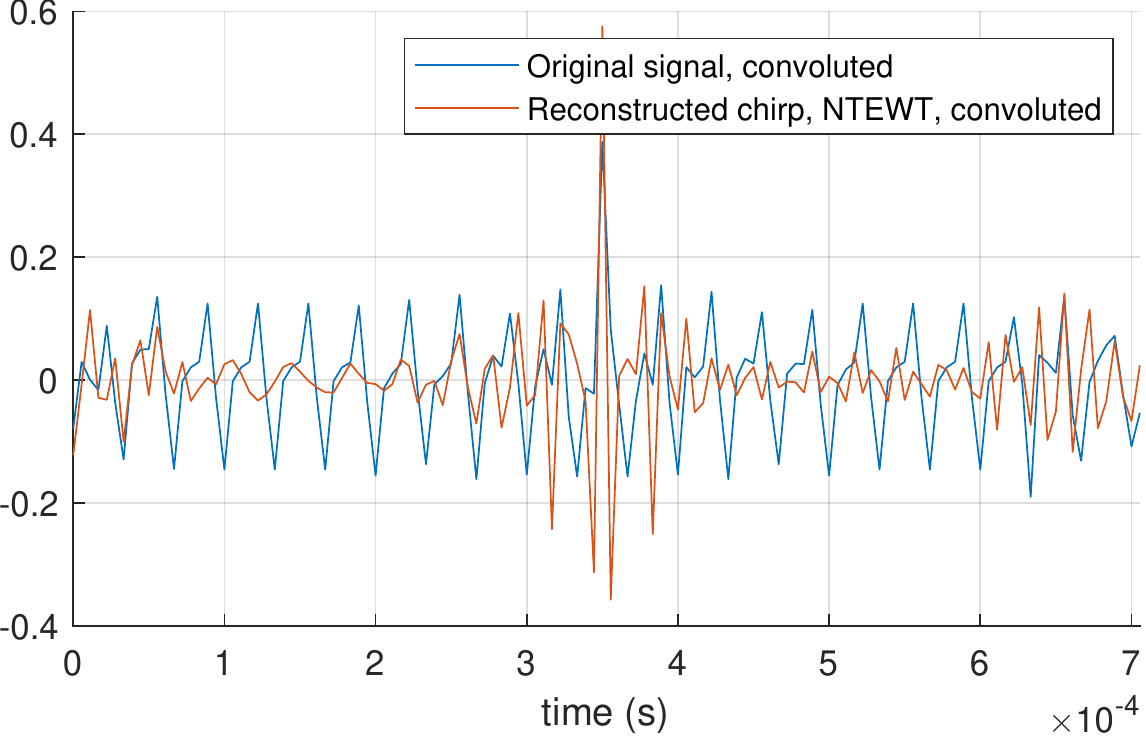}
		\caption{Matched filter output. No Gaussian noise.}
	\end{subfigure}
\end{figure}
\clearpage
\begin{figure}[H]\ContinuedFloat
	\begin{subfigure}{0.45\textwidth}
		\includegraphics[width=\textwidth]{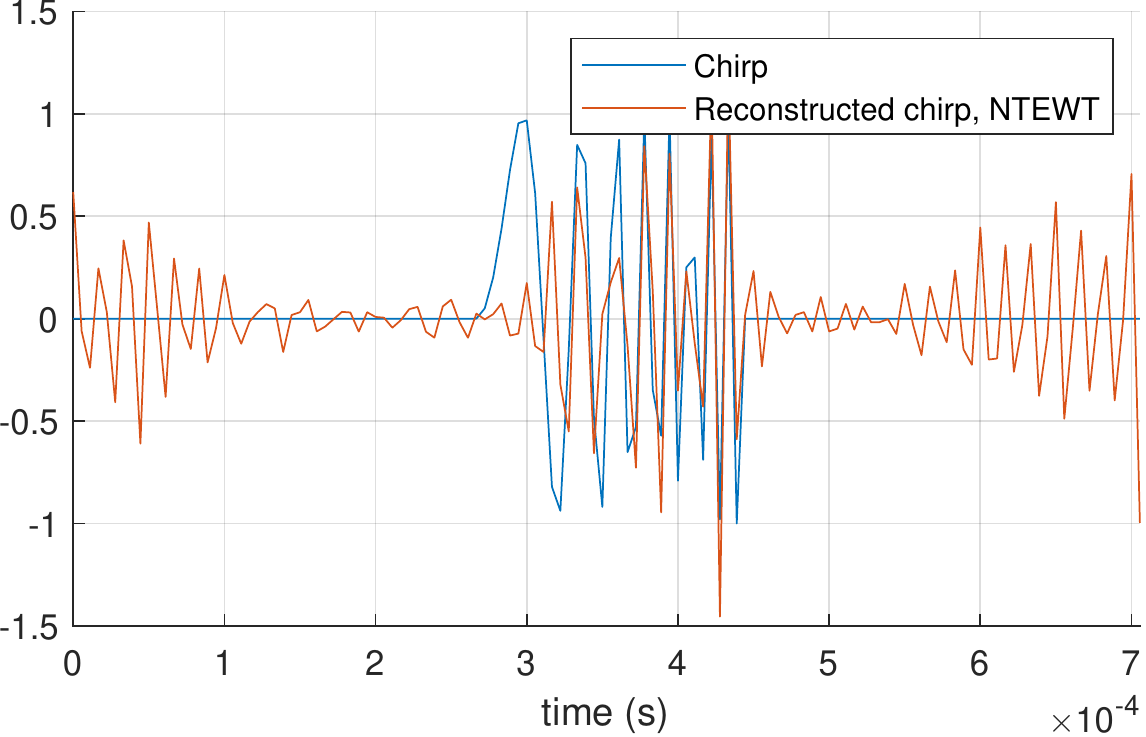}
		\caption{$x_{\text{fil}}[j]$. Gaussian noise with $s=0.2$.}
	\end{subfigure}
	\hfill
	\begin{subfigure}{0.45\textwidth}
		\includegraphics[width=\textwidth]{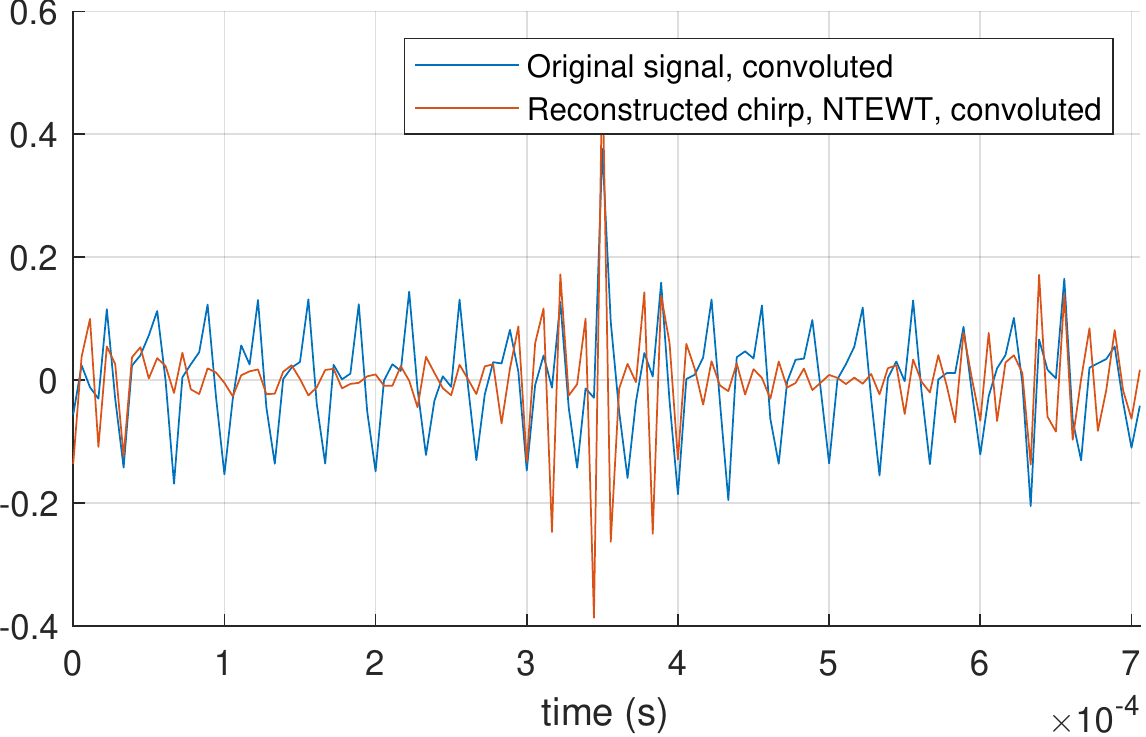}
		\caption{Matched filter output. Gaussian noise with $s=0.2$.}
	\end{subfigure}
	\begin{subfigure}{0.45\textwidth}
		\includegraphics[width=\textwidth]{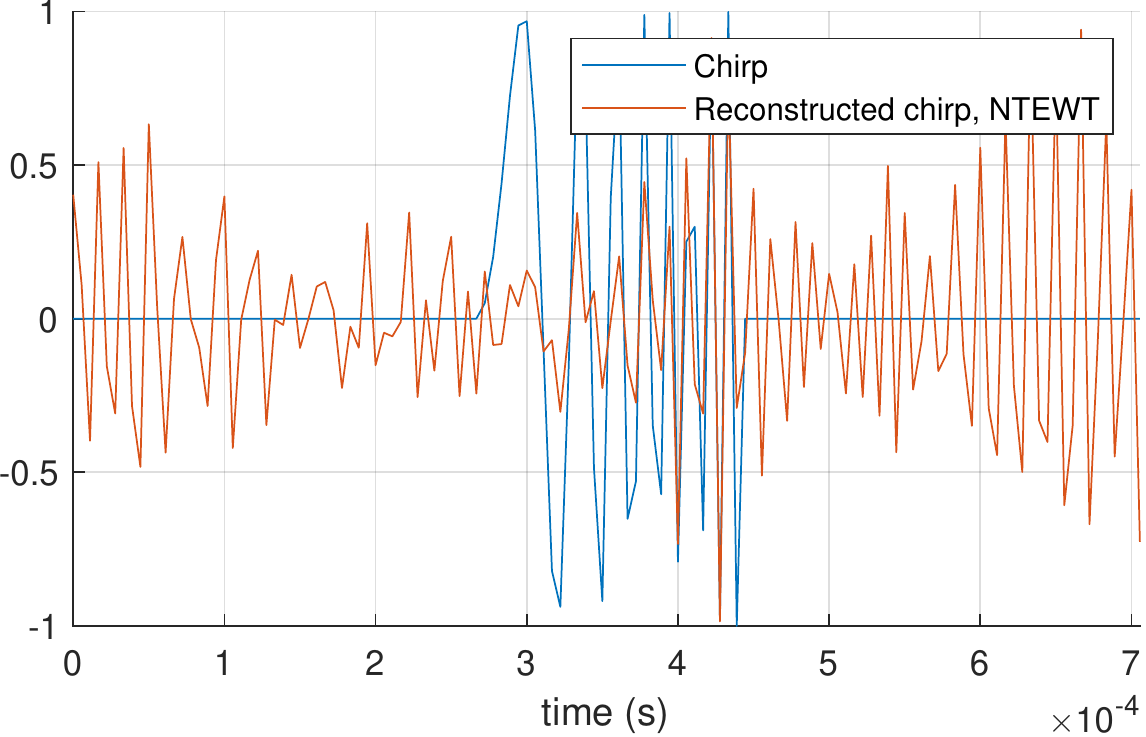}
		\caption{$x_{\text{fil}}[j]$. Gaussian noise with $s=0.4$.}
	\end{subfigure}
	\hfill
	\begin{subfigure}{0.45\textwidth}
		\includegraphics[width=\textwidth]{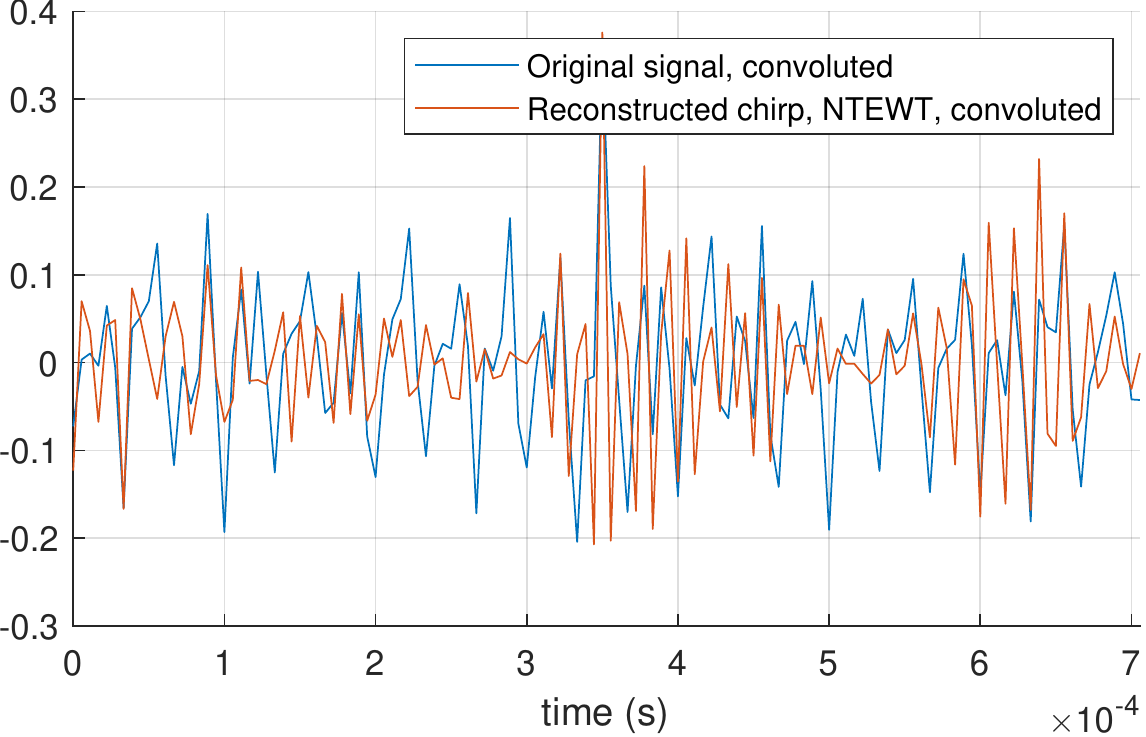}
		\caption{(Test 3) Matched filter output. Gaussian noise with $s=0.4$.}
	\end{subfigure}
	\caption{(a), (c) and (d) The outputs $x_{\text{fil}}[j]$ of the NTEWT-based filter (reconstructed chirp) compared with the noiseless chirp pulse. (b), (d) and (f) Outputs of a matched chirp filter applied on the signals in Fig.~\ref{Test3Signal} (in blue) and on the NTEWT-filter outputs (in red). Analytic Morlet with $\sigma=3$.}
	\label{Test3fil}
\end{figure}

\begin{figure}[H]
	\centering
	\begin{subfigure}{0.45\textwidth}
		\includegraphics[width=\textwidth]{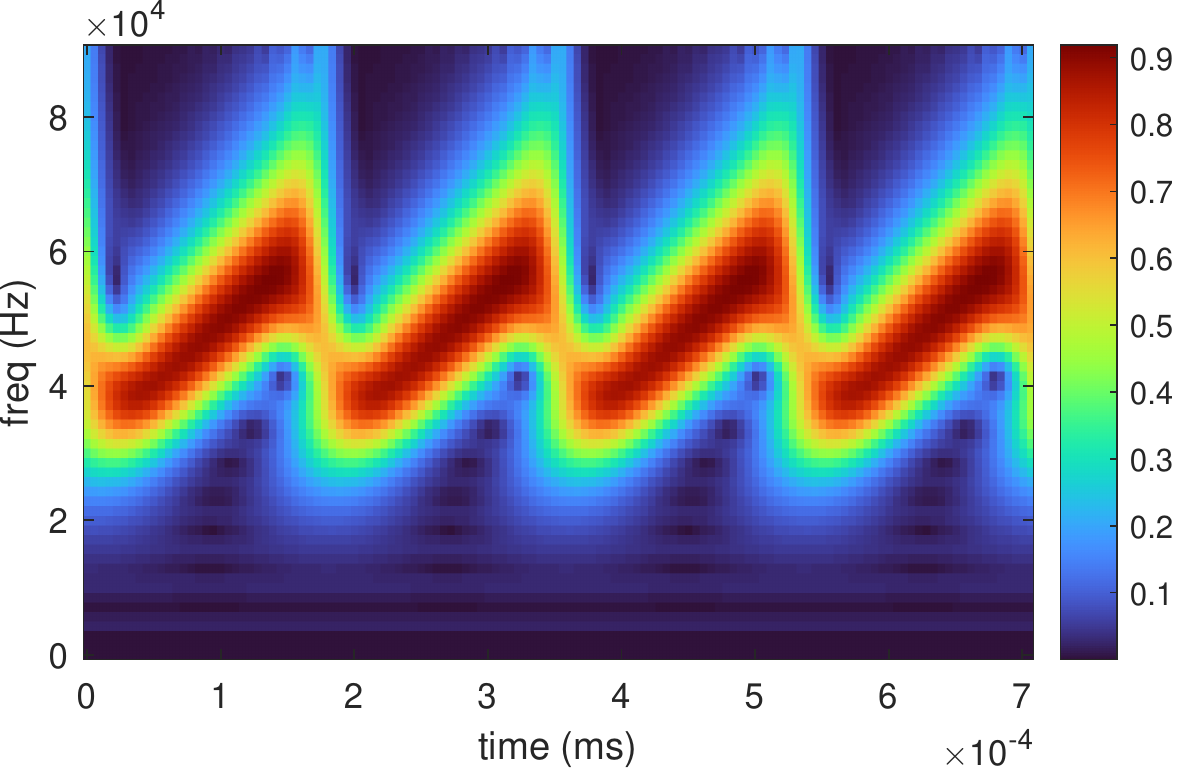}
		\caption{CWT scalogram. $\sigma=1$.}
	\end{subfigure}
	\hfill
	\begin{subfigure}{0.45\textwidth}
		\includegraphics[width=\textwidth]{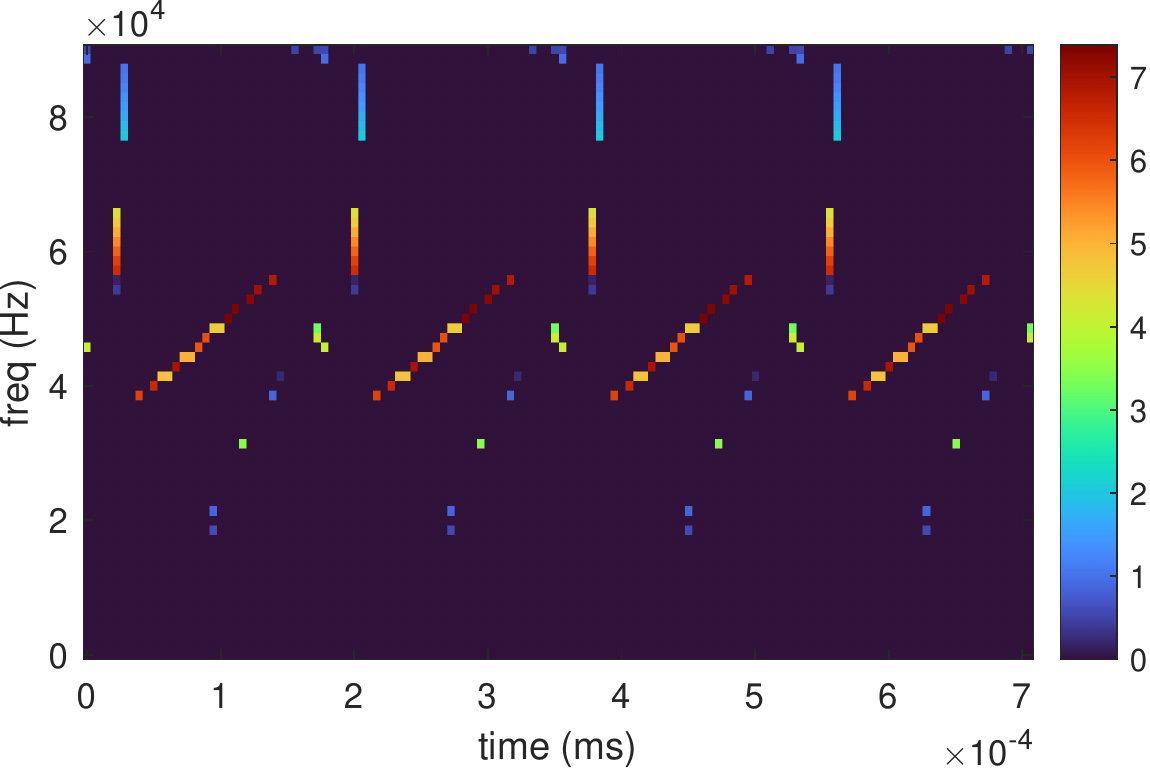}
		\caption{NTEWT scalogram. $\sigma=1$. $\epsilon=\num{5e-3}$.}
		\label{Test4NTEWTb}
	\end{subfigure}
	\begin{subfigure}{0.45\textwidth}
		\includegraphics[width=\textwidth]{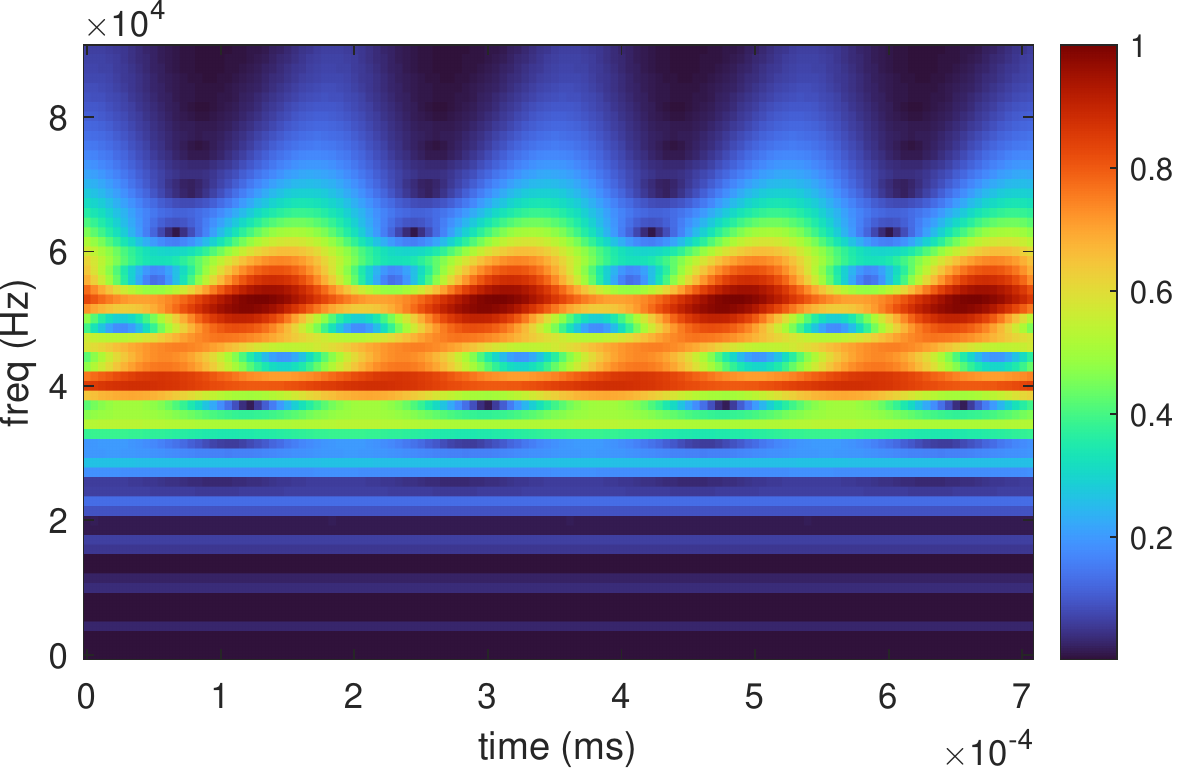}
		\caption{CWT scalogram. $\sigma=3$.}
	\end{subfigure}
	\hfill
	\begin{subfigure}{0.45\textwidth}
		\includegraphics[width=\textwidth]{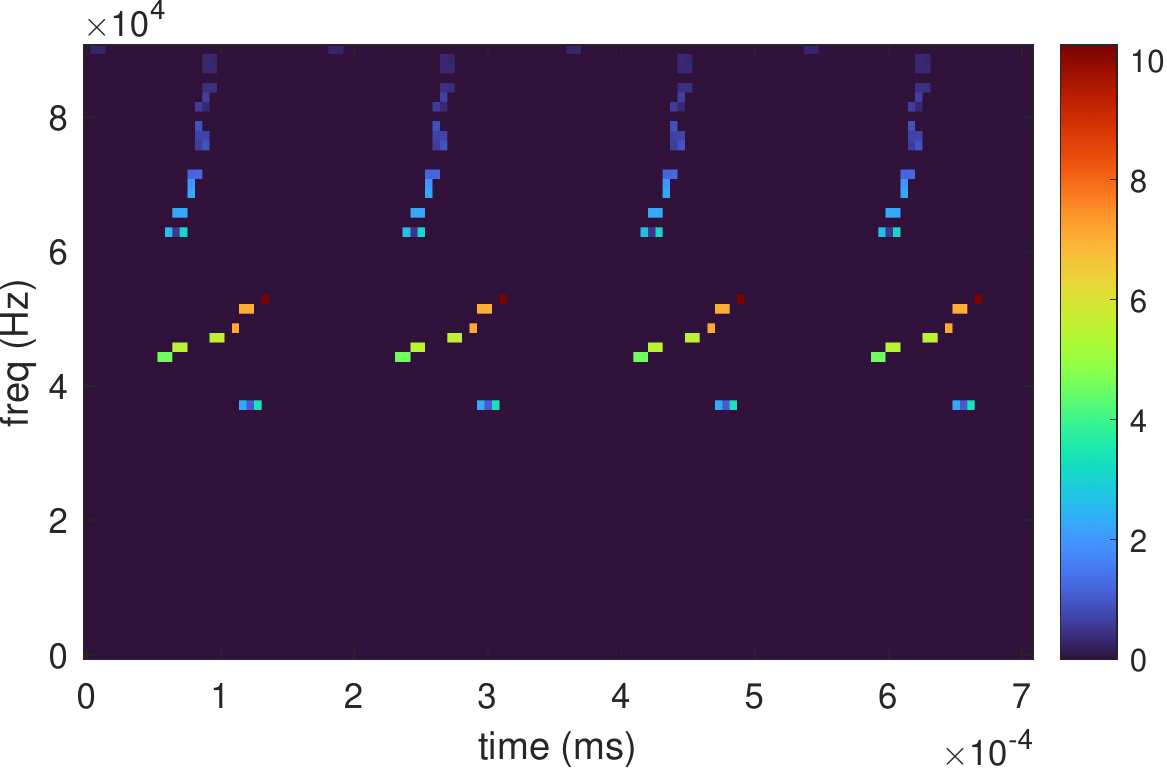}
		\caption{NTEWT scalogram. $\sigma=3$. $\epsilon=\num{1e-2}$.}
		\label{Test4NTEWTd}
	\end{subfigure}
	\begin{subfigure}{0.45\textwidth}
		\includegraphics[width=\textwidth]{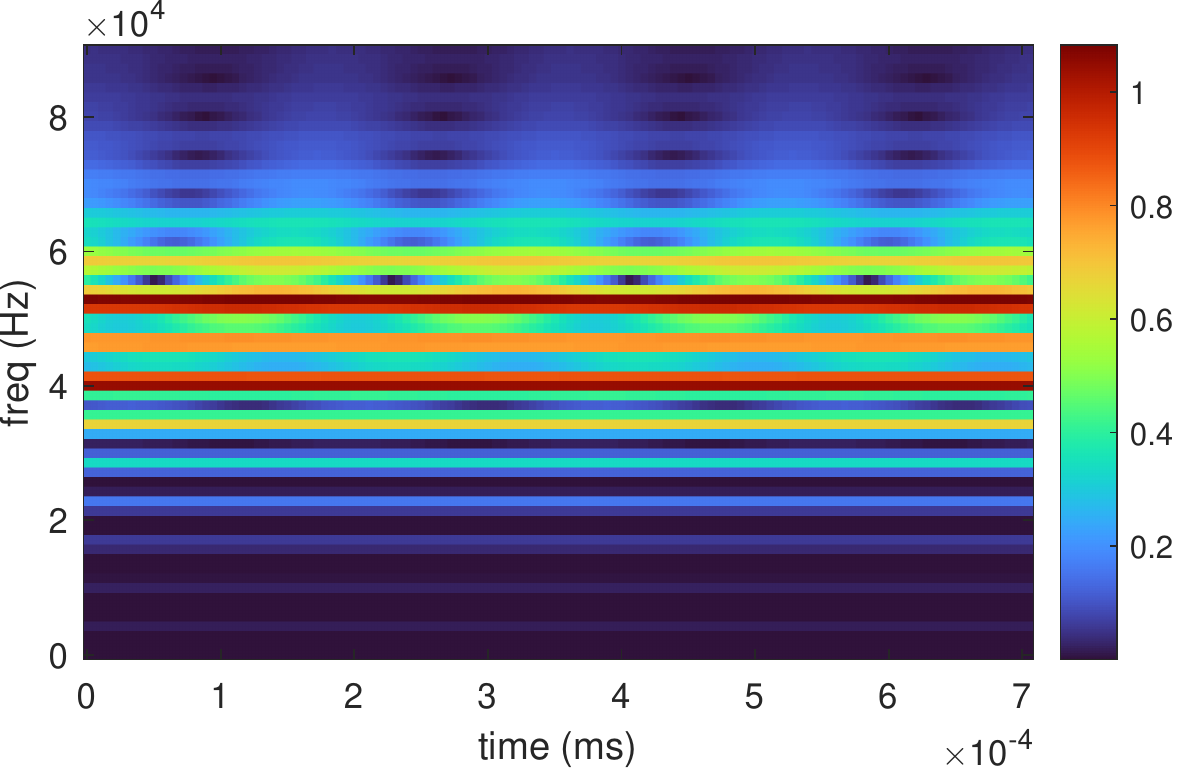}
		\caption{CWT scalogram. $\sigma=5$.}
	\end{subfigure}
	\hfill
	\begin{subfigure}{0.45\textwidth}
		\includegraphics[width=\textwidth]{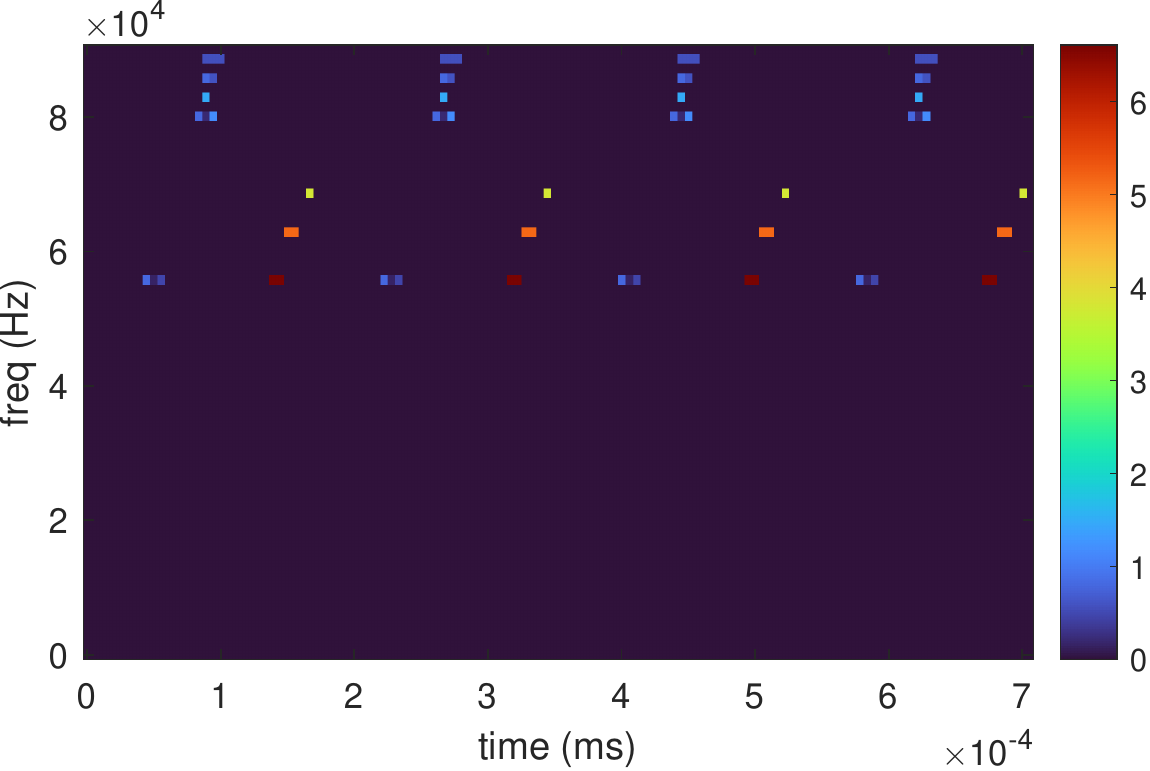}
		\caption{NTEWT scalogram. $\sigma=5$. $\epsilon=\num{1e-2}$.}
		\label{Test4NTEWTf}
	\end{subfigure}
	\caption{(Test 4) CWT and NTEWT scalograms of the signal in Fig~\ref{Test4aSignal} with different values of $\sigma$.}
	\label{Test4NTEWT}
\end{figure}

\begin{figure}[H]
	\centering
	\begin{subfigure}{0.45\textwidth}
		\includegraphics[width=\textwidth]{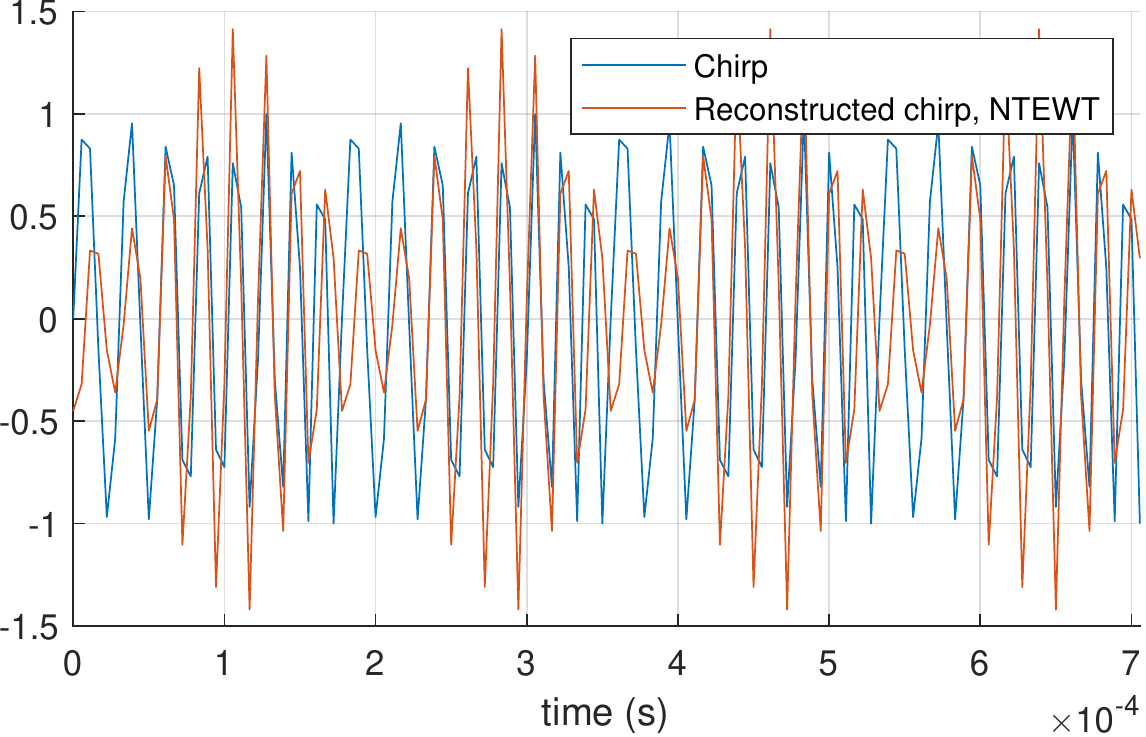}
		\caption{$x_{\text{fil}}[j]$. No Gaussian noise. $\epsilon=\num{1e-2}$.}
	\end{subfigure}
	\hfill
	\begin{subfigure}{0.45\textwidth}
		\includegraphics[width=\textwidth]{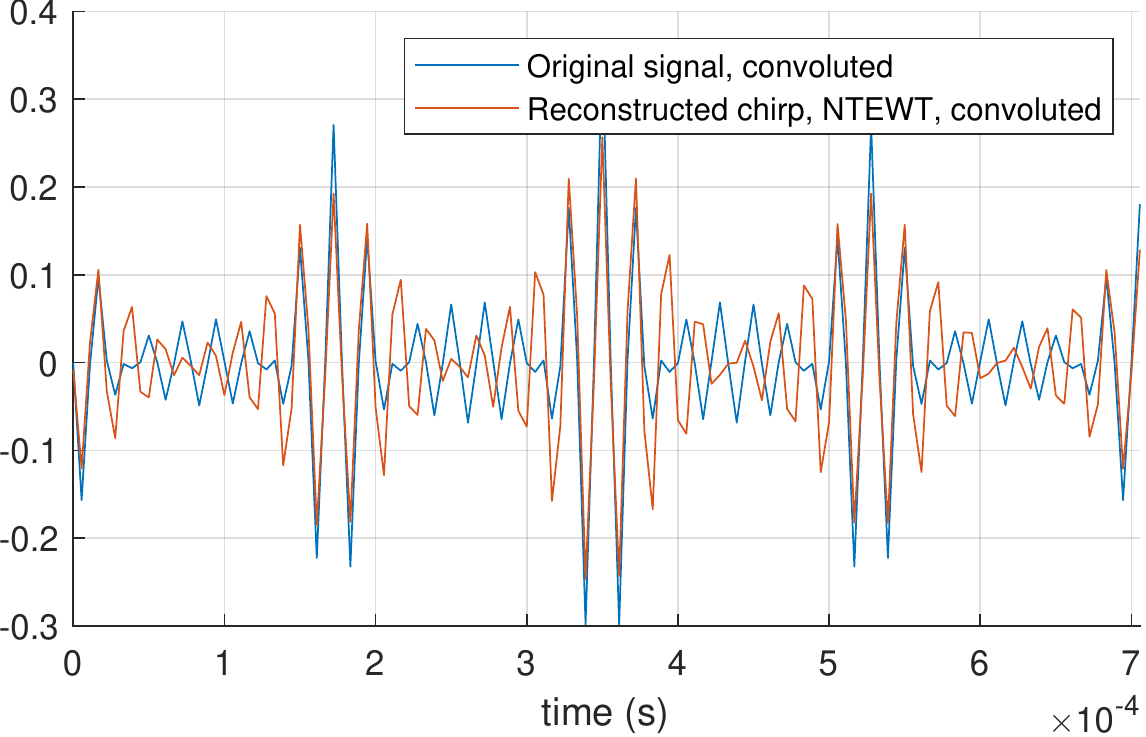}
		\caption{Matched filter output. No Gaussian noise. $\epsilon=\num{1e-2}$.}
	\end{subfigure}
	\begin{subfigure}{0.45\textwidth}
		\includegraphics[width=\textwidth]{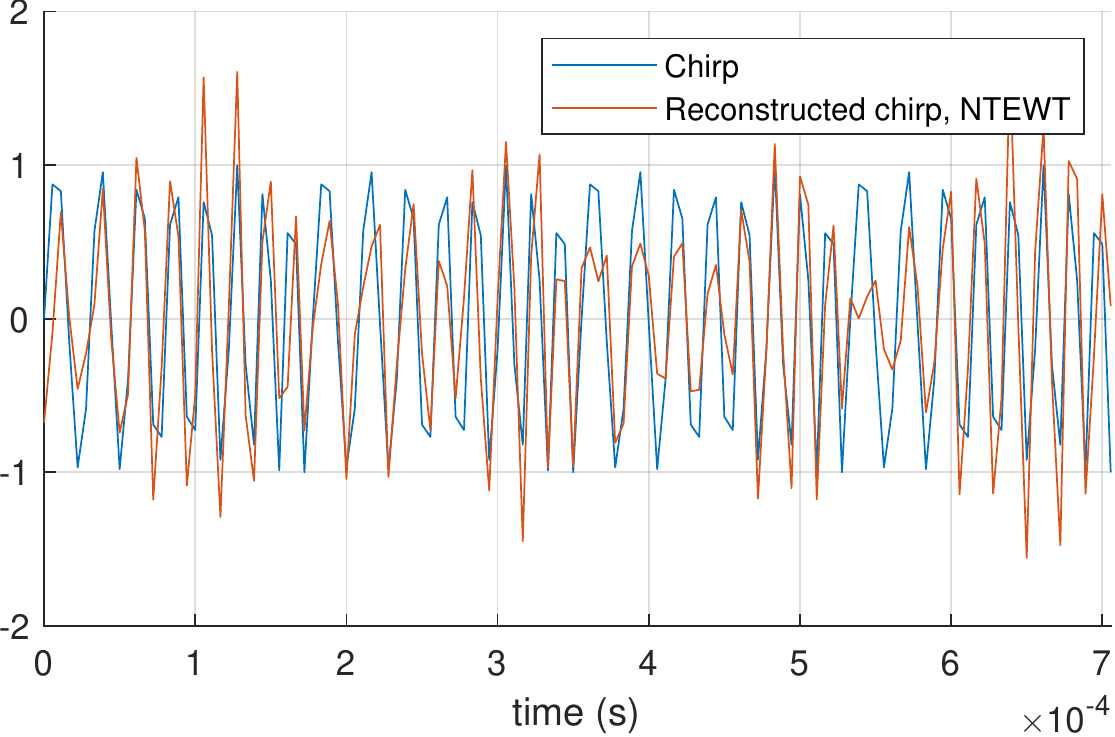}
		\caption{$x_{\text{fil}}[j]$. Gaussian noise with $s=0.2$. $\epsilon=\num{2e-2}$.}
	\end{subfigure}
	\hfill
	\begin{subfigure}{0.45\textwidth}
		\includegraphics[width=\textwidth]{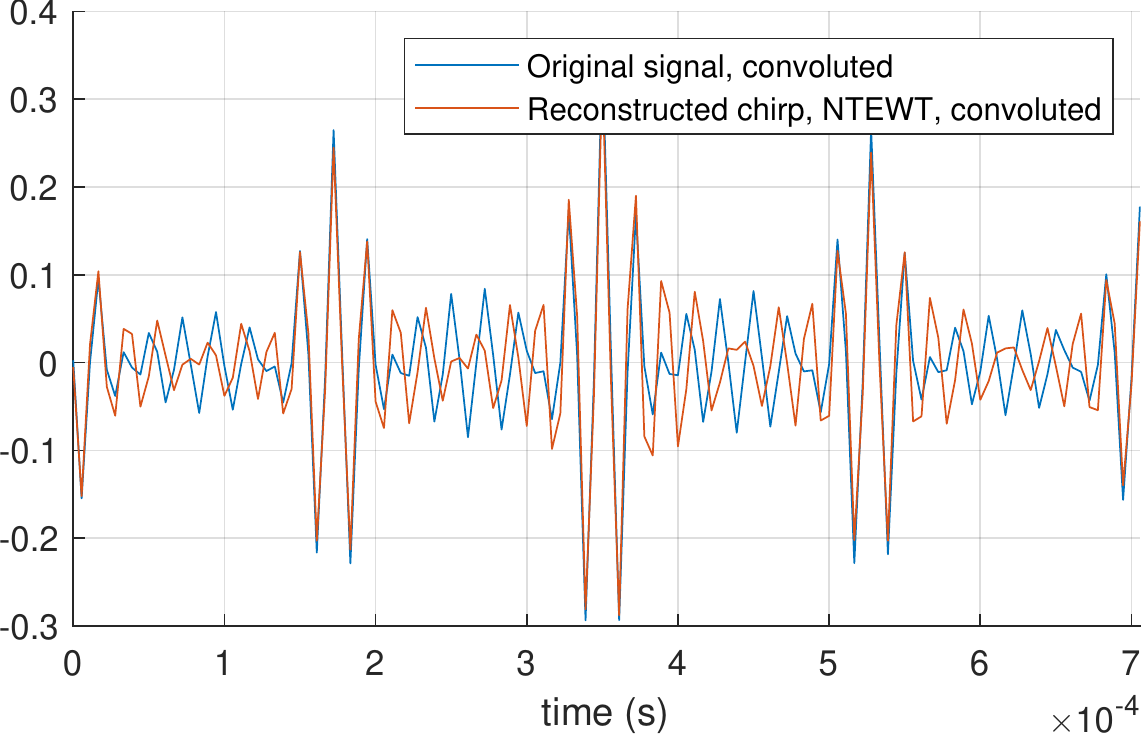}
		\caption{Matched filter output. Gaussian noise with $s=0.2$. $\epsilon=\num{2e-2}$.}
	\end{subfigure}
	\begin{subfigure}{0.45\textwidth}
		\includegraphics[width=\textwidth]{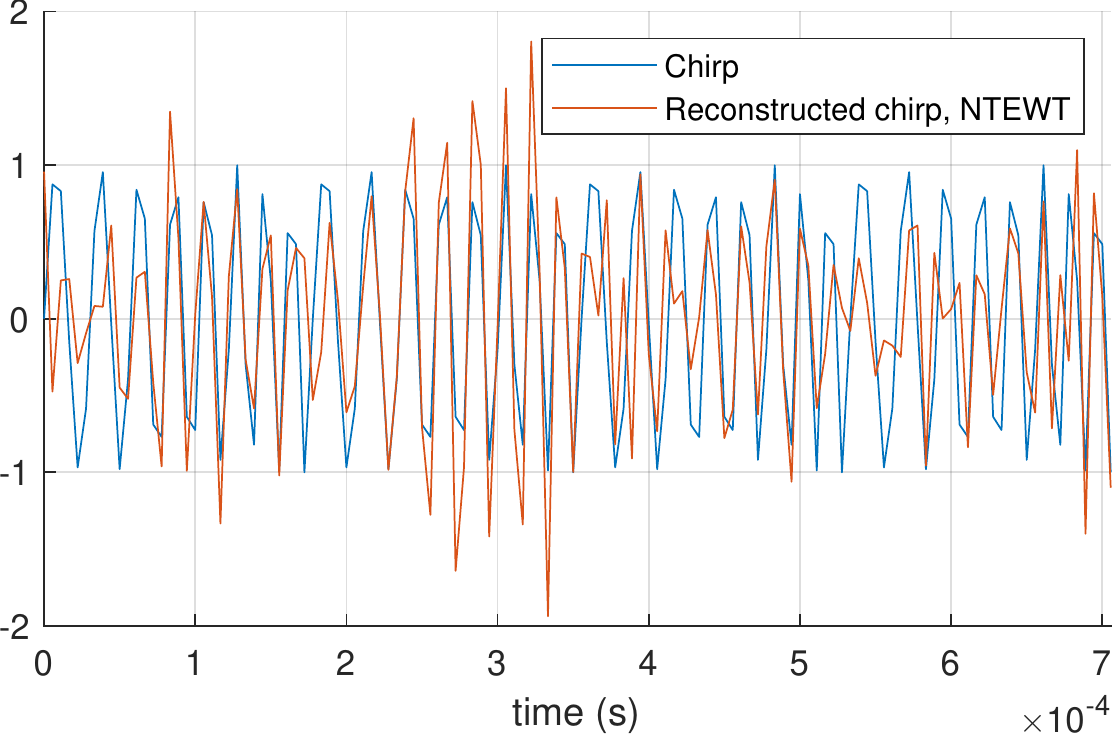}
		\caption{$x_{\text{fil}}[j]$. Gaussian noise with $s=0.4$. $\epsilon=\num{2e-2}$.}
	\end{subfigure}
	\hfill
	\begin{subfigure}{0.45\textwidth}
		\includegraphics[width=\textwidth]{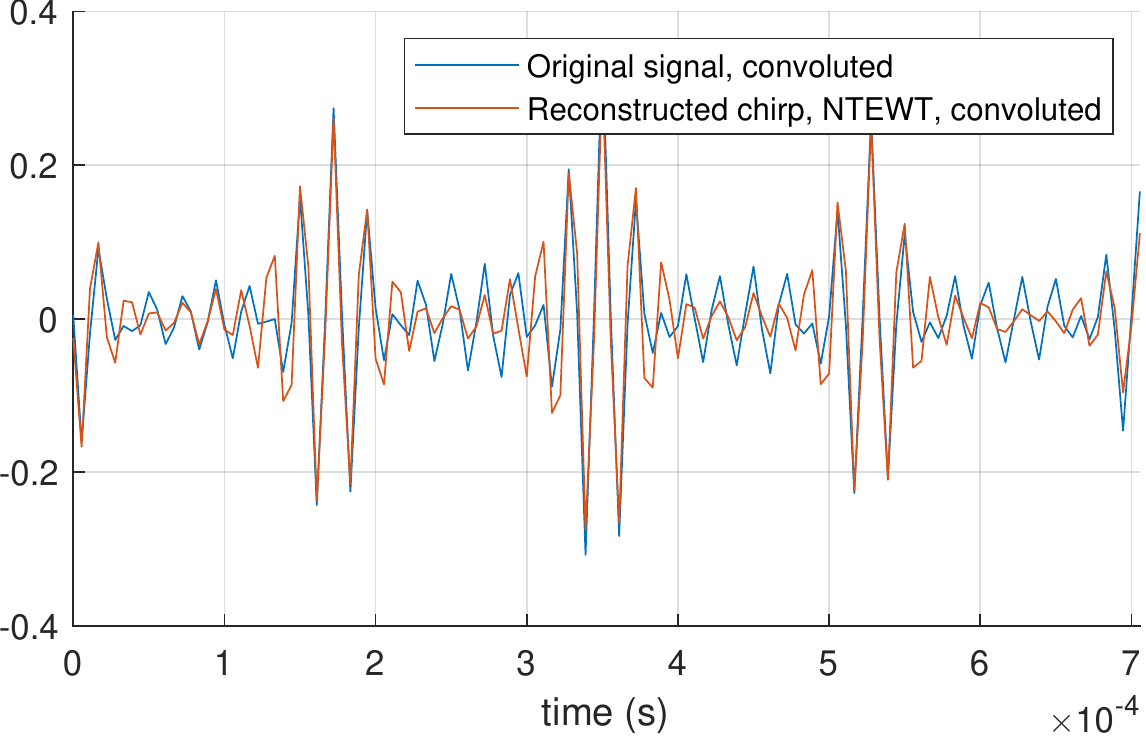}
		\caption{Matched filter output. Gaussian noise with $s=0.4$. $\epsilon=\num{2e-2}$.}
	\end{subfigure}
	\caption{(Test 4) (a), (c) and (d) The outputs $x_{\text{fil}}[j]$ of the NTEWT-based filter (reconstructed chirp) compared with the noiseless chirp train. (b), (d) and (f) Outputs of a matched chirp filter applied on the signals in Fig.~\ref{Test4Signal} (in blue) and on the NTEWT-filter outputs (in red). Analytic Morlet with $\sigma=3$.}
	\label{Test4fil}
\end{figure}

An evaluation of the NTEWT filter performance for different values of the wavelet's width parameter $\sigma$ is shown in Fig.~\ref{Test4NTEWT}. Optimal values of the reassignment tolerance parameter $\epsilon$ were found for each value of $\sigma$, and NTEWT scalograms of the noiseless signal (Fig.~\ref{Test4aSignal}) were calculated for those values (Fig.~\ref{Test4NTEWTb}, \ref{Test4NTEWTd} and \ref{Test4NTEWTf}). Due to the lower time-bandwidth product of the chirp pulses, the resolution of the CWT ridge-points is lower that in the former tests. According to the analysis, an optimal time-frequency resolution of the CWT scalogram in the chirp bandwidth is achieved with $\sigma=1$, but the reassignment is more performing with $\sigma=3$, with fewer artifacts appearing at the discontinuities between consecutive pulses. Fig.~\ref{Test4fil} compares the noiseless chirp train in Fig.~\ref{Test4aSignal} with the outputs of the NTEWT-based filter with $\sigma=3$ and shows the results of matched filtering on the original synthetic signals and the NTEWT filter outputs. As a template signal for matched filtering convolution, a single chirp pulse (Fig.~\ref{Test3aSignal}) was used. Again, it is observed that NTEWT-based filtering can potentially improve the resolution of chirp detection in signals with low and moderate noise levels. Border effects are lower than in previous tests because of the low value chosen for the parameter $\sigma$.

\subsection{Computational speed test: NTEWT-filtering time vs signal length}

\begin{figure}[H]
	\centering
	\includegraphics[width=0.45\textwidth]{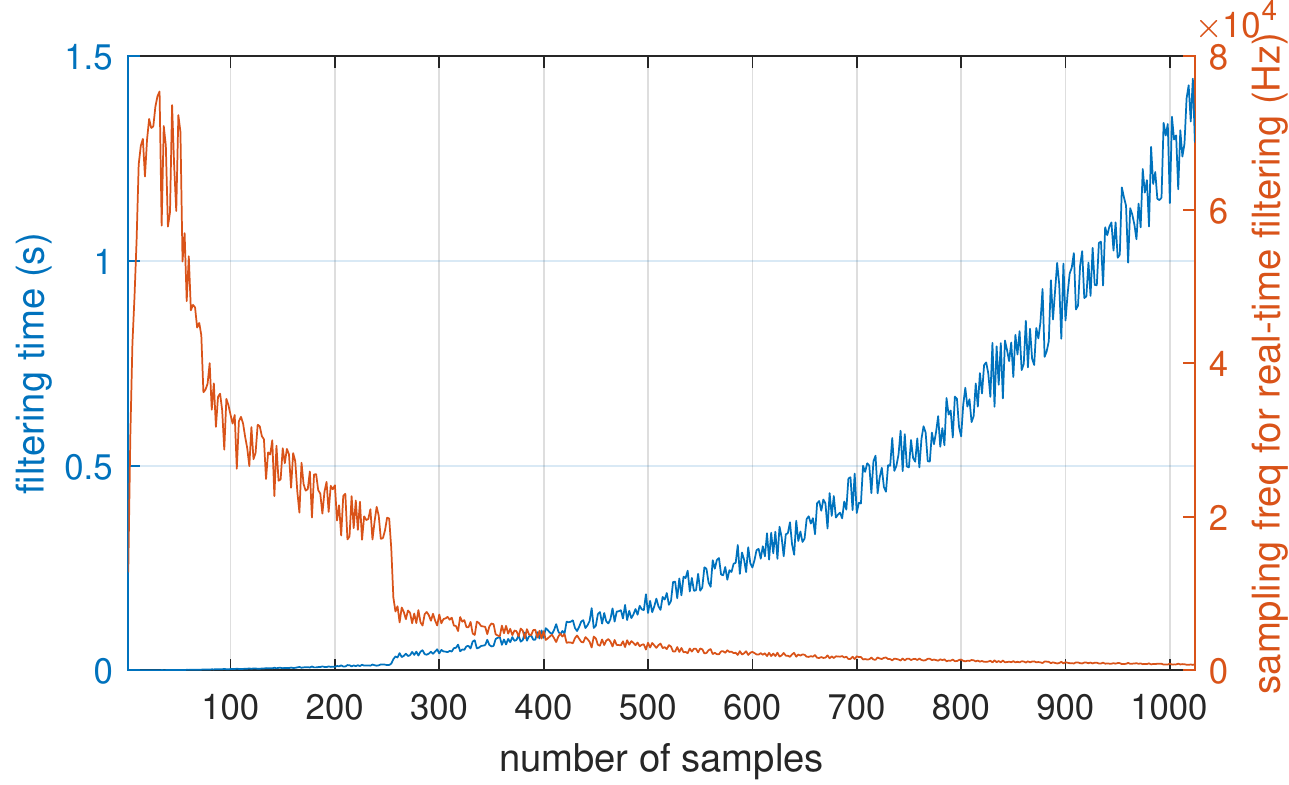}
	\caption{Relation between the number of samples of the input signal $x[j]$, the NTEWT-based filtering runtimes and the highest sampling frequencies guaranteeing feasibility of real-time NTEWT-based filtering.}
	\label{speedtest}
\end{figure}

A real-time applicability of a signal processing algorithm may be very advantageous in many fields. This section shows average runtimes of the proposed NTEWT-filtering algorithm (Alg.~\ref{alg:one}) for randomly-generated signals with even lengths ranging from 2 to 1024. For each even length in that range, Alg.~\ref{alg:one} was executed 50 times and the measured execution times were averaged. The results are shown in Fig.~\ref{speedtest} (blue curve). As expected, the average execution time increases with the square of the signal length. The red curve of Fig.~\ref{speedtest} shows the highest sampling frequency that would guarantee the feasibility of real-time NTEWT filtering with each signal length. It is calculated as the quotient between the numbers of samples (horizontal axis) and the filtering times (blue curve). It is evident from the results that a balance between signal length (which improves frequency resolution) and sampling frequency (which increases filter bandwidth) must be reached when applying the proposed filtering scheme to real-time processing.

As a side comment, we note that the NTEWT's execution time obtained for a signal length of 600 samples (around 0.3~s) is of the order of ten times lower than the computation time reported in Li et al \cite{li2023newton} for a signal with the same length. Therefore, it can be asserted that the proposed Algorithms~\ref{alg:one} and \ref{alg:two} represent a step forward in the quest for an efficient calculation of a signal's NTEWT with respect to the state of the art.

\section{Conclusions}\label{concls}

The results of the numerical experiments reported in this manuscript prove that the NTEWT, a TFA method developed by Li et al \cite{li2023newton} that combines the WT with the application of a fixed point operator for reassignment in the time direction, can be applied as a chirp filter with remarkable performance. This is because its fixed point reassignment operator discriminates weakly separated components that locally resemble linear chirps in the frequency domain, and impulsive time-domain signals such as chirp bursts are examples of such signals. Efficient numerical algorithms were proposed for the calculation of NTEWT TFRs and scalograms and for NTEWT-based filtering.

The feasibility of the proposed algorithm as a real-time filter was put to the test by measuring computation times for a range of signal lengths. Even if the algorithm is fast enough to enable its use on many applications, future works should address this issue even further by developing faster chirp reconstruction methods not necessarily requiring computation of an inverse WT through TFR integration. For example, a template chirp function could be defined and subsequently fitted to the signal's TFR fixed points using regression. Lower computation times would allow for a real-time application of NTEWT-based filtering in fields requiring ultra-high sampling frequencies and chirp rates for good performance, such as radar and sonar.

\section*{Funding}

This research was carried out during a post-doctoral research stay financed by the Spanish Ministry of Universities through the decree RD 289/2021 (Ayudas Margarita Salas 2021, grant number 28-1247030-89).

\printbibliography

\vfill

\minipage{0.24\textwidth}
	\includegraphics[width=\linewidth]{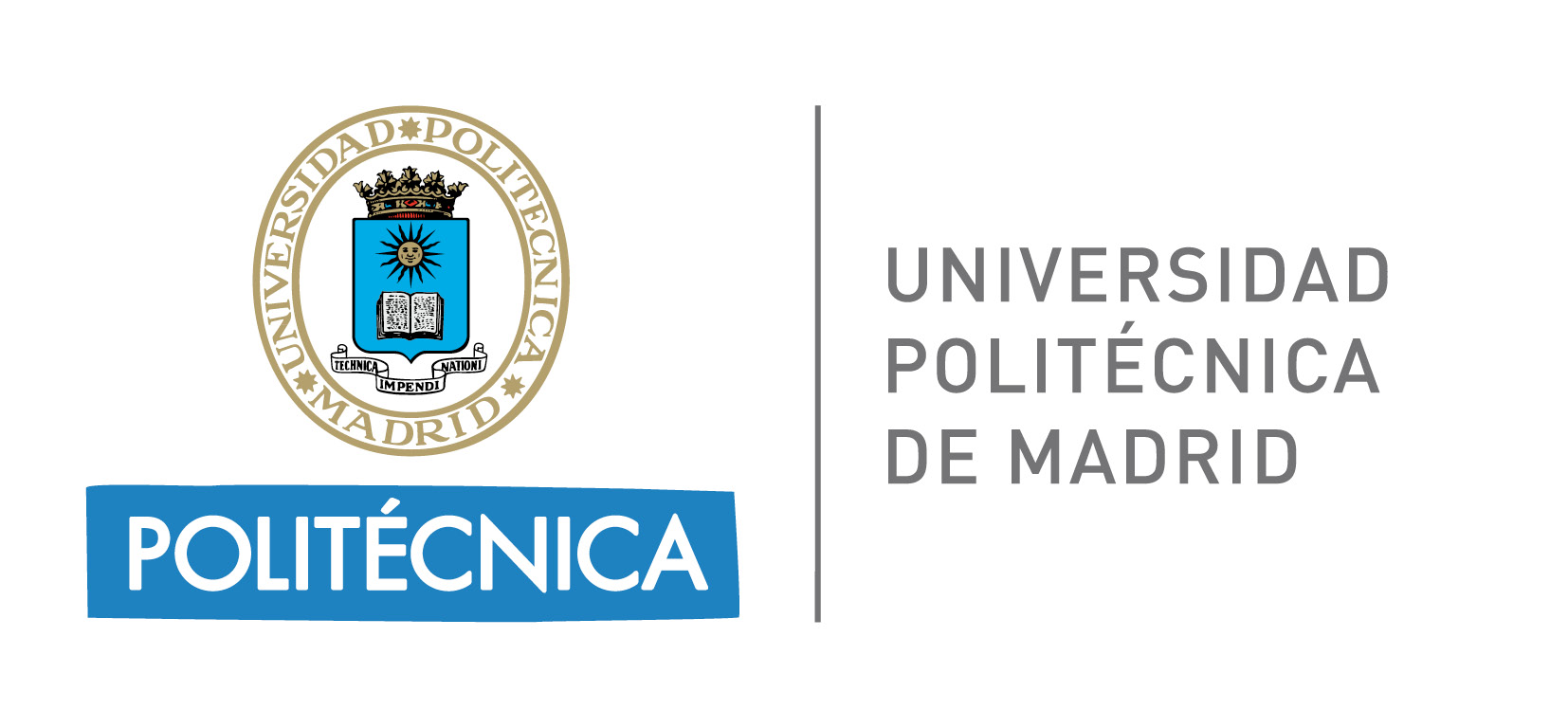}
\endminipage\hfill
\minipage{0.24\textwidth}
	\includegraphics[width=\linewidth]{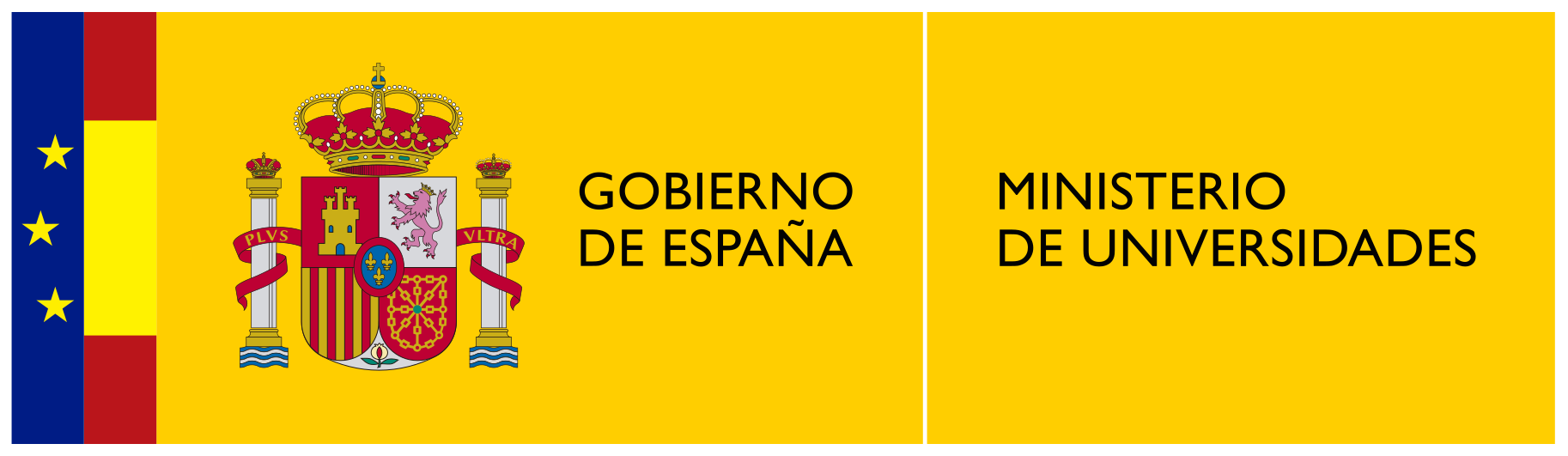}
\endminipage\hfill
\minipage{0.24\textwidth}
	\includegraphics[width=\linewidth]{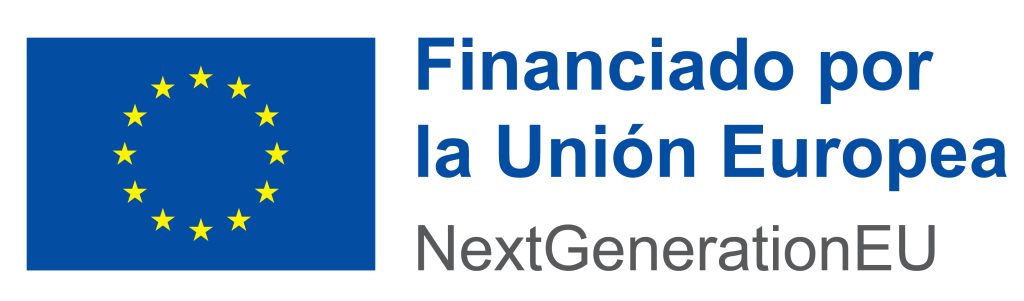}
\endminipage\hfill
\minipage{0.24\textwidth}
	\includegraphics[width=\linewidth]{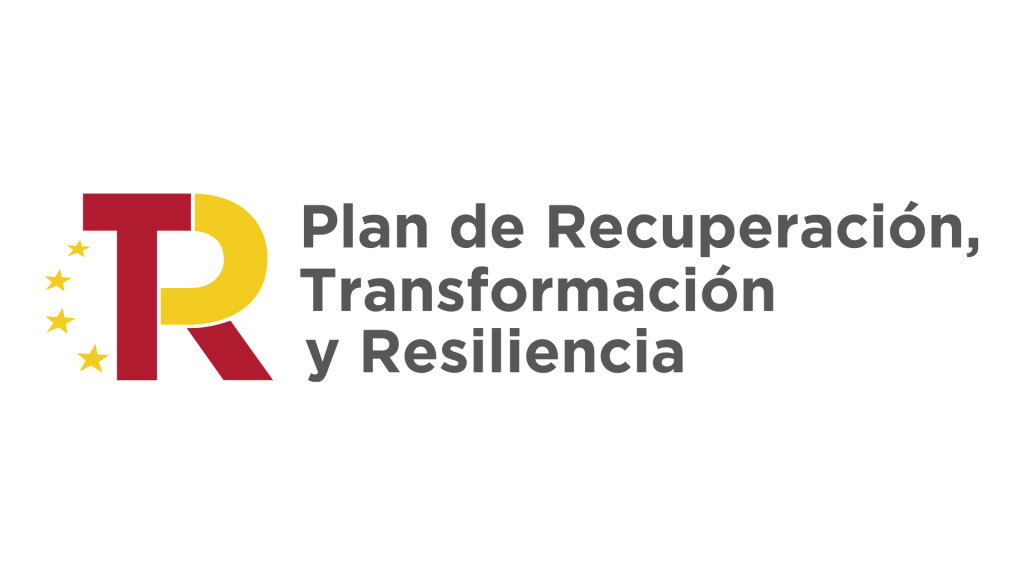}
\endminipage\hfill

\end{document}